\documentclass[fleqn,usenatbib]{mnras}

\usepackage{newtxtext,newtxmath}
\usepackage{tabularx}
\usepackage{lineno}

\usepackage{bm}
\usepackage{scalerel}
\usepackage{tikz}
\usetikzlibrary{svg.path}

\definecolor{orcidlogocol}{HTML}{A6CE39}
\tikzset{
  orcidlogo/.pic={
    \fill[orcidlogocol] svg{M256,128c0,70.7-57.3,128-128,128C57.3,256,0,198.7,0,128C0,57.3,57.3,0,128,0C198.7,0,256,57.3,256,128z};
    \fill[white] svg{M86.3,186.2H70.9V79.1h15.4v48.4V186.2z}
                 svg{M108.9,79.1h41.6c39.6,0,57,28.3,57,53.6c0,27.5-21.5,53.6-56.8,53.6h-41.8V79.1z M124.3,172.4h24.5c34.9,0,42.9-26.5,42.9-39.7c0-21.5-13.7-39.7-43.7-39.7h-23.7V172.4z}
                 svg{M88.7,56.8c0,5.5-4.5,10.1-10.1,10.1c-5.6,0-10.1-4.6-10.1-10.1c0-5.6,4.5-10.1,10.1-10.1C84.2,46.7,88.7,51.3,88.7,56.8z};
  }
}

\newcommand\orcidicon[1]{\href{https://orcid.org/#1}{\mbox{\scalerel*{
\begin{tikzpicture}[yscale=-1,transform shape]
\pic{orcidlogo};
\end{tikzpicture}
}{|}}}}

\usepackage[T1]{fontenc}
\usepackage{gensymb}
\usepackage{booktabs}
\usepackage{hyperref}

\usepackage{rotating}
\usepackage{rotfloat}

\DeclareRobustCommand{\VAN}[3]{#2}
\let\VANthebibliography\thebibliography
\def\thebibliography{\DeclareRobustCommand{\VAN}[3]{##3}\VANthebibliography}

\usepackage{graphicx}	
\usepackage{amsmath}	
\usepackage{algorithm}
\usepackage{algpseudocode}

\title[The Lite Method]{Rapid Construction of Joint Pulsar Timing Array Datasets:\\ The Lite Method}

\author[Larsen, Mingarelli, Baker et al.]{\parbox{\textwidth}{
B.~Larsen$^{\orcidicon{0000-0001-6436-8216}}$, $^{1}$\thanks{E-mail: bjorn.larsen@yale.edu} 
C.~M.~F.~Mingarelli$^{\orcidicon{0000-0002-4307-1322}}$,$^{1}$ 
P.~T.~Baker$^{\orcidicon{0000-0003-2745-753X}}$,$^{2}$ 
J.~S.~Hazboun$^{\orcidicon{0000-0003-2742-3321}}$,$^{3}$ 
S.~Chen$^{\orcidicon{0000-0002-3118-5963}}$,$^{4,5}$ 
L.~Schult$^{\orcidicon{0000-0001-6425-7807}}$,$^{6}$
S.~R.~Taylor$^{\orcidicon{0000-0003-0264-1453}}$,$^{6}$ 
J.~Simon$^{\orcidicon{0000-0003-1407-6607}}$,$^{7}$ 
J.~Antoniadis$^{\orcidicon{0000-0003-4453-3776}}$,$^{8}$ 
J.~Baier$^{\orcidicon{0000-0002-4972-1525}}$,$^{3}$ 
R.~N.~Caballero$^{\orcidicon{0000-0001-9084-9427}}$,$^{9}$ 
A.~Chalumeau$^{\orcidicon{0000-0003-2111-1001}}$,$^{10}$ 
Z.~Chen$^{\orcidicon{0000-0001-7016-9934}}$,$^{11,12}$ 
I.~Cognard$^{\orcidicon{0000-0002-1775-9692}}$,$^{13,14}$ 
D.~Deb$^{\orcidicon{0000-0003-4067-5283}}$,$^{15}$
V.~Di Marco$^{\orcidicon{0000-0003-3432-0494}}$,$^{16,17,18}$ 
T.~Dolch$^{\orcidicon{0000-0001-8885-6388}}$,$^{19}$ 
I.~O.~Eya$^{\orcidicon{0000-0002-9693-7804}}$,$^{20}$ 
E.~C.~Ferrara$^{\orcidicon{0000-0001-7828-7708}}$,$^{21,22,23}$ 
K.~A.~Gersbach$^{\orcidicon{0009-0009-5393-0141}}$,$^{6}$ 
D.~C.~Good$^{\orcidicon{0000-0003-1884-348X}}$, $^{24}$ 
H.~Hu$^{\orcidicon{0000-0002-3407-8071}}$,$^{25}$
A.~Kapur$^{\orcidicon{0009-0001-5071-0962}}$,$^{18,26}$ 
S.~Kala$^{\orcidicon{0000-0003-2379-0204}}$,$^{15}$ 
M.~Kramer$^{\orcidicon{0000-0002-4175-2271}}$,$^{24}$ 
M.~T.~Lam$^{\orcidicon{0000-0003-0721-651X}}$,$^{27,28,29}$ 
W.~G.~Lamb$^{\orcidicon{0000-0003-1096-4156}}$,$^{6}$ 
T.~J.~W.~Lazio,$^{30}$ 
K.~Liu$^{\orcidicon{0000-0002-2953-7376}}$,$^{4,5}$ 
Y.~Liu$^{\orcidicon{0000-0003-0713-6640}}$,$^{4}$
M.~McLaughlin$^{\orcidicon{0000-0001-7697-7422}}$,$^{31}$ 
D.~J.~Nice$^{\orcidicon{0000-0002-6709-2566}}$,$^{32}$ 
B.~B.~P.~Perera$^{\orcidicon{0000-0002-8509-5947}}$,$^{33}$ 
A.~Petiteau$^{\orcidicon{0000-0002-7371-9695}}$,$^{34}$ 
S.~M.~Ransom$^{\orcidicon{0000-0001-5799-9714}}$,$^{35}$ 
D.~J.~Reardon$^{\orcidicon{0000-0002-2035-4688}}$,$^{17,36}$ 
C.~J.~Russell$^{\orcidicon{0000-0002-1942-7296}}$,$^{37}$ 
G.~M.~Shaifullah$^{\orcidicon{0000-0002-8452-4834}}$,$^{38,39,40}$
L.~Speri$^{\orcidicon{0000-0002-5442-7267}}$,$^{41}$ 
A.~Srivastava$^{\orcidicon{0000-0003-3531-7887}}$,$^{42}$ 
G.~Theureau$^{\orcidicon{0000-0002-3649-276X}}$,$^{13,14}$ 
J.~Wang$^{\orcidicon{0000-0001-9782-1603}}$,$^{43}$ 
J.~Wang$^{\orcidicon{0000-0003-1933-6498}}$,$^{44}$ 
and L.~Zhang$^{\orcidicon{0000-0001-8539-4237}}$$^{45}$} 
\vspace{0.4cm} \\
Affiliations are at the end of the paper
}

\date{Accepted 2025 August 26. Received 2025 August 25; in original form 2025 March 26}

\pubyear{2025}

\begin{document}
\label{firstpage}
\pagerange{\pageref{firstpage}--\pageref{lastpage}}
\maketitle


\begin{abstract}
The International Pulsar Timing Array (IPTA)'s second data release (IPTA DR2) combines decades of observations of 65 millisecond pulsars from 7 radio telescopes. IPTA datasets should be the most sensitive datasets to nanohertz gravitational waves (GWs), but take years to assemble, often excluding valuable recent data. To address this, we introduce the IPTA ``Lite'' analysis, where a Figure of Merit is used to select an optimal PTA dataset to analyze for each pulsar, enabling immediate access to new data and preliminary results prior to full combination. We test the capabilities of the Lite analysis using IPTA DR2, finding that ``DR2 Lite’’ can be used to detect the common red noise process with an amplitude of $A = 4.8^{+1.8}_{-1.8} \times 10^{-15}$ at $\gamma = 13/3$. This amplitude is slightly large in comparison to the combined analysis, and likely biased high as DR2 Lite is more sensitive to systematic errors from individual pulsars than the full dataset. Furthermore, although there is no strong evidence for Hellings-Downs correlations in IPTA DR2, we still find the full dataset is better at resolving Hellings-Downs correlations than DR2 Lite. Alongside the Lite analysis, we also find that analyzing a subset of pulsars from IPTA DR2, available at a hypothetical ``early’’ stage of combination (EDR2), yields equally competitive results as the full dataset. Looking ahead, the Lite method will enable rapid synthesis of the latest PTA data, offering preliminary GW constraints before the superior full dataset combinations are available.
\end{abstract}




\section{Introduction} \label{sec:intro}

Pulsar Timing Arrays (PTAs) are experiments for detecting low-frequency gravitational waves (GWs), offering unprecedented access to nanohertz GWs \citep{Sazhin1978, Detweiler1979, HellingsDowns1983}. By monitoring the ultra-stable arrival times of radio pulses from millisecond pulsars (MSPs)—nature's most precise clocks—we can build a galaxy-scale GW detector. PTAs use deviations in the time of arrivals (TOAs) induced by GWs to infer their presence, enabling the detection of signals from supermassive black hole binaries (SMBHBs; \citealt{Begelman+1980}), the stochastic gravitational wave background (GWB) which should arise from their cosmic merger history \citep{Rajagopal1995, Volonteri+2003}, and potentially new physics \citep{Lasky+2016, Caprini2018, Afzal+2023}.  

Currently, PTA collaborations around the globe, including the European PTA (EPTA; \citealt{EPTADR1}), the North American Nanohertz Observatory for Gravitational Waves (NANOGrav; \citealt{Ransom2019}), the Parkes PTA (PPTA; \citealt{PPTADR1}), the Chinese PTA (CPTA; \citealt{Lee2016}), the Indian PTA (InPTA; \citealt{Joshi2018}), the MeerKAT PTA (MPTA; \citealt{Miles2023}), and the Fermi $\gamma$-ray PTA ($\gamma$PTA; \citealt{FermiPTA_2021}) provide PTA datasets of varying sensitivity and duration. The International Pulsar Timing Array (IPTA) is a consortium of PTA collaborations, with the first and second data releases, IPTA DR1 \citep{Verbiest+2016} and IPTA DR2 \citep{Perera+2019}, each combining pulsar observations from EPTA, NANOGrav, and PPTA into one coherent dataset. A third IPTA data release, IPTA DR3, is currently in development, with the potential to include data from all current PTAs.

Data combination improves a PTA's sensitivity to GWs by increasing the pulsars' effective observation timespan, cadence, and radio frequency coverage, as well as the PTA's overall number of pulsars and sky coverage. This was expected by \citet{siemens2013} and verified in the IPTA DR2 GWB search \citep{Antoniadis+2022}, which found evidence for a common red noise (CRN; also known as CURN, \citealt{Chen+2021, 3P+2024}) process, a signature of an emerging GWB signal, with greater significance than the individual constituent datasets comprising IPTA DR2. This provided a robust confirmation of the CRN first found by the regional PTA collaborations \citep{Arzoumanian2020, PPTADR2, Chen2021}.

\begin{table*}
    \vspace{-0.5\baselineskip}
    \centering
    \begin{tabular}{| c | c |}
        \hline Name & Short Description \\
        \hline Full DR2 & The fully-combined IPTA DR2 from \citet{Perera+2019}, with $T_{\mathrm{obs}} > 3$ years filter from \citet{Antoniadis+2022} for 53 pulsars in total. \\
        DR2 Lite & The Lite dataset presented in this work, with a total of 53 pulsars selected using the FoM from single-PTA data subsets of Full DR2. \\
        EDR2 & An ``early'' subset of Full DR2 which includes fully-combined data for only 22 pulsars corresponding to the highest FoM in DR2 Lite. \\
        \hline
    \end{tabular}
    \caption{Names and short descriptions of the three datasets we analyze in this work. \S\ref{section:DR2} provides further details on each of the datasets.}
    \label{tab:datasets}
    \vspace{-\baselineskip}
\end{table*}

While it did not produce the first published measurement of a CRN process, we point out that IPTA DR2 could have presented the earliest opportunity to detect the CRN, as regional PTAs needed to collect approximately $\sim$2 more years of data than was used for IPTA DR2 in order to sufficiently resolve the signal. The delay in the IPTA DR2 analysis largely resulted from the resource-intensive process of data combination, which requires meticulous handling of different kinds of data, alignment of timing models, fitting for instrumental offsets, and iterative noise modeling across datasets \citep{Verbiest+2016, Perera+2019}. This effort typically takes several years to complete, meaning by the time a combined dataset is released, portions of the underlying data are already outdated. To illustrate the timescales, IPTA DR2 contains the NANOGrav 9-yr dataset \citep{NG9}, but IPTA DR2 was not published until 3 years later \citep{Perera+2019}. By the time the IPTA DR2 GWB search was carried out \citep{Antoniadis+2022}, the NANOGrav 12.5-yr dataset was already used to detect the CRN \citep{Arzoumanian+2020}. Thus, the CRN could theoretically have been measured 3 years in advance of \citet{Arzoumanian+2020} if IPTA DR2 was constructed immediately. Given the long timescales required for new GW signals to emerge in PTA datasets, this motivates the need to either improve the speed of data combination or explore alternative methods for analyzing joint-PTA datasets, supplementing the eventual results of a fully-combined dataset.

To address this, we introduce a novel, resource-efficient approach we call the ``Lite'' method. Instead of immediately performing full data combination, we evaluate pulsar datasets which have already been produced by individual PTAs, and select the most informative dataset for each pulsar using a Figure of Merit (FoM), which quantifies sensitivity to a GW signal based on the dataset properties using the theoretical scaling laws for the signal-to-noise ratio (S/N). For example, the GWB S/N from \citet{siemens2013} suggests a FoM which increases as total observing time increases, as observation cadence increases, and as RMS white noise residual decreases. By selecting each pulsar's data based on the FoM, the Lite dataset achieves the maximum theoretical sensitivity to GW signals possible among all available data prior to performing data combination. As such, a Lite dataset analysis may provide an early look into what may result from a fully-combined analysis. A Lite dataset will also be less computationally intensive to analyze than its fully-combined counterpart due to the reduced data volume.

Data combination is a slow, intensive process, with combined datasets built up one pulsar at a time. It is tempting then to also consider the result of analyzing an \emph{early} or \emph{intermediate} combined dataset, which includes just the first set of pulsars which have had their data combined. The FoM suggests which pulsars to combine first: combining pulsars in order, starting from highest FoM to lowest, will maximize the GWB sensitivity of any intermediate combined dataset. This practice already has precedent within PTA analyses \citep{Babak+2016, Speri+2023}. For example, the creation of EPTA DR2 with 25 pulsars \citep{EPTADR2} was preceded by a version of EPTA DR2 using 6 pulsars \citep{EPTA_6psr}, which were originally selected based on their expected S/N for continuous GWs \citep{Babak+2016}. The 25 pulsars used for the EPTA DR2 GWB search \citep{EPTADR2, EPTADR2_gwb} were then selected to optimize the theoretical S/N of the Hellings-Downs curve, following the method from \citet{Speri+2023}.

Here we use IPTA DR2 as a test case to assess the benefits of performing GW searches using a \emph{Lite} dataset, an \emph{early-combined} dataset, and a \emph{fully-combined} dataset, reflecting the stages in which future combined data may be analyzed. Table~\ref{tab:datasets} provides short names and descriptions of each of these datasets for ease of reference. Specifically, we test how the detection statistics and upper limits for a GWB evolve as more data are combined. This analysis framework thus quantifies the benefits and drawbacks of a rapid, on-the-fly Lite analysis, as well as the superior sensitivity offered by the full data combination.

The information from this analysis will also be valuable for the interpretation of present-day datasets. \citet{3P+2024} performed comparisons and joint-analyses of the NANOGrav 15 year dataset \citep{Agazie2023_data, Agazie+2023_gwb}, EPTA+InPTA DR2 \citep{EPTADR2, EPTADR2_gwb}, and PPTA DR3 \citep{PPTA2023_data, PPTADR3_gwb}. The factorized likelihood cross-PTA analyses in \citet{3P+2024} are similar in spirit to the Lite analysis method we present here, and the results suggest that IPTA DR3 will place the most decisive detection to date of the Hellings-Downs curve, which is the definitive signature of an isotropic GWB imprinted in the cross-correlations between pulsar timing residuals \citep{HellingsDowns1983}. Our Lite analysis of IPTA DR2 will therefore be useful to calibrate expectations for IPTA DR3.

Our paper is laid out as follows: In \S\ref{section:DR2}, we detail IPTA DR2, the FoM, and our method of creating DR2 Lite from IPTA DR2 as a starting point. In \S\ref{section:Methods}, we describe the PTA likelihood, models, and parameters used in our Bayesian analysis of each dataset. In \S\ref{section:Results}, we assess how the statistics for both a CRN and an Hellings-Downs cross-correlated GWB evolve throughout each stage of data combination, as well as the impact of data combination on single pulsar noise characterization and ensemble noise properties. In \S\ref{section:Discussion}, we discuss our results and future directions for the Lite method.

\section{IPTA DR2 \& the Lite dataset}
\label{section:DR2}

IPTA DR2 is the most recent IPTA-combined dataset, fully detailed in \citet{Perera+2019}. Steps required to create the combined dataset include standardization of TOA flags (metadata), fitting for instrumental offsets, implementation of comprehensive timing and noise models, and simultaneous/iterative fits to the pulsar timing and noise model parameters \citep{Verbiest+2016, Perera+2019}. IPTA DR2 includes TOAs from the NANOGrav 9-yr dataset \citep{NG9}, EPTA DR1 \citep{EPTADR1}, and PPTA DR1 \citep{PPTADR1, Reardon+2016}, as well as legacy NANOGrav timing data for PSRs J1713+0747, J1857+0943, and J1939+2134 \citep{Kaspi+1994, Zhu+2015} and extended PPTA data for PSRs J0437--4715, J1744--1134, J1713+0747, and J1909--3744 \citep{Shannon2015}. In total IPTA DR2 includes a total of 65 millisecond pulsars, with datasets spanning 0.5 to 30 years, measured across 7 different telescopes. IPTA DR2 also features two versions, designated VersionA and VersionB, which were each created using different noise models. Throughout this work we use only VersionB. A few pulsars in IPTA DR2 are known by their Besselian names in NANOGrav datasets, though in this work we use their the Julian names: J1857+0943 (B1855+09), J1939+2134 (B1937+21), and J1955+2908 (B1953+29).


\begin{figure*}
    \centering
    \includegraphics[width=\linewidth]{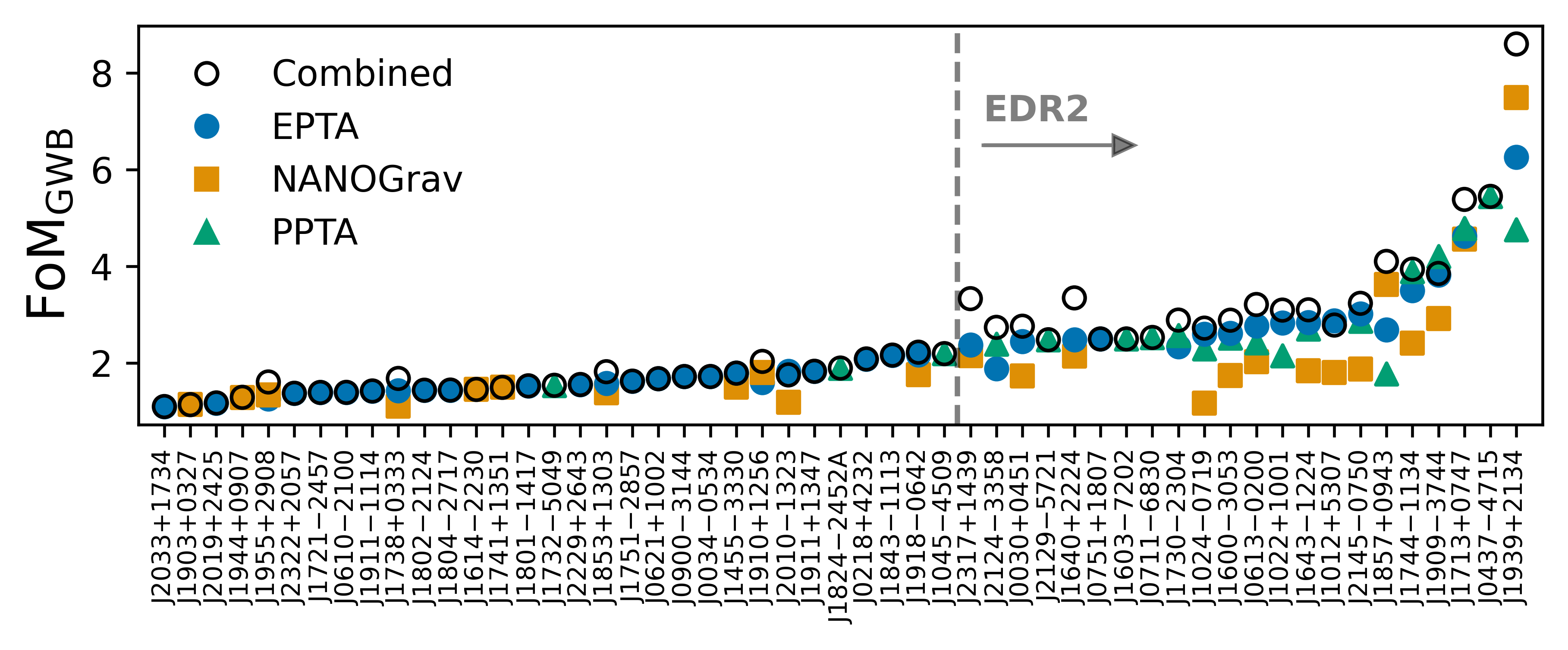}
    \vspace{-1.5\baselineskip}
    \caption{Figures of Merit (FoMs; Equation~\ref{eq:FoM}) computed for each pulsar in IPTA DR2. The different markers indicate whether the FoM is computed either using the EPTA (blue circles), NANOGrav (orange squares), or PPTA (green triangles) TOAs, while the open black circles indicate the pulsar's FoM computed using all TOAs together. The PTA data with the largest FoM for each pulsar is used in DR2 Lite. From left to right, the pulsars are ranked in order from lowest to highest FoM using whichever PTA's data are included in the Lite dataset. All pulsars to the right of the dashed vertical line are included in EDR2.}
    \vspace{-0.5\baselineskip}
    \label{fig:FoM}
\end{figure*}

\citet{Antoniadis+2022} carried out a GWB search on a subset of 53 pulsars from IPTA DR2 with $>3$ years of data. Pulsars with shorter timespans do not resolve the GWB at low frequencies, and their timing models may not yet be converged \citep{AndrewsLamDolch2021}. This dataset, designated here as Full DR2, is our benchmark against which to compare the results of the Lite analysis. The search from \citet{Antoniadis+2022} yielded a strong detection of a CRN process (the auto-correlated component of a GWB). Detection statistics for Hellings-Downs correlations (the cross-correlated component of a GWB) were also computed, but the values were considered insufficient for a detection. 

IPTA DR2's very long observation timespan of 30.2 years results from the inclusion of legacy data no longer used in some more recent PTA data releases \citep{EPTADR2, Agazie2023_data, Zic+2023}. Among these legacy data include TOAs which have been observed only at single radio frequencies, which are suboptimal for accurately characterizing DM variations \citep{ShannonCordes2017, Lam+2018_bandwidths, Sosa+2024}. It has since been shown empirically with EPTA DR2 \citep{EPTAnoise2023, EPTADR2_gwb} and simulations \citep{Ferranti+2024} that including these types of single-frequency data can reduce the sensitivity of the PTA to cross-correlations between pulsar pairs, which must be used to resolve the Hellings-Downs curve. For simplicity and consistency with \citet{Antoniadis+2022}, we include these TOAs in all versions of our analysis, but highlight that their presence should be considered during the interpretation of our results. We reserve an analysis assessing the impacts of legacy data in IPTA datasets for future work.

\subsection{Selecting pre-combined pulsar data using a Figure of Merit}

We next detail the methods of creating an IPTA Lite dataset composed of \emph{pre}-combined data from individual PTA datasets. In the presence of multiple datasets for a given pulsar, we select whichever data maximizes a FoM. The FoM encodes the theoretical sensitivity of a pulsar to a particular GW signal, based solely on the properties of the pulsar's TOAs. For this work we define our FoM according to the scaling laws for a GWB with spectral index $\gamma = 13/3$ in the intermediate regime where the lowest frequencies of the GWB have risen above the white noise level \citep{siemens2013},
\begin{equation}
    \label{eq:FoM}
    \mathrm{FoM}_\mathrm{GWB} = \left(\frac{T_{\mathrm{obs}}}{\left(\langle{\sigma_\mathrm{TOA}}\rangle^2\langle\Delta t\rangle\right)^{3/13}}\right)^{1/2},
\end{equation}
where $T_{\mathrm{obs}}$ is the pulsar's total observation timespan, $\langle{\sigma_\mathrm{TOA}}\rangle$ is the average (harmonic mean) TOA error, and $\langle\Delta t\rangle$ is the average (geometric mean) time between observations. Intuitively, Equation~\ref{eq:FoM} rewards datasets with high data quantity, i.e. long timespans $T_{\mathrm{obs}}$ and high data cadence $c = 1/\Delta t$, as well as datasets with high data quality, i.e. smaller errors $\sigma_{\mathrm{TOA}}$. The factor of $3/13$ results from the predicted spectral index of the emerging GWB. The intermediate regime GWB S/N is also proportional to the number of pulsars, but this is not included in the FoM to select which data to use for a single pulsar. Different Lite datasets can be also curated for different nHz GW searches as the theoretical S/N for each type of GW signal will follow a different scaling law. We additionally present the FoM for continuous GWs and for GW bursts with memory in Appendix~\ref{appendix:alt_FoM}, but we do not explore these in further in this work.

\begin{figure*}
    \centering
    \includegraphics[width=\textwidth]{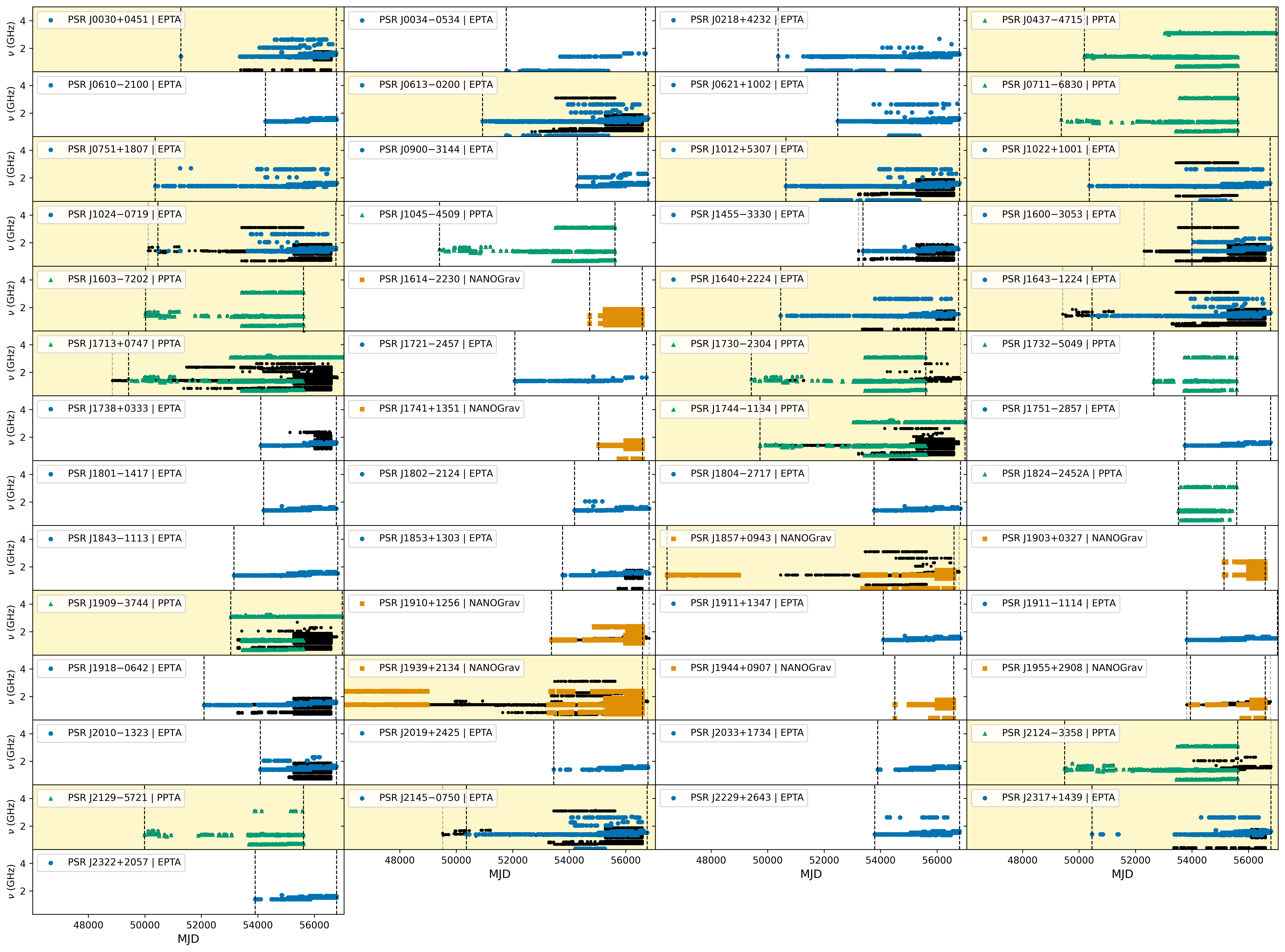}
    \vspace{-\baselineskip}
    \caption{All TOAs in DR2 Lite (colored), compared against all TOAs in Full DR2 (black; \citealt{Perera+2019}), plotted in terms of observation frequency in GHz vs time in MJD for each pulsar. The DR2 Lite TOAs from different PTAs are distinguished as follows: EPTA (blue circles), NANOGrav (orange squares), PPTA (green triangles). Vertical lines show the start and end times of each pulsar's Lite dataset (dashed black), and each pulsar's combined dataset (dashed grey). The 22 pulsar datasets used in EDR2 are highlighted with a gold background. IPTA DR2 contains an additional 44 pulsars not shown, 32 of which with $>3$ years of data were used in the \citet{Antoniadis+2022} GWB search.}
    \vspace{-0.5\baselineskip}
    \label{fig:TOAs}
\end{figure*}

To create a Lite dataset, first decide which GW signal to search for and select the FoM. Next, iterate through all pulsars of interest. Pulsars timed by a single PTA require no extra work to include them in the Lite dataset, aside from ensuring terrestrial clock references and solar system ephemeris versions are consistent and up to date among all pulsars. If a pulsar is timed by multiple PTAs, compute the FoM from each PTA's data for that pulsar, then take whichever FoM is largest and add the corresponding PTA's data to the Lite dataset. The Lite dataset is therefore the bespoke composition of uncombined pulsar datasets across different PTAs, which may be used from there to perform a joint GW search. The FoM-based selection approach has the advantage that it is purely based on the statistical properties of the dataset itself and is agnostic to which PTA timed it. Lite datasets may be created immediately from the latest PTA data releases.

\subsection{Creating Intermediate Datasets from IPTA DR2}

We next detail the specifics behind the curation of our IPTA DR2-based data subsets, which together with Full DR2 are summarized Table~\ref{tab:datasets}. IPTA DR2 is the most recent fully-combined IPTA dataset and is therefore an excellent dataset to test the performance of the Lite analysis. Here we choose to create DR2 Lite \emph{starting} from the IPTA DR2 release. This choice ensures the timing models are identical and thus any difference in GW sensitivity between DR2 Lite and Full DR2 results purely from the difference in data volume. In principle, one should also refit the timing models after reducing the data volume, or at least check to ensure the remaining timing residuals are within the regime of the linear timing model used during GW analyses. We empirically found the latter assumption to hold for the Lite version of IPTA DR2. Additionally, a maximally sensitive ``early'' version of a fully-combined dataset will start with a subset of pulsars that maximize the FoM. We use IPTA DR2 to create this hypothetical ``early'' dataset, which we call EDR2, drawing inspiration from the \textit{Gaia} data releases \citep{Gaia2021}.

To create DR2 Lite starting in IPTA DR2, we first isolate each PTA's TOAs in each pulsar and compute the FoM from equation~\ref{eq:FoM}. Each pulsar in DR2 Lite then keeps only the PTA data with the largest FoM. Figure~\ref{fig:FoM} visualizes the results of this process, by comparing the FoM computed for each PTA and each pulsar. DR2 Lite in total uses EPTA data for 33 pulsars, NANOGrav data for 8 pulsars, and PPTA data for 12 pulsars. Figure~\ref{fig:FoM} also shows each pulsar's FoM computed from the fully-combined dataset -- in nearly all cases the combined data result in a higher FoM, as expected from Equation~\ref{eq:FoM}. The FoM is slightly lower only for PSRs J1012+5307 and J1909--3744 using the combined data. Following Equation~\ref{eq:FoM}, this results if the newly combined TOAs have much larger errors on average than the TOAs included in the Lite dataset and the timespan and data cadence do not appreciably increase in contrast.

We also use the FoM distribution in Figure~\ref{fig:FoM} to select which pulsars to include in EDR2. Specifically, we rank the pulsars in order of highest to lowest FoM, as computed from DR2 Lite. This ranking represents an optimal order for combining the data to maximize the sensitivity of the early-combined dataset to a GWB. For EDR2, we choose the 22 highest ranked pulsars, represented by all pulsars to the right of the vertical line in Figure~\ref{fig:FoM}. We choose to use a 22 pulsar cutoff for this analysis for a number of reasons, though the exact number is ultimately arbitrary. Namely, this number should be large enough to avoid bias in GWB statistics due to a finite number of pulsars \citep{Johnson+2022}, but still represents less than half of the total pulsars intended for the full analysis. This also cleanly cuts off the dataset at PSR J2317+1439, which sees a large boost in the FoM post-combination based on Figure~\ref{fig:FoM}. Finally, this was also selected as a rough match for the number of pulsars with data combined for the upcoming IPTA DR3 at the time of performing this analysis \citep{Good+2023}.

Figure~\ref{fig:TOAs} further visualizes each dataset by displaying the observation times and radio frequencies of each TOA. Colored markers represent the single-PTA data used in DR2 Lite, while colored + black markers represent all data used in Full DR2. EDR2 pulsars are highlighted in gold. Several pulsars (e.g. PSR J0437$-$4715) are timed only by a single PTA, therefore their Lite and combined datasets are identical. Other pulsar's combined data (e.g. PSR J1713+0747) present a clear advantage in total radio band coverage and observation cadence, especially in the latter half of the dataset. For reproducibility, code for creating Lite datasets from IPTA DR2 can be found in the public IPTA github repository.\footnote{\url{https://github.com/ipta/IPTA_DR2_analysis/blob/master/gen_FoM_dataset.ipynb}}

\section{Analysis Methods}
\label{section:Methods}

We closely follow the methods and conventions in \citet{Antoniadis+2022} for the Bayesian analysis of IPTA DR2 and its corresponding Lite dataset. This will be review for readers familiar with the prior work, although in \S\ref{subsection:noise_models} we also update the pulsar noise models over those used in \citet{Antoniadis+2022}. We use a multivariate Gaussian likelihood to represent our timing residual vector $\bm{\delta t}$ under the full signal model $\mathcal{M}$, expressed compactly as
\begin{align}
    \mathcal{L}(\bm{\delta t}|\bm{\eta},\mathcal{M}) &= \frac{\exp\left(-\frac{1}{2}\left(\bm{\delta t} - \bm{s}\right)^T\mathbf{C}^{-1}\left(\bm{\delta t} - \bm{s}\right)\right)}{\sqrt{\det(2\pi\mathbf{C})}},
\end{align}
where $\bm{\eta}$ is a vector of GW signal and noise model (hyper)parameters, $\bm{s}$ is the mean model accounting for any signals we choose to model deterministically, and $\mathbf{C}$ is the $N_{\mathrm{TOA}} \times N_{\mathrm{TOA}}$ covariance matrix \citep{vanHaasterenVallisneri2014}. The covariance matrix is expressed $\mathbf{C} = \mathbf{N} + \mathbf{TBT}^T$, where $\mathbf{N}$ is a block diagonal matrix accounting for TOA errors and additional white noise parameters. All remaining signals are modeled as rank-reduced Gaussian processes in $\mathbf{TBT}^T$, where $\mathbf{T}$ is the concatenation of several $N_{\mathrm{TOA}} \times N_b$ design matrices of $N_b$ basis functions, and $\mathbf{B} = \langle\bm{b}\bm{b}^T\rangle$ encodes the variance over weights $\bm{b}$. Subvectors of $\bm{b}$ belong to the weight space $\mathcal{W}$ and submatrices of $\mathbf{T}$ belong to the space of design matrices $\mathcal{T}$. The posterior distribution over $\bm{\eta}$ under $\mathcal{M}$ is expressed using Bayes theorem,
\begin{align}
    \label{eq:posterior}
    \mathcal{P}(\bm{\eta}|\bm{\delta t},\mathcal{M}) &= \frac{\mathcal{L}(\bm{\delta t}|\bm{\eta},\mathcal{M})\pi(\mathcal{\bm{\eta}}|\mathcal{M})}{\mathcal{Z}(\bm{\delta t}|\mathcal{M})},
\end{align}
where $\pi(\mathcal{\bm{\eta}}|\mathcal{M})$ is our prior probability over $\bm{\eta}$ and $\mathcal{Z}(\bm{\delta t}|\mathcal{M})$ is the model evidence (or marginal likelihood). Equation~\ref{eq:posterior} is evaluated numerically using Markov Chain Monte Carlo (MCMC) or nested sampling. Given two different models $\mathcal{M}_1$ and $\mathcal{M}_0$, the ratio of model evidences is the Bayes factor,
\begin{align}
    \label{eq:BF}
    \mathcal{B}^{\mathcal{M}_1}_{\mathcal{M}_0} &= \frac{\mathcal{Z}(\bm{\delta t}|\mathcal{M}_1)}{\mathcal{Z}(\bm{\delta t}|\mathcal{M}_0)},
\end{align}
interpreted as the probability ratio for the data $\bm{\delta t}$ under $\mathcal{M}_1$ vs $\mathcal{M}_0$, or equivalently an odds ratio for $\mathcal{M}_1$ vs $\mathcal{M}_0$ given the data $\bm{\delta t}$ (assuming equal prior odds for both models). $\mathcal{B}^{\mathcal{M}_1}_{\mathcal{M}_0}$ may be used as a detection statistic for a signal represented by $\mathcal{M}_1$ if $\mathcal{M}_0$ is the signal's null hypothesis. For nested models, Equation~\ref{eq:BF} is easily approximated using the Savage-Dickey density ratio \citep{Dickey1971}.

We construct the likelihood and priors using \texttt{enterprise} \citep{enterprise} and \texttt{enterprise\_extensions} \citep{enterprise_extensions}. We perform parameter estimation using \texttt{PTMCMCSampler} (MCMC with parallel tempering; \citealt{ptmcmcsampler}) as well as \texttt{nautilus} (nested sampling; \citealt{nautilus2023}). We next describe the models used to construct the likelihood and their parameters.

\subsection{Timing model}

We start with the best-fit timing model for each pulsar from \citet{Perera+2019}. The timing model accounts for deterministic delays to a given pulsars TOAs accounting for effects such as pulsar spindown, astrometry, binary orbits, dispersion, frequency-dependent pulse profile evolution, and instrumental offsets. However, the presence of time-correlated noise will introduce perturbations to the best-fit values of the timing model parameters. As such, each pulsar's timing model is varied using an approximate linearized timing model design matrix $\mathbf{M} \in \mathcal{T}$, with elements defined
\begin{align}
    \label{eq:linear_timing_model}
    M_{ij} &= -\left.\frac{\partial t_i}{\partial\beta_j}\right|_{\beta_{0,j}},
\end{align}
where $t_i$ is the $i$th TOA, $\beta_j$ is the $j$th timing model parameter, and $\beta_{0,j}$ is the best-fit value of the $j$th timing model parameter \citep{vanHaasterenLevin2013, Taylor2021}. The timing model coefficients $\bm{\mathbf{\epsilon}} = \bm{\beta} - \bm{\beta_0} \in \mathcal{W}$ are then assigned improper uniform priors, which are implemented numerically as Gaussian priors with (near-)infinite variance, and then marginalized over when computing the likelihood following \citet{Johnson+2024}.

\subsection{Noise models}
\label{subsection:noise_models}

Pulsar noise models account at minimum for white noise, low-frequency red noise, and low-frequency chromatic noise. Here we update the pulsar noise models for IPTA DR2 from those used in \citet{Antoniadis+2022} to reflect recent advances in pulsar noise modeling. In particular, \citet{Falxa+2023} recently performed a search for continuous GWs from individual SMBHBs in IPTA DR2 and found that the detailed noise models from \citet{Chalumeau+2022}, which account for higher frequency sources of noise, were required to mitigate a spurious detection of a continuous GW.

An optimal treatment of pulsar noise would necessitate the creation of fully customized pulsar noise models tailored to IPTA DR2 (e.g., \citealt{Lentati+2016}), but this is beyond the scope of this work. Instead we use effective pulsar noise models informed by published analyses from individual PTA datasets \citep{Goncharov2021, Chalumeau+2022, EPTAnoise2023, Agazie2023_detchar, Reardon2023, Larsen2024}. We always use $\log_{10}$-uniform priors on the amplitude parameters of each noise process. These have been shown to be equivalent to spike and slab priors which enable noise model averaging \citep{vanHaasteren2025}. For each version of the dataset, we perform one round of noise analysis in each pulsar for parameter estimation and model validation prior to full-PTA analyses, using the same noise models for each dataset. Furthermore, it is also possible to improve these priors by using hierarchical modeling to create a ``population prior,'' which represents the ensemble noise properties of millisecond pulsars \citep{vanHaasteren2024, GoncharovSardana2025}. In particular, hierarchical priors have been shown \citep{vanHaasteren2024, Goncharov+2024} to reduce bias in GWB parameter estimation under a scenario where pulsars with similar intrinsic red noise properties become misattributed to the autocorrelations of the GWB (see \citealt{Zic+2022, Goncharov+2022} for discussions). While this scenario could be relevant for the analysis of IPTA DR2, we do not consider a GWB analysis using hierarchical priors, as the ensemble noise properties obtained from two datasets (i.e. DR2 Lite and Full DR2) will not be equivalent, and comparing results obtained under two different datasets with different priors is not straightforward (best left for future work). Nonetheless, it is useful to compare the ensemble noise properties obtained under different datasets with hierarchical modeling. These results are isolated to \S\ref{subsection:HBMs}, while the remainder of this work uses the standard uninformative priors.

\subsubsection{Achromatic red noise}

Each pulsar's red noise is modeled as a rank-reduced Gaussian process using a $N_{\mathrm{TOA}} \times 2N_f$ sine-cosine Fourier design matrix $\mathbf{F} \in \mathcal{T}$ with elements
\begin{align}
    \label{eq:Fmat}
    F_{ij} &= \begin{cases}
        \cos(2\pi f_{j/2}t_i) & \text{for even } j, \\
        \sin(2\pi f_{(j-1)/2}t_i) & \text{for odd } j,
    \end{cases}
\end{align}
where we use a linearly spaced frequency basis $\bm{f}$, and $N_f$ is the number of frequencies used \citep{Lentati+2013}. We place a power-law prior on the variance of the Fourier coefficients $\bm{a} \in \mathcal{W}$ at each frequency, parameterized in terms of the power spectral density
\begin{align}
    \label{eq:rn}
    S(f) &= \frac{A_{\mathrm{RN}}^2}{12\pi^2}\left(\frac{f}{\text{yr}^{-1}}\right)^{-\gamma_{\mathrm{RN}}}\text{yr}^3,
\end{align}
with uniform priors on the $\log_{10}$ spectral amplitude at $f=1/$yr $\log_{10}A_{\mathrm{RN}} \sim \mathcal{U}(-20, -11)$ and spectral index $\gamma_{\mathrm{RN}} \sim \mathcal{U}(0, 7)$, while the Fourier coefficients $\bm{a}$ are marginalized over \citep{vanHaasterenLevin2013}. Red noise processes may be reconstructed in the time domain as $\bm{\delta t}_{\mathrm{RN}}=\mathbf{F}\bm{a}$ by repeated draws from the posterior distribution over $\bm{a}$ \citep{Meyers+2023}. During a full-PTA analysis, the frequency basis for intrinsic pulsar red noise is defined to be equivalent to the CRN basis, $\bm{f} = \left(1/T_{\mathrm{DR2}}, \ldots, 30/T_{\mathrm{DR2}}\right)$, where $T_{\mathrm{DR2}} \cong 30$ yr is the timespan of IPTA DR2, and the number of frequencies are spaced in integer steps of $1/T_{\mathrm{DR2}}$. During single pulsar noise analyses, the frequency basis is tailored to the pulsar's timespan, $T_{\mathrm{obs}}$, such that $\bm{f} = \left(1/T_{\mathrm{obs}}, \ldots, 30/T_{\mathrm{DR2}}\right)$. This is chosen because any noise below $1/T_{\mathrm{obs}}$ in a given pulsar will be degenerate with pulsar spindown parameters. Meanwhile, the truncation frequency $30/T_{\mathrm{DR2}}$ is chosen to make sure each pulsar's white noise properties (which could depend on the cutoff if the spectrum is shallow) are consistent across both phases of the analysis. This red noise model is left consistent across all pulsars. However, PSR J1012+5307 also exhibits red noise up to very high Fourier modes \citep{Chalumeau+2022, Falxa+2023, EPTAnoise2023}. As such, we add an additional high-frequency power-law red noise process for PSR J1012+5307 with $\bm{f} = \left(1/T_{\mathrm{obs}}, \ldots,  150/T_{\mathrm{DR2}}\right)$ during all stages of analysis.

\subsubsection{Chromatic noise}

Any time-correlated noise processes depending on the radio-frequency of the pulsar, $\nu$, are chromatic. The primary type of chromatic noise is DM noise, varying as $\delta t_{\mathrm{DM}} \propto \nu^{-2}$. Similarly to achromatic red noise, we use a Fourier-basis Gaussian process with a power-law prior to model DM noise using hyperparameters $\log_{10}A_{\mathrm{DM}} \sim \mathcal{U}(-20,-11)$ and $\gamma_{\mathrm{DM}} \sim \mathcal{U}(0, 7)$, with an additional scaling $(\nu/\text{1400 MHz})^{-2}$ applied to the Fourier design matrix (equation~\ref{eq:Fmat}). Following \citet{Falxa+2023}, we allow the power law frequencies to extend to higher frequencies than achromatic red noise, here using $\bm{f} = \left(1/T_{\mathrm{obs}}, \ldots,  150/T_{\mathrm{DR2}}\right)$. An additional fit for linear and quadratic variations in DM$(t)$ are included in all timing models by default. 

The solar wind also induces annual quasi-periodic DM variations which we model separately from the Fourier-basis DM Gaussian process. Assuming a spherically-symmetric, $r^{-2}$ density profile surrounding the Sun, the DM induced by the solar wind is
\begin{align}
    \label{eq:sw}
    \mathrm{DM}_{\mathrm{solar}}(t) &= 4.85\times 10^{-6}\left(\frac{n_{\mathrm{Earth}}(t)}{\mathrm{cm}^{-3}}\right)\frac{\pi - \theta(t)}{\sin\theta(t)}\mathrm{\;pc\;cm}^{-3},
\end{align}
where $\theta(t)$ is the angle between the Earth-Sun and Earth-pulsar lines of sight, and $n_{\mathrm{Earth}}(t)$ is the time-dependent solar wind electron density measured at 1 AU from the Sun \citep{You+2007, Hazboun+2022, Nitu2024}. The mean, time-independent component of the electron density is included as a timing model parameter for every pulsar and marginalized over. We additionally fit for time-dependent density perturbations $\bm{\Delta n}_{\mathrm{Earth}} \in \mathcal{W}$ along each pulsar's line-of-sight as a Gaussian process using the model from \citet{Nitu2024}, with $N_b$ equal to the number of pulsar-Sun conjunctions in the pulsar's dataset, and a separate variance parameter sampled for each pulsar using the prior $\log_{10}\sigma_{n_{\mathrm{Earth}}} \sim \mathcal{U}(-4, 2)$ electrons cm$^{-3}$. Since $\theta(t)$ is bounded by the ecliptic latitude (\texttt{ELAT}) of each pulsar, many pulsars with large \texttt{ELAT} will be less sensitive to the solar wind, though there may be exceptions depending on radio-frequency coverage and TOA precision \citep{Susarla+2024}. We only include the time-dependent model in pulsars for which \texttt{ELAT} $< 35\degree$ and the model is favored with Savage-Dickey Bayes factor $\mathcal{B}^{\mathrm{SW}(t)}_{0} > 1$ using IPTA DR2.

Pulsars may also experience non-dispersive chromatic noise due to effects such as interstellar scattering or pulse profile variability. Scattering results from pulse propagation through an inhomogeneous refractive medium \citep{CordesRickett1998, HembergerStinebring2008}. A simple model for time-delays introduced by scattering is that $\delta t \propto \nu^{-\chi}$ with $\chi = 4.4$. However, this makes several assumptions, including that the refractive medium is described by Kolmogorov turbulence, the medium is isolated to a thin screen, the pulse is Gaussian, and the pulse broadening function is exponential \citep{Geiger2024}. Violations of these assumptions can and do result in alternative values for $\chi$ in millisecond pulsars \citep{Turner2021}, especially once transforming from estimates of the scattering delay to the timing residual \citep{Geiger2024}.

Here we account for some of this excess chromatic noise using the same Fourier-basis Gaussian process model as DM noise, except the radio-frequency scaling of the Fourier basis follows $(\nu/\text{1400 MHz})^{-\chi}$, with $\chi$ as a fit parameter. We incorporate the uncertainty on $\chi$ in our priors by using a truncated normal distribution, $\chi \sim \mathcal{N}(4, 0.5)\times\mathcal{U}(2.5,10)$, where the lower-bound at $\chi = 2.5$ prevents degeneracy with DM noise. We include this model for PSRs J0437-4715, J0613-0200, J1600-3053, J1643-1224, J1713+0747, J1903+0327, J1939+2134 based on the likely influence of scattering variations in these pulsars' timing residuals from prior work \citep{Alam2021_wb, NG15-detchar, Reardon2023, Srivastava+2023}. We also model a chromatic event in PSR J1713+0747 using the following deterministic signal \citep{Lam2018},
\begin{align}
    \label{eq:dip}
    s_d(t_i) = -A_d\Theta(t_i-t_0)\exp\left(-\frac{t_i-t_0}{\tau_d}\right)\left(\frac{\nu_i}{1400\text{ MHz}}\right)^{-\chi_d},
\end{align}
with uniform priors $\log_{10}A_d \sim \mathcal{U}(-10,-2)$ s, $\log_{10}\tau_d \sim \mathcal{U}(0.7, 2.7)$ d, $t_0 \sim \mathcal{U}(54742, 54768)$ MJD, $\chi_d \sim \mathcal{U}(1, 6)$ \citep{Goncharov2021, Antoniadis+2022}. To improve computational efficiency, all chromatic parameters $\chi, \chi_d$ are varied during single pulsar noise analyses, but held fixed to their maximum a posteriori (MAP) values during subsequent full-PTA analyses.

\subsubsection{White noise}

Many white noise parameters are included in IPTA-combined datasets to account for different systematic errors which may be unique to particular observing systems. We apply the same prescription as \citet{Antoniadis+2022} for fitting white noise. Two parameter types are diagonal in the $\textbf{N}$ matrix: EFAC, which applies a net scaling to the estimated TOA uncertainties, and EQUAD, which adds an additional net uncertainty in quadrature. We also apply ECORR parameters to NANOGrav TOAs, which are intended to model pulse jitter in sub-banded TOAs measured during the same observation epoch using uniform blocks along the diagonal band of the $\textbf{N}$ matrix. Separate white noise parameters are applied to TOAs from different receiver and backend combinations in each pulsar, where these combinations are specified by each TOA's \texttt{-group} flag. All white noise parameters are varied during single pulsar noise analyses, and then held fixed to their MAP values during subsequent full-PTA analyses.

\subsection{Common signals}
\label{subsec:common_signals}

In full-PTA analyses, we search for an additional common red noise (CRN) process on top of all components in each pulsar's noise model. The CRN is modeled with the same power-law spectral density as the individual red noise models, equation~\ref{eq:rn}, using new parameters $A_{\mathrm{CRN}}$ and $\gamma_{\mathrm{CRN}}$ which are fit for simultaneously in all pulsars at once. In a single pulsar analysis, the achromatic red noise includes contributions from both common and intrinsic pulsar noise. Switching from the single pulsar to full-PTA analysis decouples the total achromatic red noise into the separate intrinsic and common channels. The CRN also uses the same frequency basis $\bm{f}$ for all pulsars. To be consistent with \citet{Antoniadis+2022}, we use $N_{\mathrm{freqs}} = 13$ components for a frequency grid $\bm{f} = \left(1/T_{\mathrm{DR2}}, \ldots,  13/T_{\mathrm{DR2}}\right)$ for each analysis.

To model cross-correlations in a full-PTA analysis we define the cross-power spectral density,
\begin{align}
    S_{ab}(f) &= \Gamma_{ab}S(f),
\end{align}
where $\Gamma_{ab}$ is the overlap reduction function (ORF) encoding the geometric cross-correlation between pulsars $a$ and $b$ as a function of their sky-separation angle, and $S(f)$ is the power spectral density in each pulsar, given by the form of equation~\ref{eq:rn} if assuming a power law spectrum. One can specify different signals by the form of the ORF: $\Gamma_{ab} = \delta_{ab}$ represents a purely auto-correlated CRN process (uncorrelated between pulsars), whereas $\Gamma_{ab}$ given by the Hellings-Downs curve is the signature of an isotropic GWB under general relativity. Alternative ORFs given by monopolar and dipolar forms in pulsar sky separation angle would result from errors in terrestrial time standards \citep{Hobbs2012} and conversion to the solar system barycenter \citep{Champion+2010}, respectively.

\section{IPTA DR2 Lite vs Combined Analysis Results}
\label{section:Results}

Here we present the results of our common signal search and analysis of DR2 Lite (Table~\ref{tab:datasets}, row 2), in comparison with the same analysis of Full DR2, which was originally carried out in \citealt{Antoniadis+2022} (Table~\ref{tab:datasets}, row 1). We also perform an analysis on the fully combined dataset with only 22 pulsars, designated here as EDR2, that could reflect an intermediate stage of the data combination process (Table~\ref{tab:datasets}, row 3).

\begin{figure}
    \centering
    \includegraphics[width=\linewidth]{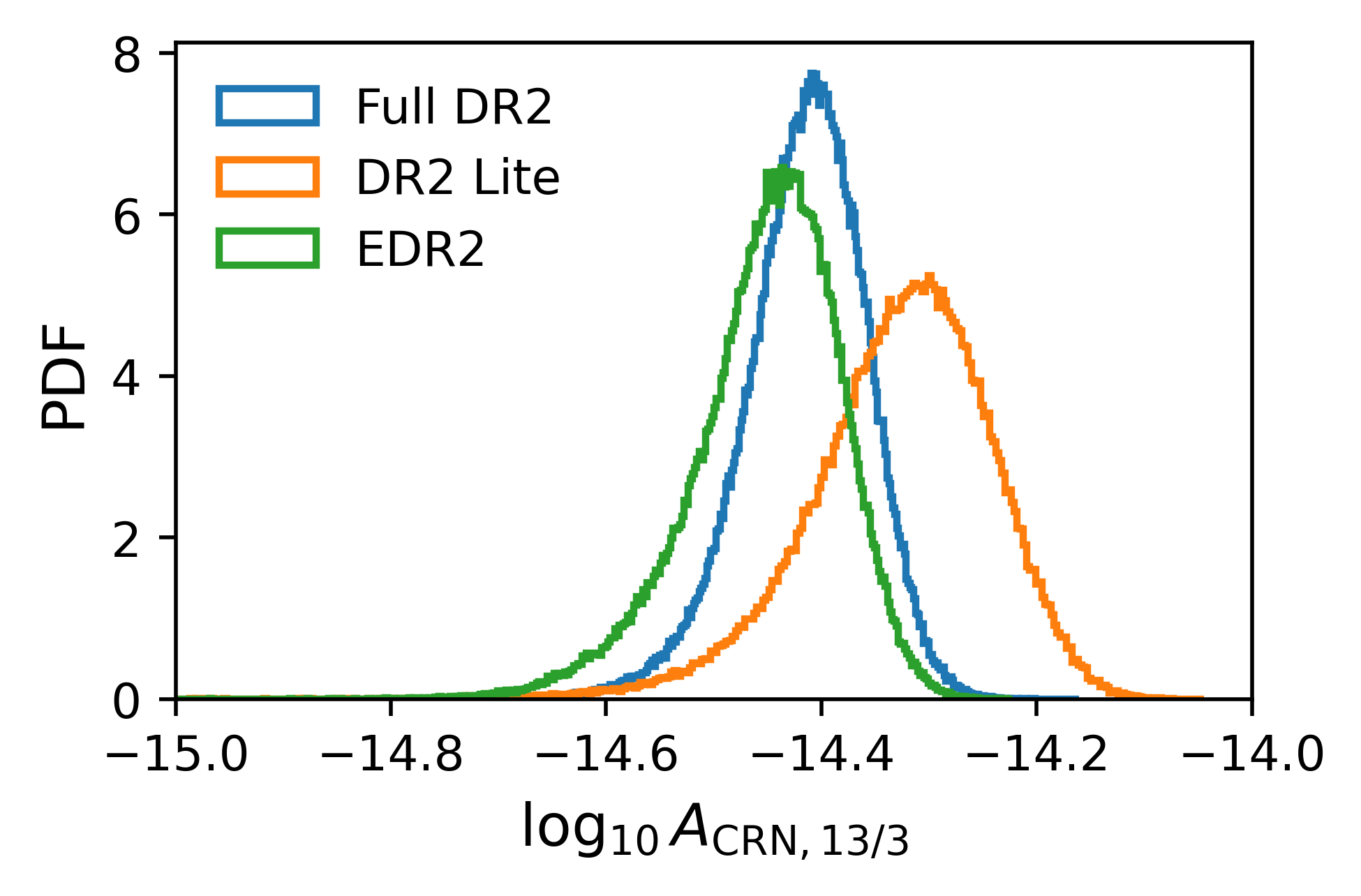}
    \vspace{-1.5\baselineskip}
    \caption{Comparison of the posterior PDF for $\log_{10}A_{\mathrm{CRN}}$ at fixed $\gamma_{\mathrm{CRN}} = 13/3$ from Full DR2 (blue), DR2 Lite (orange), and EDR2 (green). While the posteriors from all three datasets show evidence of a CRN, the amplitude distributed measured using DR2 Lite is shifted to larger values than the posteriors from the combined datasets}.
    \vspace{-\baselineskip}
    \label{fig:fixed_gamma}
\end{figure}

\begin{figure}
    \centering
    \includegraphics[width=\linewidth]{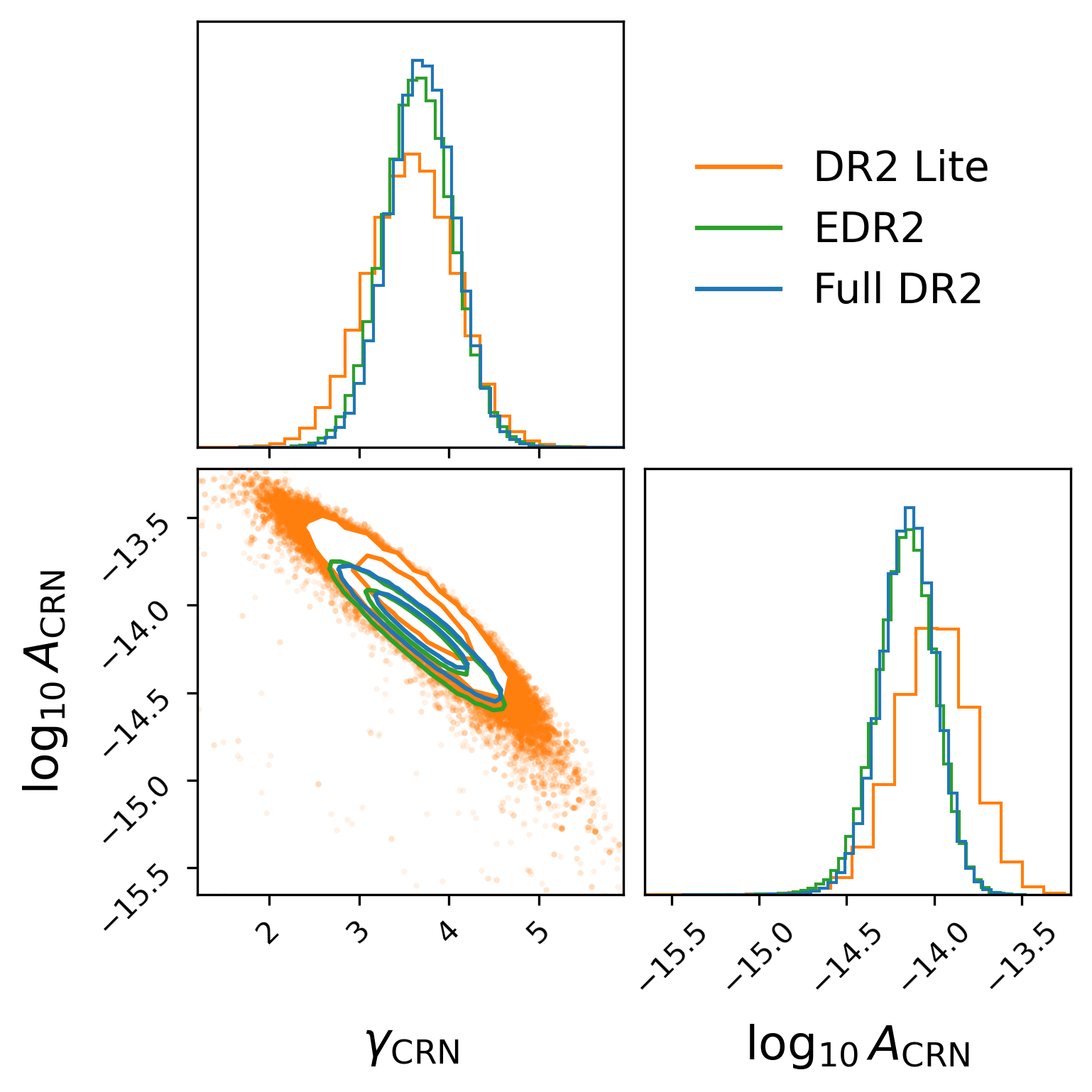}
    \vspace{-1.5\baselineskip}
    \caption{Comparison of posteriors for the CRN parameters $\log_{10}A_{\mathrm{CRN}}$ and $\gamma_{\mathrm{CRN}}$ from Full DR2 (blue), DR2 Lite (orange), and EDR2 (green). Adding more data in the combination results in tighter constraints on the CRN parameters. Contours enclose 68\% and 95\% of the posterior from each dataset (with samples outside the 95\% credible region shown only for DR2 Lite).}
    \vspace{-\baselineskip}
    \label{fig:varied_gamma}
\end{figure}

\begin{figure*}
    \centering
    \includegraphics[width=\linewidth]{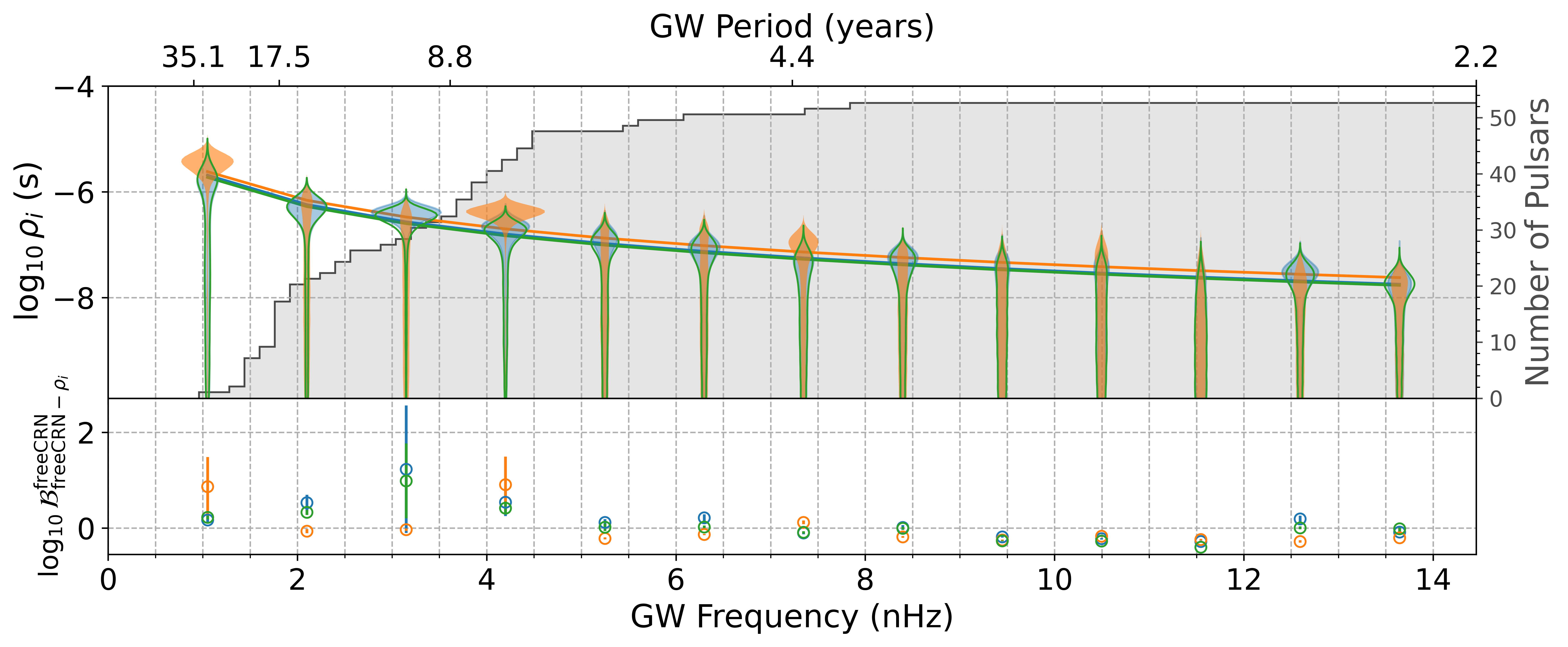}
    \vspace{-1.5\baselineskip}
    \caption{Comparison of the inferred CRN amplitude spectrum, $\rho(f) = \sqrt{S(f)/T_{\mathrm{DR2}}}$, using Full DR2 (blue), DR2 Lite (orange), and EDR2 (green). The violins in the top panel display posteriors on the $\log_{10}$ amplitude of common timing residual power $\rho_i$ at each discrete frequency $f_i$, and the solid lines indicate power law spectra corresponding to the median values of the CRN parameters from Figure~\ref{fig:varied_gamma}. To understand the number of pulsars contributing at each frequency, the violins are shown over a cumulative histogram of the pulsars in IPTA DR2 based on their observation timespans $f = 1/T_{\mathrm{obs}}$. The bottom panel quantifies the detection significance of common power at each bin via the $\log_{10}$ Bayes factor for a model containing the full CRN amplitude spectrum vs the same model with a single free spectral bin removed at the corresponding frequency. Using DR2 Lite, $\log_{10}\rho_i$ is measured most significantly at $f \sim 1$ and $f \sim 4$ nHz.}
    \vspace{-\baselineskip}
    \label{fig:free_spectrum}
\end{figure*}

\begin{table*}
    \centering
    \renewcommand{\arraystretch}{1.5}
    \begin{tabular}{ c | c c c | c c }
        \multicolumn{1}{c}{} & \multicolumn{3}{c}{Fixed $\gamma_{\mathrm{CRN}} = 13/3$} & \multicolumn{2}{c}{Varied $\gamma_{\mathrm{CRN}}$} \\
        \hline IPTA DR2 subset & $A_{\mathrm{CRN},13/3}$ & $A_{\mathrm{CRN},13/3}^{95\%}$ &  $\log_{10}\mathcal{B}^{\mathrm{CRN},13/3}_0$ & $A_{\mathrm{CRN}}$ & $\gamma_{\mathrm{CRN}}$ \\
        \hline DR2 Lite & $4.8^{+1.8}_{-1.8}\times10^{-15}$ & $6.4\times10^{-15}$ & $3.0 \pm 0.2$ & $10.0^{+15.6}_{-6.5}\times10^{-15}$ & $3.6^{+0.9}_{-1.0}$ \\
        EDR2 & $3.6^{+1.0}_{-1.2}\times10^{-15}$ & $4.5\times10^{-15}$ & $6.4 \pm 0.2$ & $6.9^{+7.3}_{-3.9}\times10^{-15}$ & $3.7^{+0.8}_{-0.8}$ \\
        Full DR2 & $3.9^{+1.0}_{-1.0}\times10^{-15}$ & $4.7\times10^{-15}$ & $9.1 \pm 0.3$ & $7.2^{+7.0}_{-3.8}\times10^{-15}$ & $3.7^{+0.7}_{-0.7}$ \\
        \hline
    \end{tabular}
    \renewcommand{\arraystretch}{1}
    \caption{CRN parameters and statistics  as a function of dataset. Using the fixed $\gamma_{\mathrm{CRN}} = 13/3$ model, we report median and 95\% credible intervals on $A_{\mathrm{CRN},13/3}$ (using $\log_{10}$-uniform priors), 95\% upper limits $A_{\mathrm{CRN},13/3}^{95\%}$ (using uniform priors), and Bayes factors on the CRN $\mathcal{B}^{\mathrm{CRN},13/3}_0$ estimated using the Savage-Dickey density ratio from a factorized likelihood analysis. Using the varied $\gamma_{\mathrm{CRN}}$ model, we report median and 95\% credible intervals on $A_{\mathrm{CRN}}$ and $\gamma_{\mathrm{CRN}}$. As data are progressively added from Lite to full, CRN detection statistics become more significant and parameter estimates become more precise, while the upper limit decreases with combined data.}
    \label{tab:crn_params}
    \vspace{-\baselineskip}
\end{table*}

\subsection{Common red noise}

We first compare our inferences on the CRN parameters using each dataset to see how much information we can learn about the common signal using the Lite method, and how much our inferences improve using the combined data.

\subsubsection{Full-PTA parameter estimation}

First we perform a simultaneous analysis of all pulsars in each dataset to estimate the CRN parameters assuming a fixed spectral index $\gamma_{\mathrm{CRN}} = 13/3$. Figure~\ref{fig:fixed_gamma} compares the posterior PDFs for $\log_{10}A_{\mathrm{CRN,13/3}}$ from each dataset. We also perform the same comparison using a varied $\gamma_{\mathrm{CRN}}$ model, for which we show the CRN posteriors in Figure~\ref{fig:varied_gamma}. The median and 95\% credible intervals on CRN parameters from both models are reported in Table~\ref{tab:crn_params}. We further report in Table~\ref{tab:crn_params} the upper limit $A_{\mathrm{CRN},13/3}^{95\%}$ for each dataset, estimated as the $95\%$ one-sided Bayesian credible interval after replacing the $\log_{10}$-uniform priors on $A_{\mathrm{CRN},13/3}$ with uniform priors.

We find that all three datasets, including DR2 lite, are able to detect a CRN, however, there are differences in spectral characterization. In particular, the amplitude measured using DR2 Lite is very large -- systematically higher than amplitude measured from the combined datasets, with an median amplitude $A_{\mathrm{CRN},13/3}$ measured 23\% larger using DR2 Lite than it is using Full DR2. This implies that DR2 Lite is allowing excess noise intrinsic to the pulsars to leak into the common channel \citep{Zic+2022}. The upside then is that performing data combination apparently helps to mitigate this effect. We explore this discrepancy further in \S\ref{subsection:discrepancy}.

Additionally, Figure~\ref{fig:varied_gamma} shows that the CRN parameter constraints become more precise as more data are added. To quantify the improvement in precision, we estimate the area $\mathcal{A}$ of the 2D region in $\log_{10}A_{\mathrm{CRN}}$, $\gamma_{\mathrm{CRN}}$ enclosed within 95\% of the posterior from each dataset. We find the ratio of areas with respect to Full DR2 to be $\mathcal{A}_{\mathrm{Lite}}/\mathcal{A}_{\mathrm{Full}} = 2.25$ and $\mathcal{A}_{\mathrm{EDR2}}/\mathcal{A}_{\mathrm{Full}} = 1.23$, i.e. Full DR2 is 2.25 times \emph{more} precise at spectral characterization than DR2 Lite, and 1.23 times more precise than EDR2.

Figure~\ref{fig:free_spectrum} further compares each dataset using a more generic ``free-spectral'' model for the CRN, where we drop the assumption of a power-law spectrum and sample the timing residual power at each discrete frequency $f_i$ as independent parameters with prior $\log_{10}\rho_i \sim \mathcal{U}(-10,-4)$ in units of seconds \citep{Lentati+2013}. The top panel shows the posteriors on each $\log_{10}\rho_i$ from each dataset, which plotted vs frequency represent the amplitude spectrum of the CRN. Here it is valid to compare amplitude spectra from the different datasets as they share the same baseline $T_{\mathrm{DR2}}$, otherwise the amplitude spectral density would be the relevant quantity to compare. Overplotted lines depict median \emph{power law} spectra from each dataset, for comparison. The bottom panel of Figure~\ref{fig:free_spectrum} shows the $\log_{10}$ Bayes factors, $\log_{10}\mathcal{B}^{\mathrm{CRN}}_{\mathrm{CRN}-\rho_i}$, for the free spectral CRN model vs the same model without the inclusion of common power in the frequency bin at frequency $f_i$, as measured using the Savage-Dickey density ratio. These quantify the detection significance of the CRN at each frequency bin, where $\log_{10}\mathcal{B}^{\mathrm{CRN}}_{\mathrm{CRN}-\rho_i} > 0$ indicates the data do favor the inclusion of additional common power at frequency $f_i$.

EDR2's amplitude spectrum is comparable to Full DR2, with small deviations in the posteriors at at higher frequencies. Both spectra are consistent with a power law form and display strongest detections of power in the $2-3$ nHz range. Meanwhile, the posterior power in DR2 Lite's amplitude spectrum exceeds that from the combined datasets at several frequencies, which is consistent with the higher power-law amplitude measured using DR2 Lite. We focus on the 1 nHz and 4 nHz frequencies (corresponding to periods of $\sim30$ and $\sim7.5$ years respectively) where power is detected \emph{more significantly} using DR2 Lite than the combined datasets. Relatively few pulsars in IPTA DR2 have sufficiently long timespans to contribute meaningfully to constraining the posterior at 1 nHz. The Figure~\ref{fig:free_spectrum} cumulative histogram of pulsars over their observation timespan $1/T_{\mathrm{obs}}$ illustrates this effect: Only 3 pulsars (PSRs J1713+0747, J1857+0943, J1939+2134) have $T_{\mathrm{obs}} > 21$ years, 2 of which (PSRs J1857+0943 and J1939+2134) have very large data gaps spanning $\sim10$ years in DR2 Lite, as shown in Figure~\ref{fig:TOAs}. Unconstrained noise in even 1 of these pulsars using DR2 Lite could feasibly cause changes in timing residual power at 1 nHz. As for the 4 nHz posterior, it visibly deviates from the median power law fit obtained from Full DR2 and EDR2, likely producing the large amplitude measured in the power law analysis. While 41 pulsars have sufficient timespan to resolve this frequency, details in the noise characterization of a minority of especially sensitive pulsars may still produce such features in the common spectrum \citep{Hazboun+2020_slicing, Larsen2024}.

\subsubsection{Factorized likelihood analysis}
\label{subsection:FL}

\begin{figure}
    \centering
    \includegraphics[width=\linewidth]{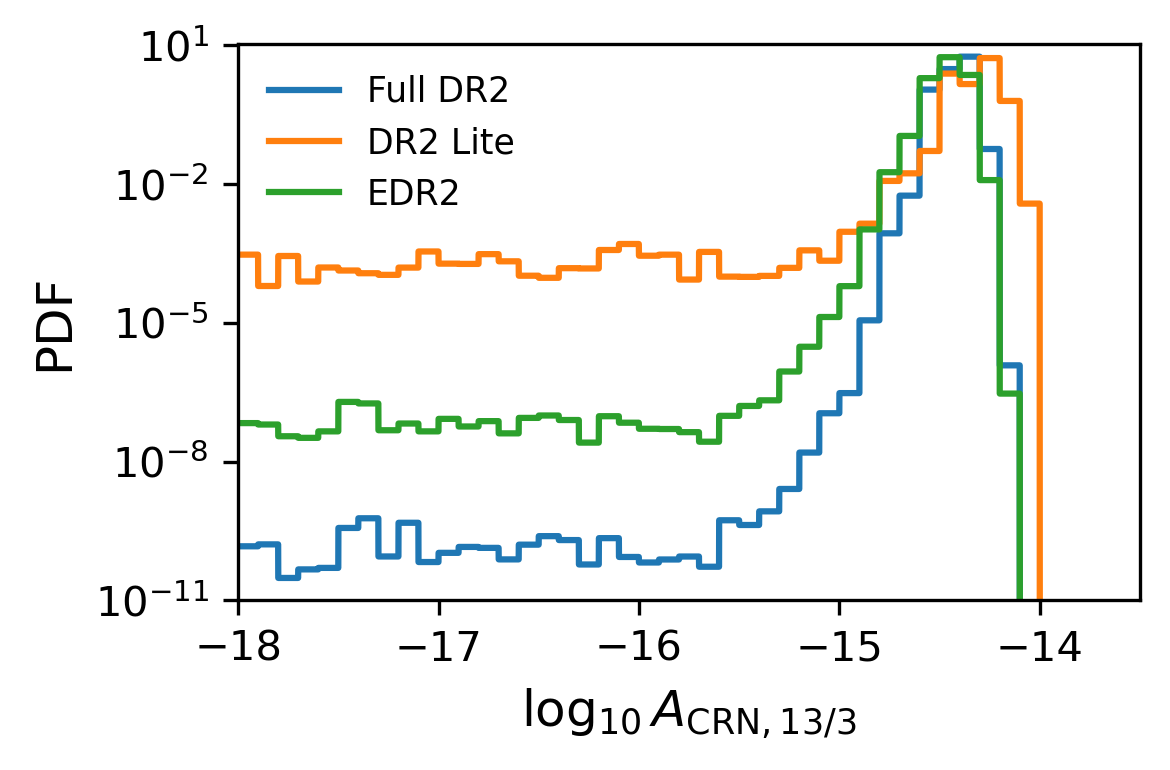}
    \vspace{-1.5\baselineskip}
    \caption{Inferences of the CRN amplitude at $\gamma = 13/3$ using the factorized likelihood with each dataset: Full DR2 (blue), DR2 Lite (orange), EDR2 (green). The factorized likelihood allows inference of the low-amplitude of each distribution, which corresponds to a null-detection. The probability of a null-detection is highest with DR2 Lite and lowest with Full DR2, as expected when using more data in the analysis. EDR2 achieves greater detection significance than DR2 Lite despite including fewer pulsars.}
    \vspace{-1\baselineskip}
    \label{fig:factlike}
\end{figure}

\begin{figure*}
    \centering
    \includegraphics[width=\linewidth]{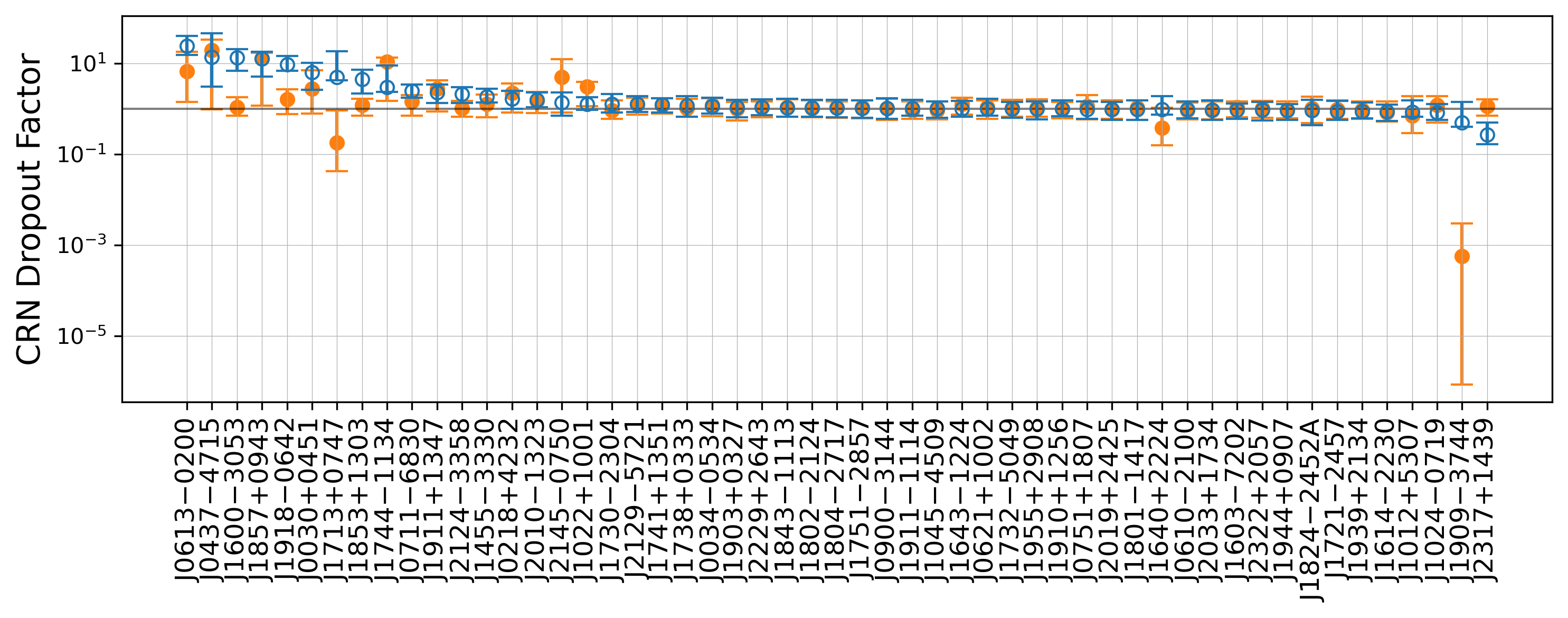}
    \vspace{-1.5\baselineskip}
    \caption{Comparison of CRN dropout factors (DFs) computed using the factorized likelihood for each pulsar, as computed using DR2 Lite (orange) and Full DR2 (blue). DFs help diagnose which pulsars are most responsible for the results of a CRN search, where pulsars with $\mathrm{DF} > 1$ support the presence of a CRN and pulsars with $\mathrm{DFs} < 1$ are in tension with a CRN. Pulsars are ranked from highest to lowest DF using Full DR2, and errors are estimated as the 95\% confidence interval of $25000$ bootstrap realizations, following \citet{Antoniadis+2022}. PSR J1909$-$3744's especially low DF signifies its total measured noise has a lower amplitude than the CRN, as further shown in Figure~\ref{fig:J1909-3744}.}
    \vspace{-1.5\baselineskip}
    \label{fig:dropout}
\end{figure*}

We gain additional information about the significance of an auto-correlated common process using the Factorized Likelihood \citep{Taylor2022}. Using the factorized likelihood, the likelihood and common noise posteriors are estimated as a product of these statistics as measured from analyses of individual pulsars. The factorized likelihood not only enables rapid PTA analyses but also allows the analysis of different Lite datasets at no additional computational cost, as demonstrated in \citet{3P+2024}. Using the factorized likelihood, we assign the CRN model a fixed spectral index $\gamma_{\mathrm{CRN}}=13/3$ in each pulsar, while the intrinsic red noise model's spectral index is allowed to vary (otherwise, the intrinsic red noise and CRN signals are completely degenerate). Since the CRN may always be assigned to the intrinsic RN model during each single-pulsar analysis, the CRN amplitude in each pulsar is always finite at $A_{\mathrm{CRN},13/3} < 10^{-17}$, therefore the Savage-Dickey density ratio can always be used to estimate the CRN Bayes factor from the product.

Figure~\ref{fig:factlike} displays the posteriors over $\log_{10}A_{\mathrm{CRN,13/3}}$ obtained from the factorized likelihood analysis, and Column 5 of Table~\ref{tab:crn_params} reports the Bayes factors for a model containing a CRN with $\gamma_{\mathrm{CRN}} = 13/3$ vs a model with no CRN. We estimate the Bayes factor means and uncertainties based on the distributions of probabilities within $\log_{10}A_{\mathrm{CRN,13/3}} < -16$ shown in Figure~\ref{fig:factlike}. All three datasets' Bayes factors are of $\mathcal{O}(10^3)$ or greater, signifying strong preference for a CRN, in agreement with the full-PTA analysis. However, the results show progressively more support for a CRN as the data combination process continues, with the odds of a CRN increasing by a factor of $\sim10^{3.4}$ going from DR2 Lite to EDR2, and again by a factor of $\sim10^{2.7}$ going from EDR2 to Full DR2. This result is unsurprising as we add more data into the analysis, but nonetheless underscores the effectiveness of data combination for improving our detections of common signals.

\subsubsection{Dropout Analysis}
\label{subsection:dropout}

The factorized likelihood can be also used to rapidly compute dropout factors (DFs), which are Bayes factors for a model with the CRN in $N$ pulsars vs a model with the CRN in $N - 1$ pulsars, while the excluded pulsar $p$ is modeled with intrinsic noise only \citep{Arzoumanian2020, Taylor2022, Antoniadis+2022}. If pulsar $p$'s $\mathrm{DF} > 1$, then pulsar $p$'s data support the CRN measurement (i.e., the CRN Bayes factor would decrease by the DF upon removing pulsar $p$ from the array), whereas if pulsar $p$'s $\mathrm{DF} < 1$ then pulsar $p$'s data are in tension with the CRN (i.e., the CRN Bayes factor would increase by the DF upon removing pulsar $p$ from the array). Comparing the DFs across pulsars is a useful way to diagnose which pulsars most heavily favor or disfavor the inclusion of a CRN. To see how individual pulsars affect \emph{parameter estimation} on $A_{\mathrm{CRN,13/3}}$, see \S\ref{subsection:discrepancy}.

Figure~\ref{fig:dropout} compares each pulsar's DF from DR2 Lite and Full DR2, as computed using the factorized likelihood. We follow the procedure from \citet{Antoniadis+2022} to compute uncertainties on the DFs: first we vary the analysis over 25 combinations of metaparameters and then for each combination, we generate $1000$ bootstrap realizations by re-sampling the chain with replacement. The Full DR2 DFs are consistent with those reported in \citet{Antoniadis+2022}, with several pulsars' $\mathrm{DFs} > 1$. Using DR2 Lite, fewer DFs are found to be larger than 1, which is unsurprising as the Bayes factor for the CRN is lower. PSR J1909--3744's value of $\mathrm{DF} = 6^{+24}_{-6}\times10^{-4}$ using DR2 Lite is extremely small (even within bootstrapping errors), as such this pulsar seems to be in major tension with the CRN process when using DR2 Lite. This is highly unusual, but may be explained if PSR J1909$-$3744 has a much lower upper limit on the CRN than the remaining pulsars, as discussed in \S\ref{subsection:discrepancy}. Using Full DR2, PSR J1909--3744's DF is $\mathrm{DF} = 0.49^{+0.91}_{-0.08}$, indicating PSR J1909--3744 is now more in line with the CRN and with the other combined pulsars. We follow up further with PSR J1909--3744 in \S\ref{subsection:Noise}. PSR J1713+0747 is also an interesting case, switching from disfavoring to supporting the CRN, going from $\mathrm{DF} = 0.18^{+0.73}_{-0.14}$ using DR2 Lite to $\mathrm{DF} = 4.93^{+13.34}_{-0.72}$ using Full DR2. Overall, five pulsars experience significant (i.e., $>2\sigma$) boosts to their DF after data combination, while only PSR J2317+1439's DF is significantly higher using DR2 Lite.

\subsubsection{Assessing the Large Common Amplitude in DR2 Lite}
\label{subsection:discrepancy}

\begin{figure*}
    \centering
    \includegraphics[width=\linewidth]{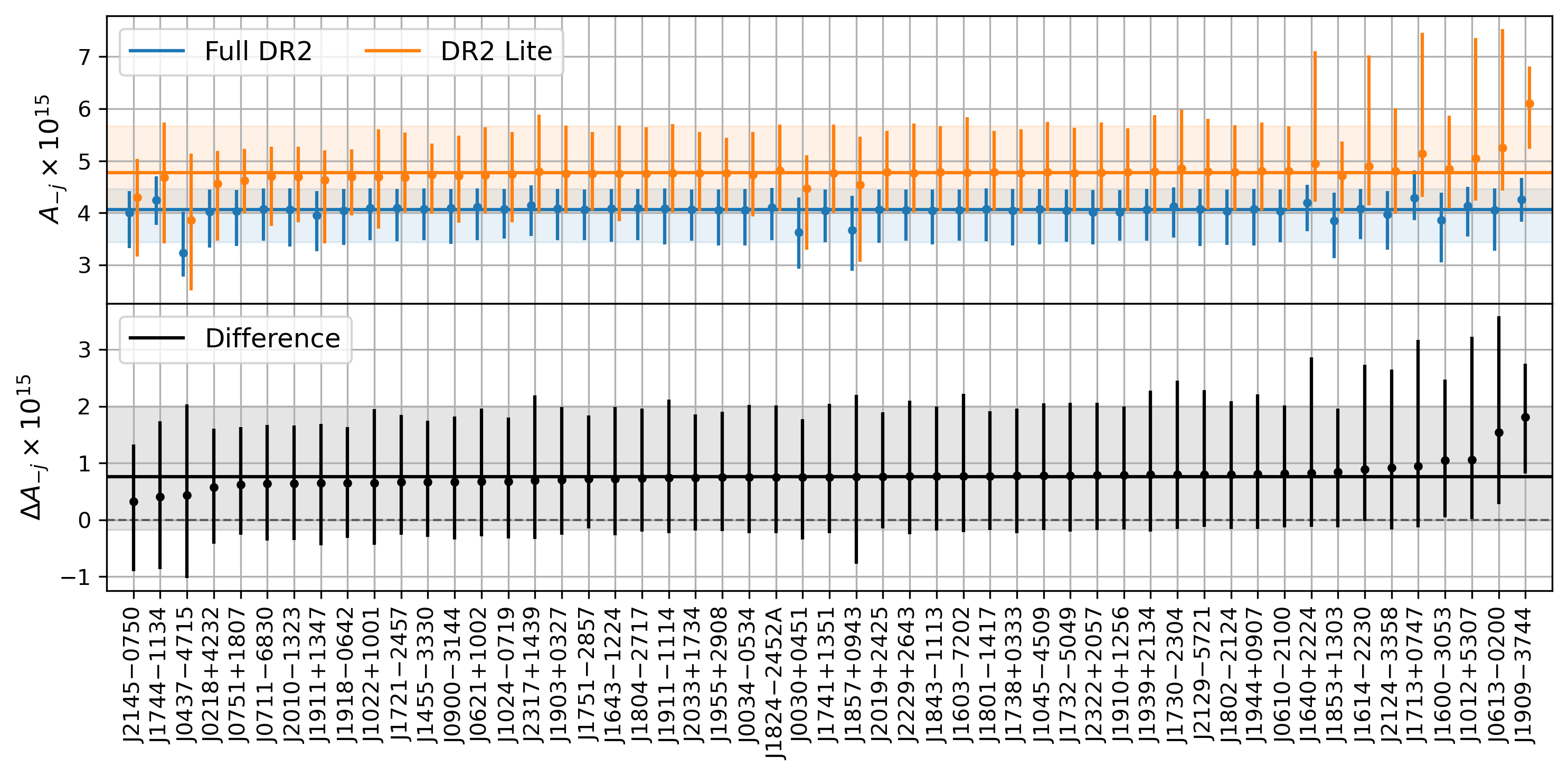}
    \vspace{-1.5\baselineskip}
    \caption{Parameter estimation on $A_{\mathrm{CRN},13/3}$ using the factorized likelihood posteriors with each pulsar $j$ dropped out one at a time by themselves. The top panel shows medians and 68\% credible intervals over $A_{-j}$ (equation~\ref{eq:dropout_A}) using each dataset, while the bottom panel shows the corresponding difference $\Delta A_{-j}$ between DR2 Lite and Full DR2 (equation~\ref{eq:diff_approx}). For comparison, the solid lines and shaded regions show the corresponding estimates $A$ and $\Delta A$ with no pulsars removed. Pulsars are ranked from the lowest to highest median values of $\Delta A_{-j}$. Several individual pulsars, when removed by themselves, skew the estimates of $\Delta A$, showing the overall discrepancy is sensitive to systematic errors in the individual pulsars.}
    \vspace{-\baselineskip}
    \label{fig:dropout_A}
\end{figure*}

While the dropout analysis in \S\ref{subsection:dropout} shows that the \emph{detection significance} of the CRN may be skewed by individual outlier pulsars using DR2 Lite, it still leaves open the question why the CRN amplitude is systematically larger when using DR2 Lite as opposed to Full DR2. To investigate further, we apply a modified version of the dropout analysis which compares how the \emph{discrepancy in the measured amplitude} depends on which pulsars are dropped. To quantify this discrepancy, we compute a distribution over the shift or difference between the values for $A_{\mathrm{CRN},13/3}$ from each dataset, $\Delta A = A_{\mathrm{Lite}} - A_{\mathrm{Full}}$, the mean of which is defined
\begin{align}
    \langle\Delta A\rangle &= \int dA_{\mathrm{Lite}}\int dA_{\mathrm{Full}}(A_{\mathrm{Lite}} - A_{\mathrm{Full}}) \times \notag \\
    &\;\;\;\;\;\;\;\;\mathcal{P}(A_{\mathrm{Lite}}|\bm{\delta t}_{\mathrm{Lite}})\mathcal{P}(A_{\mathrm{Full}}|\bm{\delta t}_{\mathrm{Full}}) \\
    &\approx \frac{1}{N_{\mathrm{samples}}}\sum_{i,j}(A_{\mathrm{Lite}})_i - (A_{\mathrm{Full}})_j. \label{eq:diff_approx}
\end{align}
For brevity we have omitted the previously used subscripts so that $A \equiv A_{\mathrm{CRN},13/3}$, and line~\ref{eq:diff_approx} makes explicit that we compute this difference simply by drawing Monte Carlo pairs of samples from the original factorized likelihood posteriors from each dataset. This quantity is related to the tension metric from \citet{3P+2024}. When including all pulsars, the median of this difference comes out to $\Delta A = 0.8^{+1.2}_{-1.0} \times 10^{-15}$, where the quantiles enclose 68\% of the distribution about the median. The value $\Delta A = 0$ is only just within the 68\% credible region, indicating a marginal tension on the level of $0.81\sigma$.

To test the dependence of this discrepancy on individual pulsar datasets, we recompute the CRN amplitude at $\gamma_{\mathrm{CRN}} = 13/3$ via the factorized likelihood method (equation 5 from \citealt{Taylor2022}) but with pulsar $j$ removed from the dataset, i.e.
\begin{align}
    p_{-j}(\log_{10}A|\{\bm{\delta t}\}_{-j}) &\equiv p\left(\log_{10}A|\prod^{N_{\mathrm{psr}}}_{i\ne j}\bm{\delta t}_i\right) \notag \\
    &\propto \prod^{N_{\mathrm{psr}}}_{i\ne j}p\left((\log_{10}A)_i|\bm{\delta t}_i\right),
    \label{eq:dropout_A}
\end{align}
where $i$ indexes each pulsar's dataset $\bm{\delta t}_i$, and the $-j$ subscripts are a shorthand to denote pulsar $j$ has been dropped. We compute $p_{-j}(\log_{10}A|\{\bm{\delta t}\}_{-j})$ for every pulsar $j$ using both DR2 Lite and Full DR2, at which points combining with Equation~\ref{eq:diff_approx} yields a new measure of the discrepancy, $\Delta A_{-j}$, with the $j$th pulsar removed from both datasets.

The bottom panel of Figure~\ref{fig:dropout_A} shows our estimates of $\Delta A_{-j}$ for each pulsar $j$ individually removed by themselves. Shaded regions show the corresponding estimates $\Delta A$ with no pulsars removed. To better understand what is causing any measured differences in $\Delta A_{-j}$, we also show the estimates on $A_{-j}$ using each dataset in the top panel of Figure~\ref{fig:dropout_A} (and corresponding shaded regions for $A$ with no pulsars removed). If $A_{-j} > A$, then pulsar $j$ is driving down the amplitude estimate, whereas if $A_{-j} < A$, then pulsar $j$ is adding power to the common amplitude.

There are several pulsars, when excluded from the analysis, that reduce the discrepancy in $\Delta A$. The most impactful pulsars on this end are PSRs J2145$-$0750, J1744$-$1134, and J0437$-$4715; their individual removal reduces the discrepancy to a $0.26\sigma$, $0.32\sigma$, and $0.30\sigma$ level respectively. Interestingly, each pulsar skews the distributions $A_{-j}$ in a different way. PSR J2145$-$0750 displays very loud red noise with a steep power law index $\gamma > 4$ in IPTA DR2 \citep{Caballero+2016, Perera+2019}; in Full DR2, this is decoupled from the CRN, whereas in DR2 Lite, the red noise is not sufficiently resolved from DM variations (see Table~\ref{tab:noise}, \S\ref{subsection:Noise}) nor is it resolved from the CRN, causing the red noise to pollute the CRN process. PSR J1744$-$1134, among the best timers in the dataset, affects the discrepancy simply by reducing $A_{\mathrm{CRN},13/3}$ using Full DR2, whereas there is less constraint on its noise using DR2 Lite (Table~\ref{tab:noise}). Finally, PSR J0437$-$4715, which only has TOAs from the PPTA in IPTA DR2, strongly drives up the CRN amplitude for both datasets. Notably, PSR J0437$-$4715 has the highest FoM behind PSR J1939+2134 and therefore has a disproportionately strong effect on the analysis, on top of very challenging noise properties to model due to its brightness, which may well contribute additional systematic error in this analysis (see e.g. \citealt{Lentati+2016, Goncharov2021, Reardon2023, Reardon+2024}). In summary, we consider the discrepancy in the CRN amplitude between DR2 Lite and Full DR2 to be statistically insignificant as it is not robust to outliers among the individual pulsars.

Finally, there are a few pulsars on the opposite end where the discrepancy \emph{widens} with their removal. This is unsurprising if an equal number of pulsars also narrow the discrepancy. However, this analysis sheds insight into the nature of PSR J1909$-$3744's extremely low DF from \S\ref{subsection:dropout}. We see on the far right hand side of Figure~\ref{fig:dropout_A} that removing PSR J1909$-$3744 from the analysis results in a much \emph{larger} CRN amplitude of $A_{-j} = 6.1^{+1.4}_{-1.5}\times10^{-15}$ (with errors enclosing 95\% quantiles). This shows PSR J1909$-$3744's DF is so low because it is uniquely suppressing a higher-amplitude mode of the CRN posterior, which is more likely to be dominated by intrinsic pulsar noise. Meanwhile, the Full DR2 amplitude $A_{-j}$ when dropping PSR J1909$-$3744 is shifted by much less using Full DR2, reflective of the fact that several more pulsars contribute constraints on the CRN measurement after performing data combination.

\subsection{Cross-correlations}

In order for the data to signify evidence of a GWB, they ought to favor a cross-correlated common signal following the Hellings-Downs curve, as opposed to a purely auto-correlated common signal. We assume that data combination will ultimately improve our ability to resolve a cross-correlated GWB over the use of uncombined data. IPTA DR2 has limited value to test this assumption, since the statistics for cross-correlations in IPTA DR2 obtained first by \citet{Antoniadis+2022} are well below the thresholds required for GW detection \citep{Allen+2023}. Thus the following comparisons of cross-correlation statistics obtained from the three datasets are merely exploratory, but still presented for completeness.

\subsubsection{Optimal Statistic}

We first assess the significance of Hellings-Downs cross-correlations with the Optimal Statistic (OS; \citealt{Anholm+2009, Chamberlin+2015}) using the \texttt{defiant} tool\footnote{\url{https://github.com/GersbachKa/defiant/tree/main}}. We specifically use the multiple component OS to measure all three ORFs (Monopole, Dipole, Hellings-Downs) simultaneously, which reduces any bias incurred while measuring one ORF due to the presence of another \citep{Sardesai2023}. To account for CRN and intrinsic pulsar noise parameter uncertainties, we apply noise marginalization to obtain a distribution of the OS (i.e., the NMOS; \citealt{Vigeland+2018}) over 3000 draws from our Bayesian posteriors. For each dataset, we then follow \citet{Vallisneri+2023} to obtain a Bayesian S/N (S/N$_{\mathrm{Bayes}}$), which can be interpreted as a probability-weighted mean over noise marginalized S/N distribution. For our purposes we assume the GWB S/N null distribution is Gaussian (noting an accurate measure of GW significance mandates use of a GX2 null distribution; see \citealt{Hazboun2023}). Our analysis also neglects to account for covariance between pulsar pairs, which naturally arises in pulsar pairs with pulsars in common \citep{AllenRomano2023, Johnson+2024}. However, this assumption is justified here as we are in the weak S/N regime of the GWB cross-correlations \citep{Sardesai2023}.

\begin{figure}
    \centering
    \includegraphics[width=\linewidth]{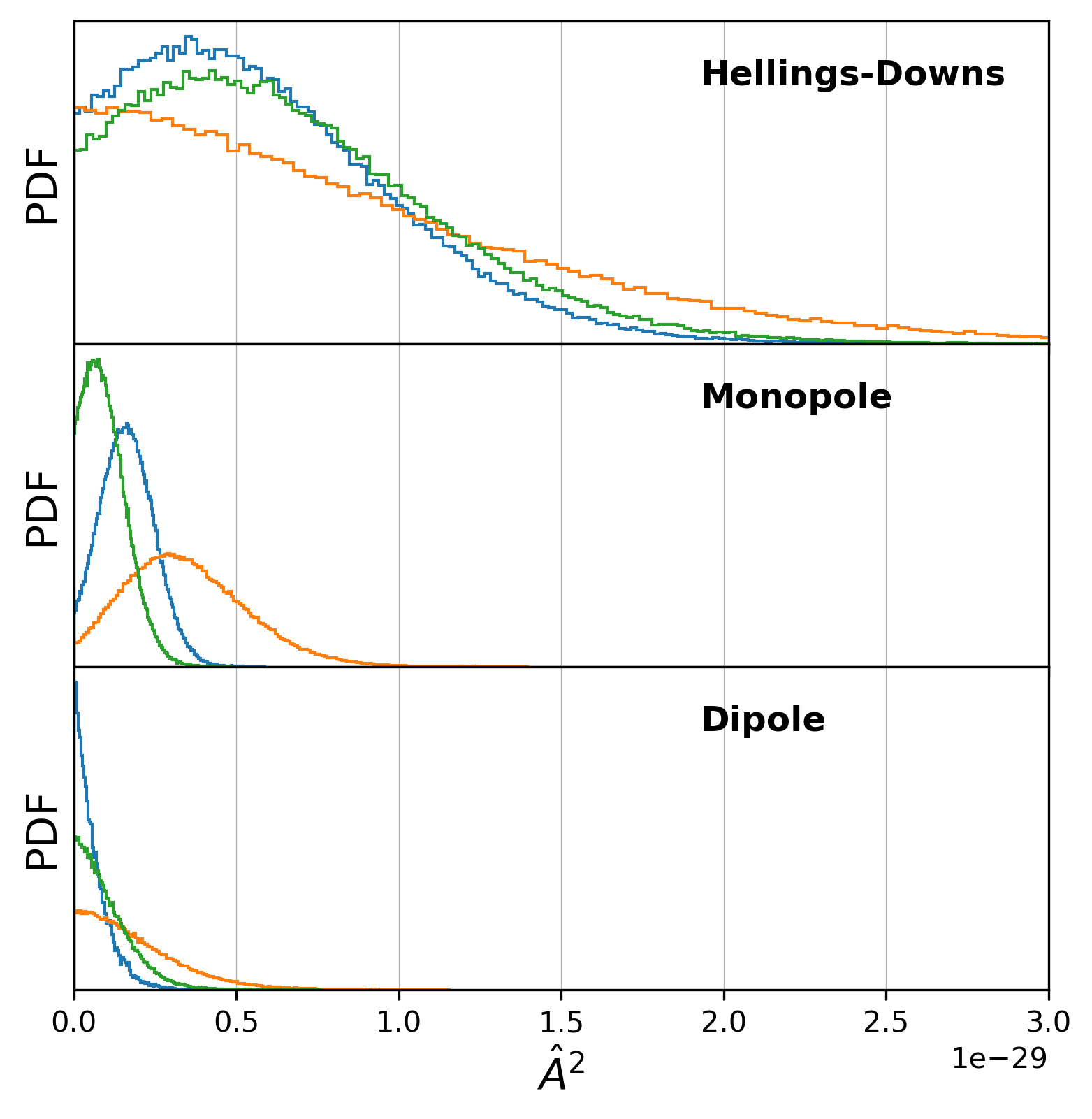}
    \vspace{-\baselineskip}
    \caption{Full distribution of $\hat{A}^2$ of Hellings-Downs, monopolar, and dipolar correlations, computed from the noise marginalized multiple component optimal statistic with uncertainty sampling for Full DR2 (blue), DR2 Lite (orange), and EDR2 (green).}
    \vspace{-\baselineskip}
    \label{fig:A_squared}
\end{figure}

We report the S/N$_{\mathrm{Bayes}}$ values we obtain for Hellings-Downs ORF on the right-most column of Table~\ref{tab:hd_BFs}. Using DR2 Lite, We are completely unable to resolve HD correlations from noise, as given by S/N$_{\mathrm{Bayes}} = 0$. The combined datasets yield higher, but still relatively low values of S/N$_{\mathrm{Bayes}} = 0.79$ from Full DR2 and S/N$_{\mathrm{Bayes}} = 0.92$ from EDR2, with EDR2 yielding slightly higher significance despite including fewer pulsars than Full DR2. These results appear to be consistent with expectations from \citet{Agazie+2025_1by1}, which found that the 30 \emph{noisiest} pulsars can be dropped from the NANOGrav 15-year dataset before the Hellings-Downs S/N experiences any appreciable drop (here EDR2 is effectively Full DR2 with the 31 noisiest pulsars dropped). In this case, a slight increase in S/N with EDR2 may imply either excess noise from a set of the 31 remaining pulsars, or it is simply a statistical fluctuation given the low values of S/N$_{\mathrm{Bayes}}$.

Figure~\ref{fig:A_squared} further shows the full distributions of the squared amplitude $\hat{A}^2$ of each correlation signature, computed using the multiple component NMOS from each dataset with uncertainty sampling \citep{Gersbach+2024}. The Hellings-Downs GWB amplitudes are consistent with each other and the reported values of S/N$_\mathrm{Bayes}$, with the combined datasets reducing the uncertainty from DR2 Lite. The presence of monopolar or dipolar correlations indicates additional systematic correlated noise (see \S\ref{subsec:common_signals}). Most notable is that DR2 Lite produces a larger monopole than Full DR2, suggesting the presence of additional unmodeled noise in the uncombined data, which is then mitigated via data combination. The monopole amplitude is further reduced in EDR2 from Full DR2, implying a set of the 31 remaining pulsars may be partially responsible for the noise contributing to this monopole. The DR2 Lite and Full DR2 monopoles both correspond to $\mathrm{S/N} \sim 2$ \citep{Antoniadis+2022}. This S/N level is not particularly significant, as it was shown in \citet{Agazie+2023_gwb} that monopolar cross-correlations with higher corresponding S/N may arise frequently in simulations containing solely HD-correlated GWs and intrinsic pulsar noise. As such, it is possible these monopole measurements are statistical fluctuations, as opposed to the result of a real systematic such as a clock error. We leave a deeper analysis to understand the emergence and nature of monopolar cross-correlations in PTA datasets for future work. The Dipole amplitude is highly consistent with zero in each case, which makes sense as the solar system ephemeris version (DE436) used in this analysis is fixed, independently of how many TOAs are combined. However, the constraints on the amplitude of the dipolar correlations improve as data is combined.

\subsubsection{Bayes factors for Hellings-Downs correlations over common noise}

IPTA DR2 does not contain enough data to detect the Hellings-Downs curve: \citet{Antoniadis+2022} reports a Bayes factor $\log_{10}\mathcal{B}^{\mathrm{HD}}_{\mathrm{CRN}} \sim 0.3$ ($\mathcal{B}^{\mathrm{HD}}_{\mathrm{CRN}} \sim 2$) for an Hellings-Downs-correlated model vs the auto-correlated CRN model. Nevertheless, this value still indicates a modest preference for an Hellings-Downs cross-correlated model in IPTA DR2. As such, it is valuable to compare the Bayes factors for Hellings-Downs correlations obtained from each dataset to find out if DR2 Lite can resolve Hellings-Downs correlations at the same level as the Full DR2.

\begin{table}
    \centering
    \renewcommand{\arraystretch}{1.5}
    \begin{tabular}{ c | c c | c }
        \multicolumn{1}{c}{} & \multicolumn{2}{c}{$\mathcal{B}^{\mathrm{HD},13/3}_{\mathrm{CRN},13/3}$} & \multicolumn{1}{c}{S/N$_{\mathrm{Bayes}}$} \\
        \hline IPTA DR2 subset & Reweighting & \texttt{HyperModel} & OS \\
        \hline DR2 Lite & $0.662 \pm 0.005$ & $0.57 \pm 0.03$ & 0.00 \\
        EDR2 & $2.841 \pm 0.006$ & $3.0 \pm 0.2$ & 0.92 \\
        Full DR2 & $1.39 \pm 0.04$ & $1.45 \pm 0.07$ & 0.79 \\
        \hline
    \end{tabular}
    \renewcommand{\arraystretch}{1}
    \caption{Detection statistics for Hellings-Downs correlations as a function of the dataset. The middle column reports Bayes factors for an Hellings-Downs cross-correlated model vs an auto-correlated CRN model, as estimated using both likelihood reweighting and the \texttt{HyperModel}. The last column reports the Bayesian S/N for Hellings-Downs correlations from a noise marginalized multiple component optimal statistic (OS) analysis.}
    \vspace{-\baselineskip}
    \label{tab:hd_BFs}
\end{table}

We report the Bayes factors we measure using each dataset in the middle column of Table~\ref{tab:hd_BFs}. To check consistency, we report the Bayes factors and sampling uncertainties measured using likelihood reweighting \citep{Hourihane2023} and using product-space sampling, also known as the \texttt{HyperModel} in \texttt{enterprise\_extensions} (\citealt{Hee+2016, Johnson+2024}). We estimate the uncertainties on the Bayes factors using the effective sample size for reweighting \citep{Hourihane2023}, and using bootstrapping methods for the \texttt{HyperModel}. The two methods agree on the Bayes factor estimates within sampling uncertainties in all cases except for DR2 Lite. However, it is likely that the uncertainty using the \texttt{HyperModel} is underestimated, as the model switch parameter usually has a large autocorrelation length. In comparison to Full DR2 ($\mathcal{B} \sim 1.4$), the cross-correlated model is less favored using DR2 Lite ($\mathcal{B} \sim 0.6$) and slightly more favored using EDR2 ($\mathcal{B} \sim 2.8$; Table~\ref{tab:hd_BFs}). Although it is unexpected and interesting that EDR2 should return the highest Bayes factor, this appears to be consistent with the OS analysis, especially given we are not considering the presence of a monopole. Overall, even though the Bayes factors are all $\mathcal{O}(1)$, these results are a promising sign regarding the potential for combined datasets to improve measurements of cross-correlations over Lite datasets.

\subsection{Single pulsar noise}
\label{subsection:Noise}

\begin{table*}
    \centering
    \label{tab:noise}
    \renewcommand{\arraystretch}{1.5}
    \begin{tabular}{c | c c c c | c c c c}
        \multicolumn{1}{c}{} & \multicolumn{4}{c}{DR2 Lite Single Pulsar Bayes Factors} & \multicolumn{4}{c}{Full DR2 Single Pulsar Bayes Factors} \\
        \hline Pulsar & $\mathcal{B}^{\mathrm{PLRN}}_0$ & $\mathcal{B}^{\mathrm{PLDM}}_0$ & $\mathcal{B}^{\mathrm{PLChr}}_0$ & $\mathcal{B}^{\mathrm{SWGP}}_0$ & $\mathcal{B}^{\mathrm{PLRN}}_0$ & $\mathcal{B}^{\mathrm{PLDM}}_0$ & $\mathcal{B}^{\mathrm{PLChr}}_0$ & $\mathcal{B}^{\mathrm{SWGP}}_0$ \\
        \hline J0030+0451 & $2.1$ & $1.0$ & $-$ & $-$ & $>\mathbf{10^3}$ & $3.6$ & $-$ & $-$ \\ 
        J0613--0200 & $3.2^*$ & $2.0^*$ & $1.2^*$ & $-$ & $>\mathbf{10^3}$ & $9.0$ & $\mathbf{19.8}$ & $-$ \\ 
        J1012+5307 & $1.1^\dagger$ & $0.6$ & $-$ & $-$ & $>\mathbf{10^3}^\dagger$ & $>\mathbf{10^3}$ & $-$ & $-$ \\ 
        J1022+1001 & $1.2^*$ & $3.0^*$ & $-$ & $0.8$ & $1.4$ & $\mathbf{56.8}$ & $-$ & $\mathbf{386.0}$ \\ 
        J1024--0719 & $1.7^*$ & $1.3^*$ & $-$ & $-$ & $>\mathbf{10^3}$ & $>\mathbf{10^3}$ & $-$ & $-$ \\ 
        J1455--3330 & $1.4$ & $1.4$ & $-$ & $-$ & $0.8$ & $0.7$ & $-$ & $-$ \\ 
        J1600--3053 & $0.8$ & $1.2^*$ & $1.6^*$ & $1.7$ & $\mathbf{15.7}$ & $3.0$ & $>\mathbf{10^3}$ & $\mathbf{292.7}$ \\ 
        J1640+2224 & $0.7$ & $0.7$ & $-$ & $-$ & $1.1$ & $>\mathbf{10^3}$ & $-$ & $-$ \\ 
        J1643--1224 & $0.7$ & $2.9^*$ & $1.0^*$ & $-$ & $>\mathbf{10^3}$ & $>\mathbf{10^3}$ & $>\mathbf{10^3}$ & $-$ \\ 
        J1713+0747 & $\mathbf{67.6}$ & $1.0$ & $0.9$ & $-$ & $>\mathbf{10^3}$ & $>\mathbf{10^3}$ & $>\mathbf{10^3}$ & $-$ \\ 
        J1730--2304 & $0.8$ & $1.1$ & $-$ & $-$ & $0.8$ & $0.7$ & $-$ & $-$ \\ 
        J1738+0333 & $0.9$ & $0.8$ & $-$ & $-$ & $1.2^*$ & $1.8^*$ & $-$ & $-$ \\ 
        J1744--1134 & $1.5^*$ & $1.1^*$ & $-$ & $-$ & $>\mathbf{10^3}$ & $\mathbf{50.0}$ & $-$ & $-$ \\ 
        J1853+1303 & $0.8$ & $0.8$ & $-$ & $-$ & $1.4$ & $1.1$ & $-$ & $-$ \\ 
        J1857+0943 & $9.6$ & $>\mathbf{10^3}$ & $-$ & $-$ & $\mathbf{1087.1}$ & $>\mathbf{10^3}$ & $-$ & $-$ \\ 
        J1909--3744 & $0.5$ & $>\mathbf{10^3}$ & $-$ & $-$ & $>\mathbf{10^3}$ & $>\mathbf{10^3}$ & $-$ & $-$ \\ 
        J1910+1256 & $0.8$ & $\mathbf{30.9}$ & $-$ & $-$ & $0.8$ & $\mathbf{13.7}$ & $-$ & $-$ \\ 
        J1918--0642 & $1.2^*$ & $1.9^*$ & $-$ & $4.1$ & $2.3$ & $>\mathbf{10^3}$ & $-$ & $4.0$ \\ 
        J1939+2134 & $>\mathbf{10^3}$ & $>\mathbf{10^3}$ & $>\mathbf{10^3}$ & $-$ & $>\mathbf{10^3}$ & $>\mathbf{10^3}$ & $>\mathbf{10^3}$ & $-$ \\ 
        J1955+2908 & $0.8$ & $\mathbf{24.6}$ & $-$ & $-$ & $0.9$ & $\mathbf{169.4}$ & $-$ & $-$ \\ 
        J2010--1323 & $1.5$ & $1.0$ & $-$ & $-$ & $1.4$ & $>\mathbf{10^3}$ & $-$ & $-$ \\ 
        J2124--3358 & $0.9$ & $0.8$ & $-$ & $3.7$ & $0.9$ & $0.9$ & $-$ & $1.0$ \\ 
        J2145--0750 & $2.2^*$ & $3.5^*$ & $-$ & $0.8$ & $>\mathbf{10^3}$ & $\mathbf{158.0}$ & $-$ & $7.0$ \\ 
        J2317+1439 & $0.7$ & $0.7$ & $-$ & $-$ & $0.6$ & $>\mathbf{10^3}$ & $-$ & $-$ \\
        \hline Total $\mathcal{B} > 10$: & 2 & 5 & 1 & 0 & 12 & 16 & 5 & 2 \\
        \hline
    \end{tabular}
    \renewcommand{\arraystretch}{1}
    \caption{Bayes factors for different processes computed from single pulsar analyses. The 24 pulsars featured here are those fom IPTA DR2 with data from 2 or more PTAs. $\mathcal{B}^{\mathrm{PLRN}}_0$ is the Bayes factor for a model with power-law red noise included vs the same model without, $\mathcal{B}^{\mathrm{PLDM}}_0$ for power-law DM noise, $\mathcal{B}^{\mathrm{PLChr}}_0$ for higher-order chromatic noise, and $\mathcal{B}^{\mathrm{SWGP}}_0$ for Gaussian perturbations to the solar wind electron density along the Earth-pulsar line of sight. These Bayes factors were computed as the Savage-Dickey density ratio using each signal's relevant amplitude parameter; where there were a lack of samples in the tail of each parameter's posterior, we place a lower bound of $\mathcal{B} > 10^3$. Bold values mark cases where $\mathcal{B} > 10$, indicating strong evidence for the noise process. The total number of bolded entries are tallied at the bottom of each column. Entries with asterisks$^*$ indicate cases of severe covariance within the noise model parameter space, such that either, but not both, marked signals are favored by the data (i.e. the data cannot distinguish the chromaticity of the process). The daggered$^\dagger$ entries for PSR J1012+5307 indicate that the high-frequency achromatic red noise process was used to compute the Bayes factor, rather than the 30 frequency red noise process.}
    \vspace{-\baselineskip}
\end{table*}

We next compare how the characterization of pulsar noise changes whether we use Full DR2 or DR2 Lite. Out of the 53 pulsars in IPTA DR2, 24 have timing data from 2 or more PTAs, while the remaining pulsars' data are the same in Full DR2 and DR2 lite. We therefore focus on how the noise properties, primarily red noise and chromatic noise, compare across these 24 pulsars. We do not assess the changes to the white noise or timing model parameters, but we acknowledge these parameters also play a large role in the total characterization of the pulsar.

First we examine how many new noise processes are detected using Full DR2 vs DR2 Lite. Table~\ref{tab:noise} shows the Bayes factors measured for each noise process and pulsar using DR2 Lite and Full DR2. These are estimated using the Savage-Dickey density ratio applied to the amplitude parameter of each noise process, e.g. $\mathcal{B}^{\mathrm{PLRN}}_0$ in PSR J0030+0451 is the Bayes factor for a model with DM noise and red noise vs a model with DM noise only. We compare the Bayes factors for red noise, DM noise, higher-order chromatic noise, and time-dependent perturbations to the solar wind density. Bold parameters indicate the noise process is measured with $\mathcal{B} > 10$, and where there are no posterior samples in the tail, we place a lower limit of $\mathcal{B} > 1000$. Dashes indicate the process was not included in that pulsar's noise model. While these Bayes factors all depend on our choice for how to construct the prior, we are applying the same model and priors to each dataset, therefore the prior-bias becomes less important for the purpose of performing a comparison.

Using the threshold $\mathcal{B} > 10$ to indicate a probable detection of the noise process, we find using Table~\ref{tab:noise} that using DR2 Lite, 2/24 pulsars detect achromatic red noise, 5/24 detect DM noise, 1/24 pulsars detect higher-order chromatic noise, and 0/24 detect solar wind density variations. Meanwhile, using Full DR2 we find 12/24 pulsars detect achromatic red noise, 16/24 detect DM noise, and 5/24 pulsars detect higher-order chromatic noise, indicating 10 new detections of red noise, 11 new detections of DM noise, and 4 new detections of chromatic noise in the combined data. 2/17 pulsars newly detect solar wind density variations. Overall, 16/24 pulsars newly detect a noise process that was not detected using DR2 Lite.

Additionally, we find some of these noise processes are not detected in the Lite dataset specifically because of source confusion. Bayes factors marked with asterisks in Table~\ref{tab:noise} signify that there is a strong case of parameter covariance between two or more processes in the model, such that the removal of one process from the model increases the detection significance of the other, and vice versa. In other words, one or more processes are favored by the data, but the data cannot distinguish which one it is (\citealt{Lentati+2016, EPTAnoise2023, Ferranti+2024} each discuss this effect in more depth). We observed this behavior in 8 pulsars using DR2 Lite. For PSRs J1600--3053 and J1643--1224 the source confusion is between DM and chromatic noise. For PSRs J1022+1001, J1024--0719, J1744--1134, J1918--0642, and J2145--0750 the source confusion is between DM and achromatic red noise. For PSR J0613--0200, the data prefer either to include only DM noise, or to include both achromatic and higher-order chromatic noise. This behavior is detailed more clearly for PSR J0613--0200 in Figure~\ref{fig:J0613-0200}. In all of these cases, using Full DR2 results in strong measurement of 1 or more of these noise processes, resolving the source confusion. Additionally, for PSR J1738+0333 there is no evidence of achromatic red noise or DM noise in DR2 Lite, but using Full DR2 it has entered into the source confusion regime of the two signals.

We further find from comparing the chromatic, DM, and achromatic red noise parameters from the single-pulsar noise analyses of all 24 pulsars that the changes in the noise parameters going from DR2 Lite to Full DR2 fall under three general categories:
\begin{enumerate}
    \item \textbf{Consistency \& improved constraints}: In this case, the noise parameter posteriors measured using Full DR2 are more constrained than the posteriors measured using DR2 Lite, but both sets of posteriors are consistent with one another. This demonstrates the expected effect that adding more data results in more constrained posteriors. The majority of pulsars (15 out of 24) best fit into this category: PSRs J0030+0451, J0613--0200, J1012+5307, J1022+1001, J1024--0719, J1455--3330, J1600--3053, J1640+2224, J1738+0333, J1744--1134, J1857+0943, J1918--0642, J2010--1323, J2145--0750, J2317+1439.
    \item \textbf{Inconsistency \& improved constraints} In this case, using the combined data changes the noise parameter posteriors such that there is noticeable tension between DR2 Lite and Full DR2 in the posteriors for one or more parameters. This could possibly arise if we have model misspecification, or if additional unmitigated chromatic noise enters into the data combination. Otherwise, there may be more complicated interactions between the different model components than expected. 3 out of 24 pulsars most cleanly fit into this category: PSRs J1643--1224, J1713+0747, J1909--3744.
    \item \textbf{Consistency \& similar constraints}: In this case, the posterior distributions are very similar and consistent with one another. This indicates the full data combination is not significantly improving noise characterization at the single pulsar level. 6 out of 24 pulsars best fit into this category: PSRs J1730--2304, J1853+1303, J1910+1256, J1939+2134, J1955+2908, J2124--3358.
\end{enumerate}
These categories summarize the effects we observe here of data combination of noise parameter characterization, though some pulsars also toe the line between these categories.

We next follow up with closer examinations of three pulsars, PSRs J0613--0200, J1909--3744, and J1910+1256, which each serve as an illustrative case from each category. Figures~\ref{fig:J0613-0200}--\ref{fig:J1910+1256} compare their noise parameter distributions obtained using DR2 Lite (orange) and using Full DR2 (blue). We also show the CRN parameters from Full DR2 overplotted over the total achromatic red noise parameters in each figure.

\subsubsection{PSR J0613--0200 | Consistency \& improved constraints}

\begin{figure}
    \centering
    \includegraphics[width=\linewidth]{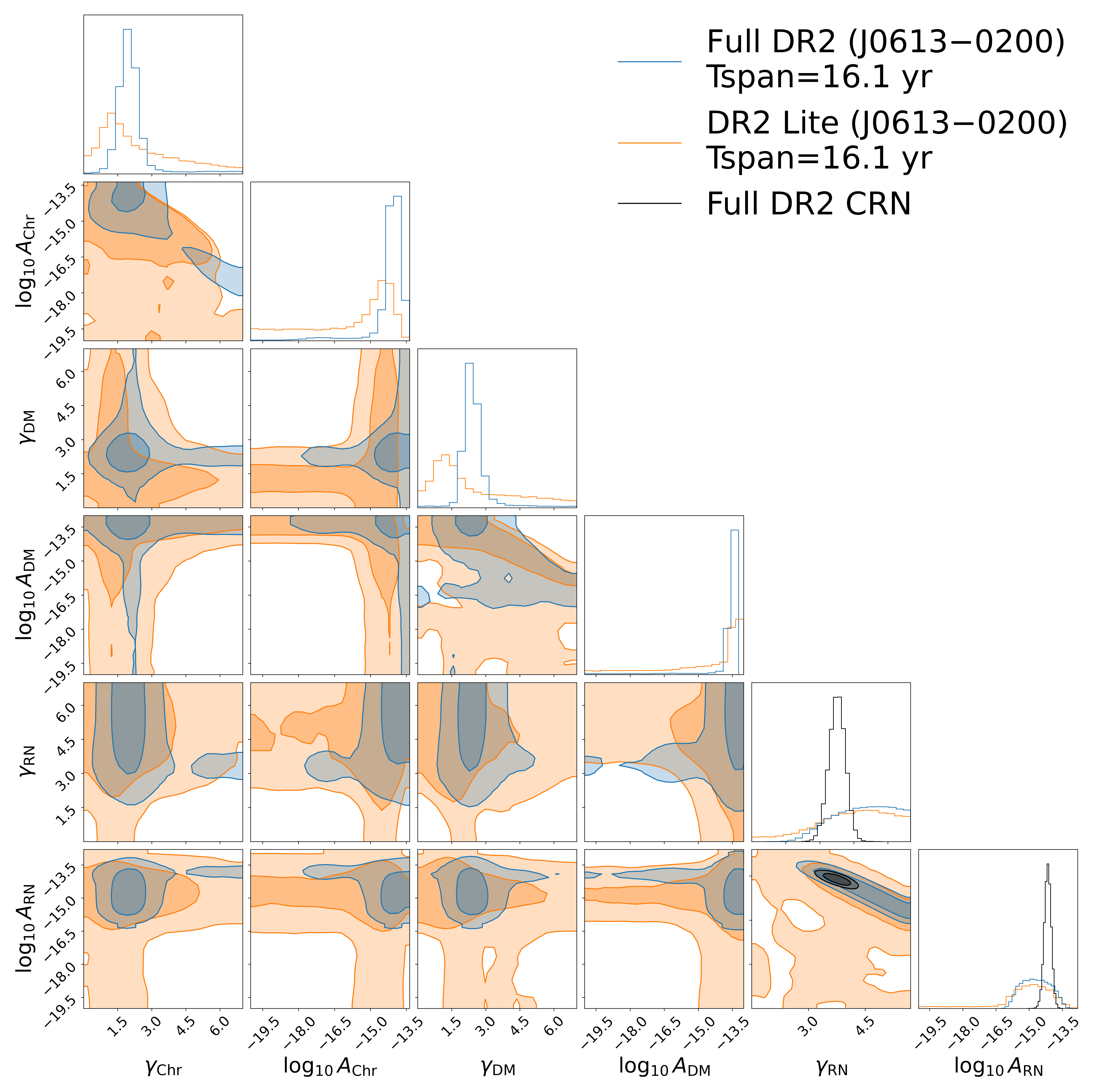}
    \vspace{-\baselineskip}
    \caption{Comparison of noise parameter posteriors for chromatic, DM, and red noise from the single pulsar noise analysis of PSR J0613--0200 using DR2 Lite (orange) and using Full DR2 (blue). Panels show the amplitude and spectral index parameters for each noise process. Overplotted in the achromatic red noise panels are the CRN parameters from the analysis of Full DR2, as this noise is a subset of the total achromatic red noise in each pulsar. Contours enclose 68\% and 95\% 2D Bayesian credible intervals.}
    \vspace{-0.5\baselineskip}
    \label{fig:J0613-0200}
\end{figure}

Figure~\ref{fig:J0613-0200} shows that the Lite dataset for PSR J0613--0200 is not sufficient to distinguish the different sources of noise from each other. The 1D posteriors over the DM and red noise amplitudes each have long tails, with posterior support near $\log_{10}A_{\mathrm{RN}} = -20$ and $\log_{10}A_{\mathrm{DM}} = -20$, indicating the signals are not detected. However, the 2D posterior over both parameters shows has a deficiency of samples where both $\log_{10}A_{\mathrm{RN}} = -20$ \emph{and} $\log_{10}A_{\mathrm{DM}} = -20$; i.e. the data does favor inclusion of at least one signal with high significance, but they cannot distinguish which. PSR J0613--0200 also includes higher-order chromatic noise in its model, and this same effect also occurs between $\log_{10}A_{\mathrm{DM}}$ and $\log_{10}A_{\mathrm{Chr}}$. This means in this case, a side effect of including the higher-order chromatic noise is to slightly \emph{increase} the detection significance for achromatic red noise, rather than to decrease it. This source confusion effect does not occur directly between $\log_{10}A_{\mathrm{Chr}}$ and $\log_{10}A_{\mathrm{RN}}$, as there are still samples in the 2D region where $\log_{10}A_{\mathrm{RN}} = -20$ \emph{and} $\log_{10}A_{\mathrm{DM}} = -20$ (instead the L-shape of the 2D 95\% credible interval in Figure~\ref{fig:J0613-0200} is just a product of their 1D posteriors). In contrast, using Full DR2 allows much more precise measurements of each of these processes, and they are better distinguished from one another. These improvements likely result from the improved data cadence and radio frequency bandwidth achieved by data combination for this pulsar.

PSR J0613--0200 also shows the strongest preference for the CRN out of the pulsars in EDR2 based on the DFs from \S\ref{subsection:dropout}, as well as the original IPTA DR2 analysis \citep{Antoniadis+2022}. This is demonstrated in Figure~\ref{fig:J0613-0200} by the excellent overlap between the achromatic red noise parameters from the single-pulsar analysis and the CRN parameters from the full-PTA analysis using Full DR2. Meanwhile, the red noise parameters measured using DR2 Lite are much less constrained, indicating weaker evidence for a detection of red noise as well as a much higher upper limit on red noise. In total, the improved characterization of pulsar noise resulting from the data combination directly translate here to improved measurement of the CRN. We can infer a similar story is at play for several of the other pulsars with improved constraints, particularly those pulsars in Table~\ref{tab:noise} with covariances recorded between noise parameters, as well as those that show improvements to their DFs in Figure~\ref{fig:dropout}. For other pulsars, such as PSR J2317+1439, the data combination improves their characterization of DM noise but not achromatic red noise.

\subsubsection{PSR J1909--3744 | Inconsistency \& improved constraints}

\begin{figure}
    \centering
    \includegraphics[width=\linewidth]{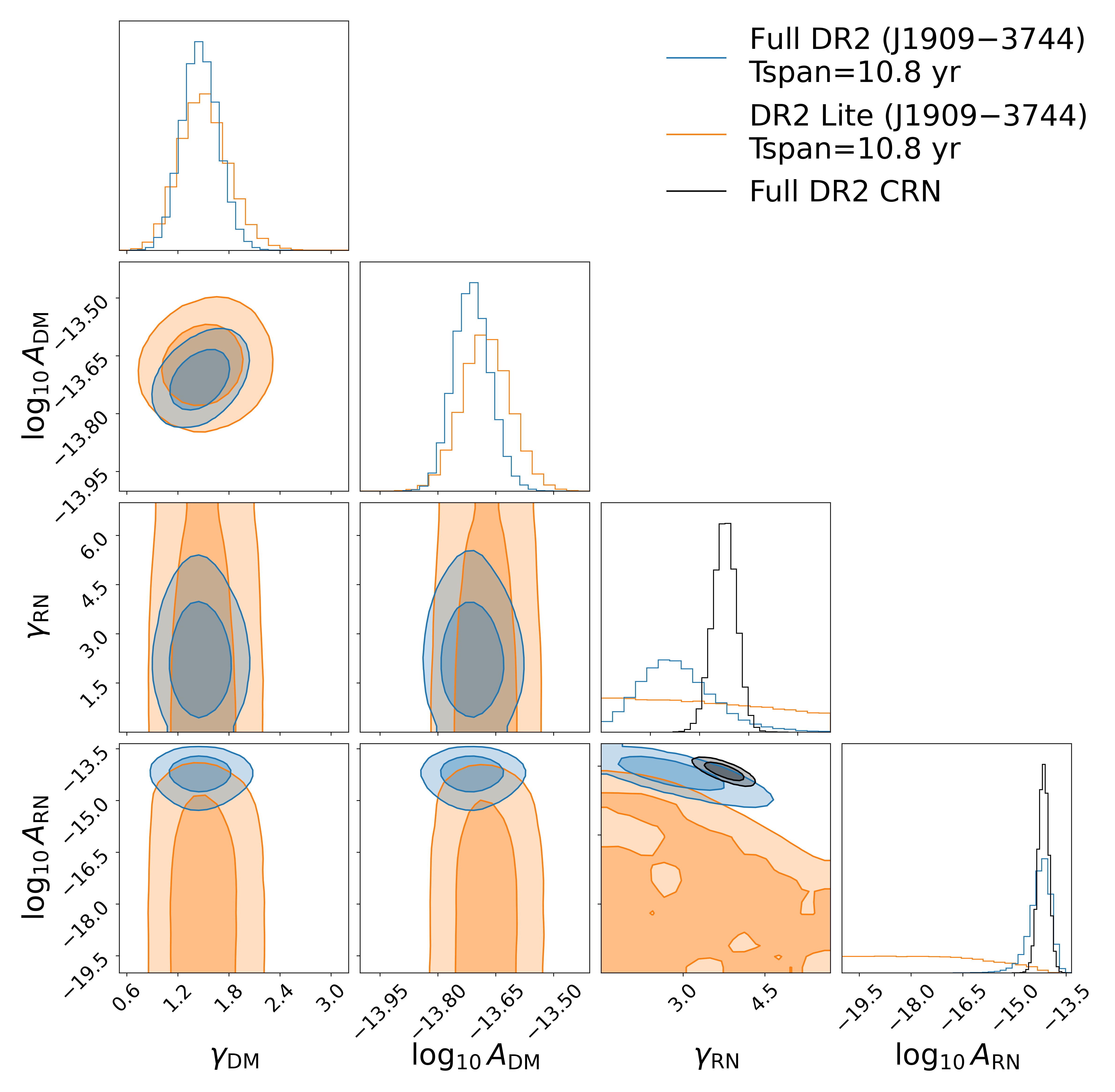}
    \vspace{-\baselineskip}
    \caption{Comparison of noise parameter posteriors for DM and red noise from the single pulsar noise analysis of PSR J1909--3744 using DR2 Lite (orange) and using Full DR2 (blue). See Figure~\ref{fig:J0613-0200} caption for more details.}
    \vspace{-0.5\baselineskip}
    \label{fig:J1909-3744}
\end{figure}

Unlike the case for PSR J0613--0200, Figure~\ref{fig:J1909-3744} shows that the achromatic red noise distributions obtained using DR2 Lite and Full DR2 are in tension with each other. Using DR2 Lite, no red noise is detected in PSR J1909-3744, however the red noise detected using Full DR2 lies above the expected region using DR2 Lite (i.e., the 68\% 2D credible regions do not overlap). This makes it appear that a red noise process has emerged in the combined data where there was none previously in DR2 Lite. There is no difference in observation timespan between DR2 Lite and Full DR2 for this pulsar, so nonstationary noise is unlikely to cause this. Furthermore, Figure~\ref{fig:J1909-3744} shows both datasets detect DM variations with similar characteristics in PSR J1909--3744, which makes the possibility of \emph{new} chromatic noise entering into the combined dataset seem unlikely. However, the DM noise parameters are measured more precisely using the fully-combined data. Improving the precision on DM variations as a result of improved cadence and radio frequency coverage (shown in Figure~\ref{fig:TOAs}) may have the effect of helping to uncover the red noise present in the fully-combined data. This effect only makes sense in tandem with the $\log_{10}$-uniform priors used on $A_{\mathrm{RN}}$, which heavily downweight the presence of noise in the dataset. Another possibility is that some level of model misspecification is at play and resulting in inconsistent noise properties across the two datasets.

PSR J1909--3744 is the pulsar in the most tension with the CRN using DR2 Lite, with the lowest DF out of all pulsars (Figure~\ref{fig:dropout}). The CRN parameters from Full DR2 overplotted in Figure~\ref{fig:J1909-3744} help explain this -- they are well above the 95\% Bayesian credible interval of the total achromatic red noise measured using the Lite dataset under the $\log_{10}$-uniform prior. Meanwhile, using EDR2, PSR J1909--3744 is more agnostic to the CRN measurement, indicated in Figure~\ref{fig:J1909-3744} by the modest overlap between the achromatic red noise using Full DR2 and the CRN. These effects are also evident in the corner plots for PSRs J1713+0747 and J1744--1134 (not shown): the achromatic red noise measured using DR2 Lite lies below the level of the red noise measured using Full DR2.

\subsubsection{PSR J1910+1256 | Consistency \& similar constraints}

\begin{figure}
    \centering
    \includegraphics[width=\linewidth]{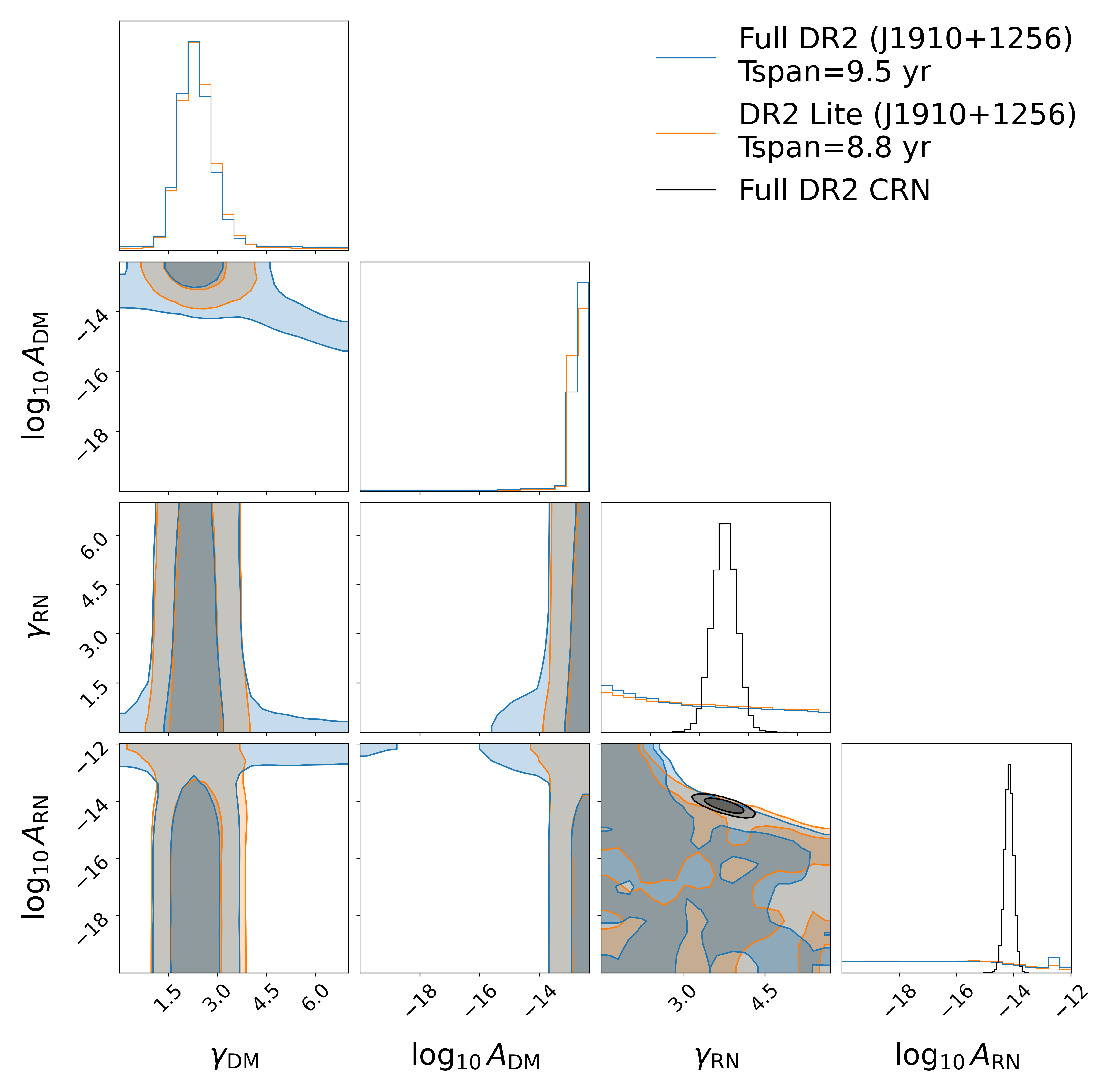}
    \vspace{-\baselineskip}
    \caption{Comparison of noise parameter posteriors for DM and red noise from the single pulsar noise analysis of PSR J1910+1256 using DR2 Lite (orange) and using Full DR2 (blue). See Figure~\ref{fig:J0613-0200} caption for more details.}
    \vspace{-0.5\baselineskip}
    \label{fig:J1910+1256}
\end{figure}

\begin{figure*}
    \centering
    \begin{tabular}{cc}
         \includegraphics[width=0.45\linewidth]{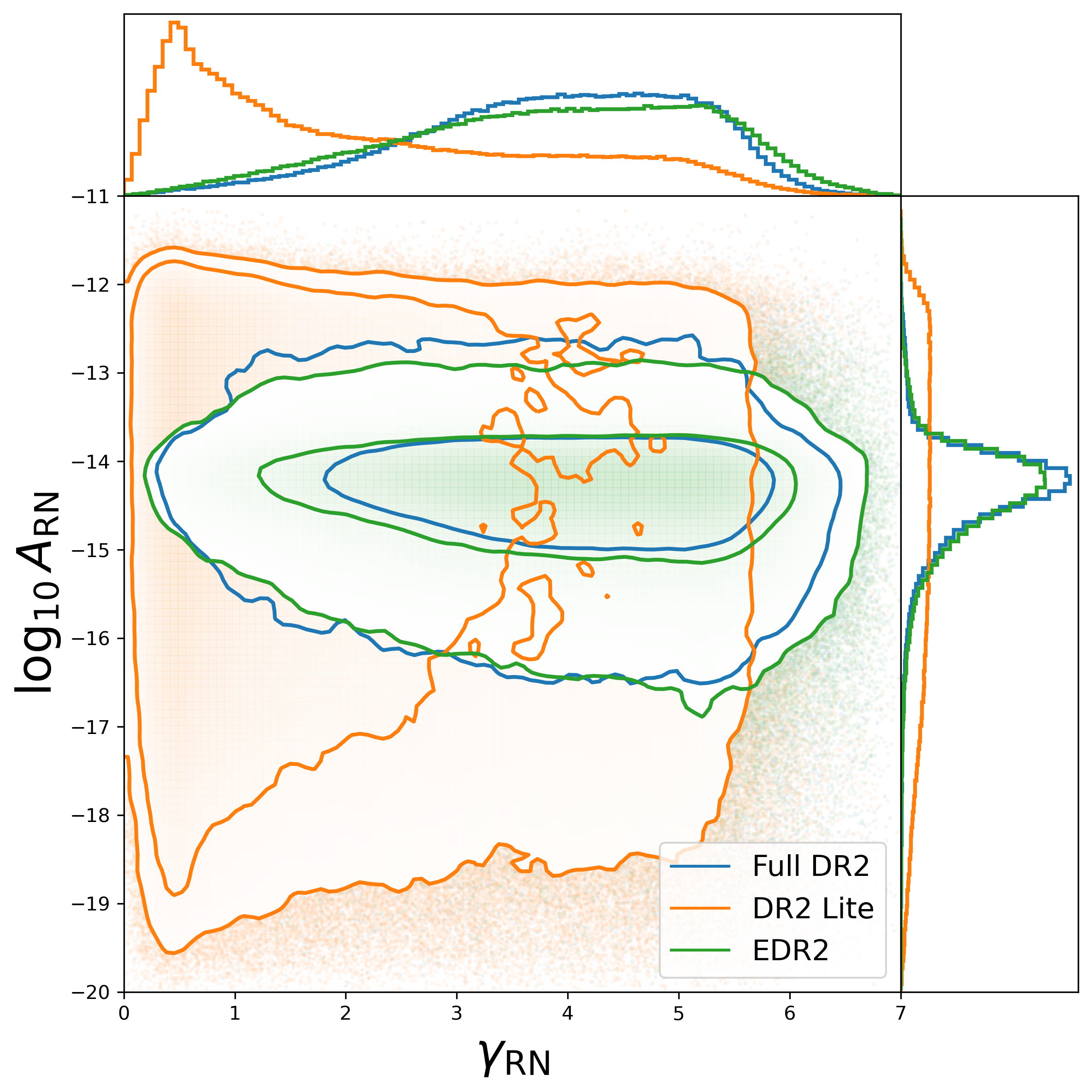} &  \includegraphics[width=0.45\linewidth]{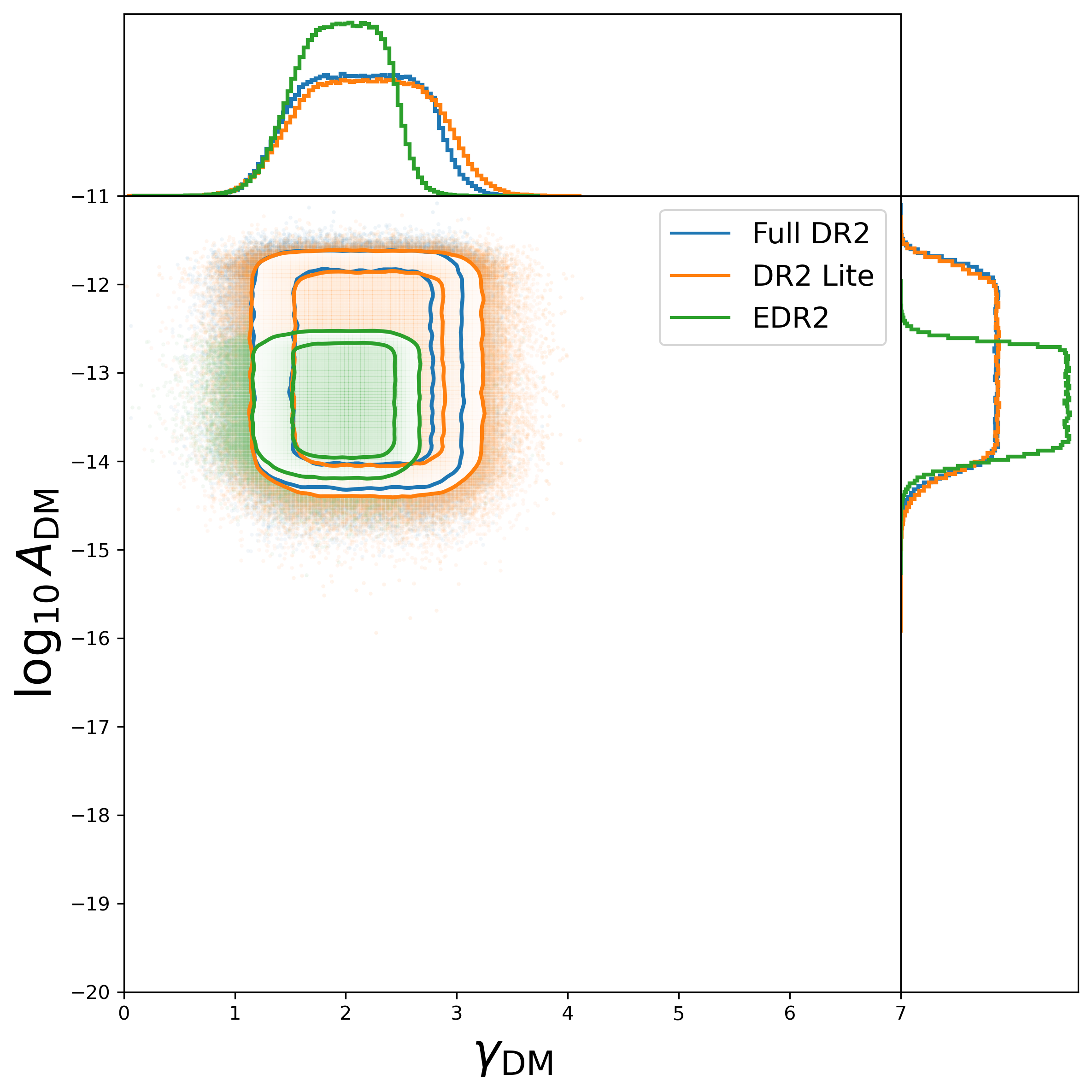}
    \end{tabular}
    \vspace{-1.0\baselineskip}
    \caption{Distributions encoding the ensemble noise properties of the IPTA DR2 pulsars as compared whether using DR2 Lite (53 pulsars), EDR2 (22 pulsars), or Full DR2 (53 pulsars). \emph{Left:} Intrinsic red noise properties of the ensemble (decoupled from a separate $\gamma = 13/3$ process in each pulsar) are substantially broader using DR2 Lite as opposed to EDR2 and Full DR2, which each place equally competitive constraints. \emph{Right:} DM noise properties are tightly constrained using all three datasets, with the EDR2 analysis producing an overly-tight distribution due to the inclusion of fewer pulsars.}
    \vspace{-1.5\baselineskip}
    \label{fig:HBM_priors}
\end{figure*}

PSR J1910+1256 is the last unique case we cover here. Unlike the previous cases, Figure~\ref{fig:J1910+1256} shows the posteriors measured for PSR J1910+1256 using DR2 Lite are similar to the posteriors measured using Full DR2, with nearly identical red noise and very similar DM noise. Thus, there apparently exist cases where data combination does not improve noise characterization at the single-pulsar level. The largest factor playing into this may simply be that not enough new data are added into the combination. Indeed, PSR J1910+1256 has one of the lowest FoMs out of the mutli-PTA pulsars (Figure~\ref{fig:FoM}), also has a $\mathrm{DF} \sim 1$ (Figure~\ref{fig:dropout}), indicating its combined dataset is not yet advanced enough to measure the CRN. While a future combination may likely yield the emergence of a red noise process in this and similar pulsars, it is also important to consider that pulsar sensitivity may eventually saturate, and therefore reduce the effect of combined data. For example, even with increasing telescope sensitivity, some pulsar's will eventually become jitter-noise limited. However, it does not seem we have reached this regime for the majority of pulsars \citep{Lam+2019}. Other pulsars, such as PSR J1939+2134, may be dominated by very strong intrinsic red noise processes, explaining why PSR J1939+2134's noise is well-measured in both DR2 Lite and Full DR2.

\subsection{Ensemble Noise Properties}
\label{subsection:HBMs}

For our final analysis, we use the framework of hierarchical Bayesian modeling \citep{LoredoHendry2019, Thranetalbot2019, vanHaasteren2024} to estimate the ensemble noise properties of intrinsic red noise and DM variations from each dataset (DR2 Lite, EDR2, and Full DR2), following closely the procedures described in \citet{GoncharovSardana2025}. The ensemble properties are encoded in the distribution $p(\bm{\theta}|\mathcal{M}_\Lambda)$, where we define $\bm{\theta}$ as our noise parameters $\bm{\theta} = \{\log_{10}A_{\mathrm{RN}},\gamma_{\mathrm{RN}},\log_{10}A_{\mathrm{DM}},\gamma_{\mathrm{DM}}\}$. $\mathcal{M}_\Lambda$ designates the hierarchical model with hyperparameters $\bm{\Lambda}$, which are learned from the data $\bm{\delta t}$, while $\mathcal{M}_\oslash$ designates the original model using uninformative priors. For the current analysis, the ensemble noise distribution $p(\bm{\theta}|\mathcal{M}_\Lambda)$ is neither a prior nor a posterior on pulsar noise parameters. Rather, it is a posterior predictive distribution for noise parameters, as it is informed by the population of millisecond pulsars in the current dataset. For an independent analysis, $p(\bm{\theta}|\mathcal{M}_\Lambda)$ could be used as a population-informed prior distribution on the noise parameters.

We estimate the ensemble properties following \citet{GoncharovSardana2025} by first inferring the hyperparameters $\bm{\Lambda}$ from single pulsar analyses using the marginalized likelihood obtained from importance sampling \citep{Thranetalbot2019}
\begin{align}
    \mathcal{L}(\bm{\delta t}|\bm{\Lambda},\mathcal{M}_\Lambda) &= \prod_i^{N_{\mathrm{psr}}}\mathcal{Z}(\bm{\delta t}_i|\mathcal{M}_\oslash)\int\frac{\pi(\bm{\theta}_i|\bm{\Lambda},\mathcal{M}_\Lambda)}{\pi(\bm{\theta}_i|\mathcal{M}_\oslash)}\mathcal{P}(\bm{\theta}_i|\bm{\delta t}_i,\mathcal{M}_\oslash)d\bm{\theta}_i \\
    &\cong \prod_i^{N_{\mathrm{psr}}}\frac{\mathcal{Z}(\bm{\delta t}_i|\mathcal{M}_\oslash)}{n_i}\sum_k^{n_i}\frac{\pi(\bm{\theta}_i^k|\bm{\Lambda},\mathcal{M}_\Lambda)}{\pi(\bm{\theta}_i^k|\mathcal{M}_\oslash)},
    \label{eq:HBM_likelihood_approx}
\end{align}
where $N_{\mathrm{psr}}$ is the number of pulsars, $n_i$ is the number of posterior samples obtained from the single pulsar $i$'s noise analysis, and $\mathcal{Z}(\bm{\delta t}_i|\mathcal{M}_\oslash)$ is the evidence from the single pulsar $i$'s noise analysis (the evidences may be treated as an unknown normalization for the purpose of parameter estimation). In Equation~\ref{eq:HBM_likelihood_approx}, each pulsar's integral is estimated by iterating over the $n_i$ posterior samples (i.e., $\bm{\theta}_i^k$ is the $k$'th posterior sample for the $i$'th pulsar). As suggested by \citet{GoncharovSardana2025}, we attempt to separate intrinsic red noise contributions from the GWB by drawing red and DM noise samples from the single pulsar posteriors obtained from our factorized likelihood analyses, which each contained an additional red noise term in each pulsar at fixed $\gamma=13/3$ which provides a channel to separate the GWB from intrinsic noise. After defining a suitable hyperprior $\pi(\bm{\Lambda}|\mathcal{M}_\Lambda)$, we can obtain samples over the posterior $\mathcal{P}(\bm{\Lambda}|\bm{\delta t},\mathcal{M}_\Lambda)$ from stochastic sampling, and numerically marginalize over $\bm{\Lambda}$ to obtain the ensemble noise distribution \citep{GoncharovSardana2025}
\begin{align}
    p(\bm{\theta}|\mathcal{M}_\Lambda) &= \int\pi(\bm{\theta}|\bm{\Lambda},\mathcal{M}_\Lambda)\mathcal{P}(\bm{\Lambda}|\bm{\delta t},\mathcal{M}_\Lambda)d\bm{\Lambda} \label{eq:marginalized_pop_prior} \\
    &\cong \frac{1}{n_p}\sum_k^{n_p}\pi(\bm{\theta}|\bm{\Lambda}_k,\mathcal{M}_\Lambda),
\end{align}
where the second line suggests we concatenate samples from the prior $\pi(\bm{\theta}|\bm{\Lambda}_k,\mathcal{M}_\Lambda)$ over $n_p$ $\bm{\Lambda}$ samples drawn from the posterior $\mathcal{P}(\bm{\Lambda}|\bm{\delta t},\mathcal{M}_\Lambda)$. Equation~\ref{eq:marginalized_pop_prior} shows how $p(\bm{\theta}|\mathcal{M}_\Lambda)$ implicitly depends on the data $\bm{\delta t}$, and as such it should be interpreted as a posterior predictive distribution for pulsar noise parameters rather than a true prior on $\bm{\theta}$.

We are left with some freedom to define the hyperparameters $\bm{\Lambda}$, the functional form of $\pi(\bm{\theta}|\bm{\Lambda},\mathcal{M}_\Lambda)$, and the hyperprior $\pi(\bm{\Lambda}|\mathcal{M}_\Lambda)$. We choose to keep the uniform priors on $\bm{\theta}$ and infer the min and max ranges of the prior as our hyperparameters $\bm{\Lambda}$, such that
\begin{align}
    \pi(\theta_j|\Lambda_{\max,j}, \Lambda_{\min,j}) = \begin{cases}
        \frac{1}{\Lambda_{\max,j} - \Lambda_{\min,j}} &\text{if } \Lambda_{\min,j} < \theta_j < \Lambda_{\max,j}, \\
        0 &\text{elsewhere},
    \end{cases}
\end{align}
for each of our four noise parameters denoted by $j$. This choice is based on \citet{GoncharovSardana2025}, who found using EPTA DR2 that this uniform distribution model was preferred with a higher Bayes factor than alternative models using a normal distribution or a mixture of normal and uniform distributions. We enforce the constraint $\Lambda_{\min,j} < \Lambda_{\max,j}$ by defining our hyperpriors to draw from the distributions
\begin{align}
    p(\Lambda_{\min,j}) &= \frac{2(u_j-\Lambda_{\min,j})}{u_j - l_j}& & \{l_j < \Lambda_{\min,j} < u_j\}, \\
    p(\Lambda_{\max,j}|\Lambda_{\min,j}) &= \frac{1}{u_j-\Lambda_{\min,j}}& & \{\Lambda_{\min,j} < \Lambda_{\max,j} < u_j\},
\end{align}
where $u_j$ and $l_j$ are the original upper and lower bounds on the red/DM noise parameters (\S\ref{subsection:noise_models}).

We use \texttt{numpyro} \citep{numpyro1, numpyro2} to define our hierarchical Bayesian model and \texttt{jaxns} \citep{jaxns} to infer the posterior distribution $\mathcal{P}(\bm{\Lambda}|\bm{\delta t},\mathcal{M}_\Lambda)$ using nested sampling. Figure~\ref{fig:HBM_priors} shows the resulting ensemble distributions of the parameters of our two noise processes, $p(\log_{10}A_{\mathrm{RN}},\gamma_{\mathrm{RN}}|\mathcal{M}_\Lambda)$ and $p(\log_{10}A_{\mathrm{DM}},\gamma_{\mathrm{DM}}|\mathcal{M}_\Lambda)$ after numerical marginalization over the hyperparameters $\bm{\Lambda}$. The left hand side shows that DR2 Lite does not produce strong constraints on the intrinsic red noise properties of the ensemble, whereas EDR2 produces equal constraints on the ensemble noise properties as Full DR2. This consistent in particular with the single pulsar results table~\ref{tab:noise} which shows only 2 pulsars out of a subset of 24 have strongly detected red noise using DR2 Lite (the list increases to 5 out 53 once including PSRs J0437$-$4715, J0621+1002, and J1824$-$2452A), while Full DR2 is capable of constraining red noise in the majority of pulsars. Meanwhile, the right hand side of Figure~\ref{fig:HBM_priors} shows that DR2 Lite can be used to place similar levels of constraints on the ensemble properties of DM variations as Full DR2, even despite detecting DM noise in fewer pulsars than Full DR2 (but still detecting more DM noise processes than red noise processes; Table~\ref{tab:noise}). The distributions are centered near $\gamma_{\mathrm{DM}} \sim 2$, but display errors consistent with the value $\gamma_{\mathrm{DM}} = 8/3$ expected for DM variations from Kolmogorov turbulence \citep{Keith+2013}. EDR2 in the meantime displays a tighter distribution of DM noise properties, centered near lower amplitude and spectral index. This is most likely because EDR2 contains only 22 pulsars, and neither PSRs J1721$-$2457 and J1903+0327, two high-DM pulsars with the highest DM noise amplitudes in IPTA DR2, are included among the 22. 

Overall, this analysis complements and reinforces our results throughout the rest of this work that DR2 Lite is not as informative as the combined datasets, while the best subset of combined pulsars comprising EDR2 produce comparable results to a full-combination of all pulsars.

\section{Summary \& Discussion}
\label{section:Discussion}

\subsection{Summary of the Lite method}

The ``Lite'' method is a novel and computationally efficient approach to create joint-PTA datasets for GW searches.  Full data combination, while essential for maximizing GW sensitivity, is a meticulous and resource-intensive process. The Lite method circumvents these challenges by focusing on individual pulsar datasets, enabling quick searches for different GW signals using the latest data. Since Lite datasets can be created on-the-fly using existing PTA data, they are particularly valuable in scenarios where timely results are required, such as when evaluating newly acquired data or gauging the potential of preliminary observations.

By using a FoM to maximize GW sensitivity, Lite datasets represent the optimal combination of single-PTA pulsar data streams that can be achieved prior to the formal combination of all pulsars. The FoM used here comes from the analytic scaling law for the GWB S/N in the intermediate-signal regime, as presented in \citet{siemens2013}. This serves as a PTA-agnostic and statistically robust metric for selecting the most sensitive pulsar data streams from each PTA, while ensuring that the initial analysis retains focus on pulsars that contribute the most to detecting the GW signal of interest, i.e. the GWB. Furthermore, the Lite method allows the flexibility to use different FoMs to construct Lite datasets optimized for different GW signals, such as continuous GWs. We emphasize, however, that the FoMs we use here also make unrealistic assumptions about the data, such as ignoring the effects of uneven data sampling, the presence of chromatic noise, or loud intrinsic red noise. Radio-frequency band coverage in particular should be an important factor for selecting maximally sensitive single-PTA datasets \citep{Sosa+2024, Ferranti+2024}. One way to overcome these limitations and create more realistic FoMs in the future might be to numerically compute a FoM using sensitivity curves \citep{Hazboun+2019, Baier+2024}. Future metrics might also take the narrow bandwidths into account by adjusting their errorbars using later measurements of DM in the pulsar \citep{Sosa+2024}. An additional improvement would be to account for sky separation between pulsars pairs when computing the S/N for a GWB, as was done in \citet{Speri+2023}.

\subsection{Summary of results using IPTA DR2}

To test the Lite method, we create a Lite version of IPTA DR2 \citep{Perera+2019}, which we call DR2 Lite. We first carry out a GWB search on DR2 Lite, next comparing the results with a truncated version of Full DR2 using the 22 most sensitive pulsars, which we call EDR2, and finally comparing with Full DR2 (first analyzed in \citealt{Antoniadis+2022}). Each version of the dataset keeps all single-frequency and legacy data.

We measure a significant CRN process in all versions of IPTA DR2, including DR2 Lite. This provides a proof of concept that a Lite dataset may improve upon the individual datasets provided by the regional PTAs while also providing a sneak peek at the signals which may be detected in a fully-combined dataset. However, the combined datasets still remain superior. Using the Factorized Likelihood, we show that the detection significance of the CRN improves by orders of magnitude progressing through each stage of the combination process. Data combination also improves the precision of CRN parameter inference. Although the evidence for Hellings-Downs correlations in IPTA DR2 is weak, we tenuously find using both a Bayesian and an OS analysis that the combined data contain more support for Hellings-Downs correlations than DR2 Lite. Finally, across the board EDR2 produces nearly identical cross-correlation statistics and spectral characterization results as Full DR2, demonstrating that indeed using the FoM or similar statistic to inform the order in which to combine pulsars may yield valuable combined dataset at the intermediate stage of data combination, before all pulsars have been combined.

Alongside the above results, we found the amplitude distribution of the CRN is shifted to higher values using DR2 Lite than using EDR2 or Full DR2, suggesting unmitigated intrinsic noise is present in some of the DR2 Lite pulsars and leaking into the common channel. This conclusion is reaffirmed using various ``dropout'' analyses that show the CRN amplitude using DR2 Lite is strongly dependent on individual pulsars. The multiple component OS analysis also finds a larger amplitude of monopolar correlations in DR2 Lite. This monopole could be partially responsible or connected to the larger CRN amplitude, or it may just be a statistical fluctuation. By using combined data, the amplitude distributions of both the CRN and the monopole shift to lower values, which shows that data combination is capable of mitigating these systematic errors with no required knowledge of their source.

While these results are encouraging, we have not fully stress-tested the Lite method or forecasted its potential for future datasets -- this would require analyses of numerous detailed simulations of the data combination process, which is beyond the scope of this work. Another caveat is we do not omit or otherwise account for legacy or single-frequency data in IPTA DR2 from any version of the analysis. As shown by \citet{EPTADR2_gwb, Ferranti+2024}, including legacy data in the analysis helps to constrain the spectrum but is less useful for measuring the Hellings-Downs curve.

Finally, the effects of data combination are also noticeable on the single pulsar level. After examining 24 pulsars in DR2 Lite which have multi-PTA data in Full DR2, we detect at least one new intrinsic noise process using Full DR2 that we could not detect using DR2 Lite in 16 out of 24 pulsars. These improvements in noise characterization appear to correlate with improvements in the effective radio frequency bandwidth and data cadence that result from data combination. Meanwhile, DR2 Lite resulted in similar constraints on pulsar noise for only 6 out of 24 pulsars. In most of these cases, it appears the data combination did not improve band coverage at low frequencies enough to improve constraints on DM variations, and/or the Lite dataset already contained the majority of the TOAs in Full DR2, which likely dominated the statistics. These conditions are unlikely to hold true for many pulsars in the next IPTA dataset, as numerous data at low radio frequencies from LOFAR \citep{LOFAR2011}, NenuFAR \citep{NenuFAR2012}, the GMRT \citep{Joshi2018}, and CHIME \citep{CHIME2021} are now becoming available for combination. We should therefore expect to see that the next data combinations will further improve pulsar noise characterization, and by extension, sensitivity to GWs. The reduced number of pulsars with significant detections of noise in DR2 Lite also translate to less informative ensemble distributions of pulsar red noise properties.

\subsection{Future Directions}

Looking to the future, the Lite method has a clear role as an intermediate step in the analysis of PTA datasets. This was demonstrated recently in \citet{3P+2024} where several pseudo-IPTA datasets were composed from the most recent EPTA, NANOGrav, and PPTA datasets using a factorized likelihood approach. While these datasets were not created using a FoM, they are similar in spirit to the Lite method we present here. \citet{3P+2024} found that adding pulsars to each PTA's dataset using this method consistently results in a higher GWB S/N and more precise constraints on $A_{\mathrm{CRN}}$ than what one obtains using each PTA's individual data release. This further affirms the capability for the Lite method to improve measurements of GW signals. Extrapolating the results of our study comparing IPTA DR2 with its Lite version, we expect the upcoming IPTA dataset, IPTA DR3, will further improve GWB spectral characterization and GW detection prospects \citep{Good+2023}. Furthermore, our analysis of EDR2 reinforces that we can capture a large amount of the information in a fully combined dataset using an intermediate version with fewer pulsars, as expected based on  \citet{Speri+2023}. This suggests the creation and analysis of an IPTA ``EDR3'' as a way to start reaping the rewards of a fully-combined IPTA DR3 at an earlier time. Though we do not explore this here, the creation of even more optimal joint-PTA datasets should also be possible with a hybrid approach by using combined data for the most sensitive pulsars and using uncombined data for the remaining pulsars.

The Lite analysis may have further unquantified benefits if curating datasets for other nHz GW searches, such as continuous GWs from individual SMBHBs, as it may be used to rapidly improve the sky coverage of the PTA. Of the regional PTAs, three are northern hemisphere (\hbox{CPTA}, \hbox{EPTA}, and NANOGrav), one is quasi-equatorial (InPTA), and two are southern hemisphere (MPTA and PPTA). The maximum sky coverage achieved by any PTA is approximately 75\%, due to telescope elevation limits, while for transit telescopes such as CHIME, the sky coverage is considerably less. This sky coverage is perhaps only a secondary concern for the initial studies of the isotropic stochastic GWB, but it is a major concern for efforts to detect and study individual GW sources, as well GWB anisotropy. These gains in sky coverage may be achieved immediately with the creation of Lite datasets. 

In conclusion, the Lite method does not replace the need for full data combination, but serves as a powerful exploratory tool for evaluating new datasets. By providing early indicators of GWB sensitivity, the Lite method may help motivate the creation of fully combined datasets, while the FoM may guide the order in which to combine the pulsar's data together. As future IPTA data releases incorporate larger and more complex datasets, the Lite method will become increasingly more useful for nHz GW searches as one balances the trade-off between computational efficiency and sensitivity. The Lite method may also complement other future avenues of analyzing joint-PTA datasets, such as the Fourier-space combination of posteriors from single-PTA datasets \citep{Laal+2024, Valtolina2024}, or the fully extended PTA analysis from \citet{3P+2024} using the OS.

\section*{Acknowledgements}

This paper is the result of the work of many people and uses data from over 3 decades of pulsar timing observations. We thank Boris Goncharov for reviewing the manuscript and providing useful suggestions, such as the hierarchical Bayesian analysis, that improved the quality of this work. BL additionally thanks members of the IPTA Gravitational Wave Analysis working group, co-chaired by Nihan Pol, Aur\'elien Chalumeau, and Paul Baker, for constructive comments and discussions, and to Rutger van Haasteren for providing useful insights on hierarchical modeling. We acknowledge support received from NSF AAG award number 2009468, and NSF Physics Frontiers Center award number 2020265, which supports the NANOGrav project. Part of this work was undertaken as part of the Australian Research Council Centre of Excellence for Gravitational Wave Discovery (project number CE230100016). The National Radio Astronomy Observatory is a facility of the National Science Foundation operated under cooperative agreement by Associated Universities, Inc. The Nan{\c c}ay Radio Observatory is operated by the Paris Observatory, associated to the French Centre National de la Recherche Scientifique (CNRS) and to the Universit{\'e} d’Orl{\'e}ans. We acknowledge financial support from “Programme National de Cosmologie and Galaxies” (PNCG), and “Programme National Hautes Energies” (PNHE) funded by CNRS/INSU-IN2P3-INP, CEA and CNES, France. We acknowledge financial support from Agence Nationale de la Recherche (ANR-18-CE31-0015), France. CMFM was supported in part by the National Science Foundation under Grants No. PHY-2020265 and AST-2106552, and the Flatiron Institute, which is part of the Simons Foundation. JSH acknowledges support from NSF CAREER award No. 2339728 and is supported through an Oregon State University start up fund. JA acknowledges support from the European Commission under project ARGOS-CDS (Grant Agreement number: 101094354). AC acknowledges financial support provided under the European Union’s Horizon Europe ERC Starting Grant ``A Gamma-ray Infrastructure to Advance Gravitational Wave Astrophysics'' (GIGA). ZC is supported by the National Natural Science Foundation of China under Grant No.~12405056 and the innovative research group of Hunan Province under Grant No.~2024JJ1006. DD acknowledges the Department of Atomic Energy, Government of India's support through `Apex Project-Advance Research and Education in Mathematical Sciences' at The Institute of Mathematical Sciences. TD acknowledges support from the NSF Physics Frontiers Center (PFC) award number 1430284. TD was partially supported through the National Science Foundation (NSF) PIRE program award number 0968296. ECF is supported by NASA under award number 80GSFC24M0006. DCG is supported by NSF Astronomy and Astrophysics Grant (AAG) award \#2406919. SMR is a CIFAR Fellow and is supported by the NSF Physics Frontiers Center award 2020265. GMS acknowledges financial support provided under the European Union's H2020 ERC Consolidator Grant ``Binary Massive Black Hole Astrophysics'' (B Massive, Grant Agreement: 818691). JBW is supported by the Major Science and Technology Program of Xinjiang Uygur Autonomous Region (No. 2022A03013-4), the Zhejiang Provincial Natural Science Foundation of China (No. LY23A030001). LZ is supported by the National Natural Science Foundation of China (Grant No. 12103069).

\section*{Data Availability}

Jupyter Notebooks and Python scripts used to produce our figures and our results are freely available at \url{https://github.com/blarsen10/IPTA_DR2_analysis}. Correspondence and requests for materials should be addressed to Bjorn Larsen.


\bibliographystyle{mnras}
\bibliography{bib} 

\begin{thebibliography}{}
\makeatletter
\relax
\def\mn@urlcharsother{\let\do\@makeother \do\$\do\&\do\#\do\^\do\_\do\%\do\~}
\def\mn@doi{\begingroup\mn@urlcharsother \@ifnextchar [ {\mn@doi@}
  {\mn@doi@[]}}
\def\mn@doi@[#1]#2{\def\@tempa{#1}\ifx\@tempa\@empty \href
  {http://dx.doi.org/#2} {doi:#2}\else \href {http://dx.doi.org/#2} {#1}\fi
  \endgroup}
\def\mn@eprint#1#2{\mn@eprint@#1:#2::\@nil}
\def\mn@eprint@arXiv#1{\href {http://arxiv.org/abs/#1} {{\tt arXiv:#1}}}
\def\mn@eprint@dblp#1{\href {http://dblp.uni-trier.de/rec/bibtex/#1.xml}
  {dblp:#1}}
\def\mn@eprint@#1:#2:#3:#4\@nil{\def\@tempa {#1}\def\@tempb {#2}\def\@tempc
  {#3}\ifx \@tempc \@empty \let \@tempc \@tempb \let \@tempb \@tempa \fi \ifx
  \@tempb \@empty \def\@tempb {arXiv}\fi \@ifundefined
  {mn@eprint@\@tempb}{\@tempb:\@tempc}{\expandafter \expandafter \csname
  mn@eprint@\@tempb\endcsname \expandafter{\@tempc}}}

\bibitem[\protect\citeauthoryear{{Afzal} et~al.,}{{Afzal}
  et~al.}{2023}]{Afzal+2023}
{Afzal} A.,  et~al., 2023, \mn@doi [\apjl] {10.3847/2041-8213/acdc91}, \href
  {https://ui.adsabs.harvard.edu/abs/2023ApJ...951L..11A} {951, L11}

\bibitem[\protect\citeauthoryear{{Agazie} et~al.,}{{Agazie}
  et~al.}{2023a}]{Agazie+2023_gwb}
{Agazie} G.,  et~al., 2023a, \mn@doi [\apjl] {10.3847/2041-8213/acdac6}, \href
  {https://ui.adsabs.harvard.edu/abs/2023ApJ...951L...8A} {951, L8}

\bibitem[\protect\citeauthoryear{{Agazie} et~al.,}{{Agazie}
  et~al.}{2023b}]{Agazie2023_data}
{Agazie} G.,  et~al., 2023b, \mn@doi [\apjl] {10.3847/2041-8213/acda9a}, \href
  {https://ui.adsabs.harvard.edu/abs/2023ApJ...951L...9A} {951, L9}

\bibitem[\protect\citeauthoryear{{Agazie} et~al.,}{{Agazie}
  et~al.}{2023c}]{Agazie2023_detchar}
{Agazie} G.,  et~al., 2023c, \mn@doi [\apjl] {10.3847/2041-8213/acda88}, \href
  {https://ui.adsabs.harvard.edu/abs/2023ApJ...951L..10A} {951, L10}

\bibitem[\protect\citeauthoryear{{Agazie} et~al.,}{{Agazie}
  et~al.}{2023d}]{NG15-detchar}
{Agazie} G.,  et~al., 2023d, \mn@doi [\apjl] {10.3847/2041-8213/acda88}, \href
  {https://ui.adsabs.harvard.edu/abs/2023ApJ...951L..10A} {951, L10}

\bibitem[\protect\citeauthoryear{{Agazie} et~al.,}{{Agazie}
  et~al.}{2024}]{3P+2024}
{Agazie} G.,  et~al., 2024, \mn@doi [\apj] {10.3847/1538-4357/ad36be}, \href
  {https://ui.adsabs.harvard.edu/abs/2024ApJ...966..105A} {966, 105}

\bibitem[\protect\citeauthoryear{{Agazie} et~al.,}{{Agazie}
  et~al.}{2025}]{Agazie+2025_1by1}
{Agazie} G.,  et~al., 2025, \mn@doi [\apj] {10.3847/1538-4357/ad93aa}, \href
  {https://ui.adsabs.harvard.edu/abs/2025ApJ...978..168A} {978, 168}

\bibitem[\protect\citeauthoryear{{Alam} et~al.,}{{Alam}
  et~al.}{2021}]{Alam2021_wb}
{Alam} M.~F.,  et~al., 2021, \mn@doi [\apjs] {10.3847/1538-4365/abc6a1}, \href
  {https://ui.adsabs.harvard.edu/abs/2021ApJS..252....5A} {252, 5}

\bibitem[\protect\citeauthoryear{{Albert}}{{Albert}}{2020}]{jaxns}
{Albert} J.~G.,  2020, \mn@doi [arXiv e-prints] {10.48550/arXiv.2012.15286},
  \href {https://ui.adsabs.harvard.edu/abs/2020arXiv201215286A} {p.
  arXiv:2012.15286}

\bibitem[\protect\citeauthoryear{{Allen} \& {Romano}}{{Allen} \&
  {Romano}}{2023}]{AllenRomano2023}
{Allen} B.,  {Romano} J.~D.,  2023, \mn@doi [\prd]
  {10.1103/PhysRevD.108.043026}, \href
  {https://ui.adsabs.harvard.edu/abs/2023PhRvD.108d3026A} {108, 043026}

\bibitem[\protect\citeauthoryear{{Allen}, {Dhurandhar}, {Gupta}, {McLaughlin},
  {Natarajan}, {Shannon}, {Thrane}  \& {Vecchio}}{{Allen}
  et~al.}{2023}]{Allen+2023}
{Allen} B.,  {Dhurandhar} S.,  {Gupta} Y.,  {McLaughlin} M.,  {Natarajan} P.,
  {Shannon} R.~M.,  {Thrane} E.,   {Vecchio} A.,  2023, \mn@doi [arXiv
  e-prints] {10.48550/arXiv.2304.04767}, \href
  {https://ui.adsabs.harvard.edu/abs/2023arXiv230404767A} {p. arXiv:2304.04767}

\bibitem[\protect\citeauthoryear{{Andrews}, {Lam}  \& {Dolch}}{{Andrews}
  et~al.}{2020}]{AndrewsLamDolch2021}
{Andrews} P.,  {Lam} M.,   {Dolch} T.,  2020, Investigating the Impact of
  Slicing on Fitted Timing Model Parameters, NANOGrav Memo \#4

\bibitem[\protect\citeauthoryear{{Anholm}, {Ballmer}, {Creighton}, {Price}  \&
  {Siemens}}{{Anholm} et~al.}{2009}]{Anholm+2009}
{Anholm} M.,  {Ballmer} S.,  {Creighton} J. D.~E.,  {Price} L.~R.,   {Siemens}
  X.,  2009, \mn@doi [\prd] {10.1103/PhysRevD.79.084030}, \href
  {https://ui.adsabs.harvard.edu/abs/2009PhRvD..79h4030A} {79, 084030}

\bibitem[\protect\citeauthoryear{{Antoniadis} et~al.,}{{Antoniadis}
  et~al.}{2022}]{Antoniadis+2022}
{Antoniadis} J.,  et~al., 2022, \mn@doi [\mnras] {10.1093/mnras/stab3418},
  \href {https://ui.adsabs.harvard.edu/abs/2022MNRAS.510.4873A} {510, 4873}

\bibitem[\protect\citeauthoryear{{Arzoumanian} et~al.,}{{Arzoumanian}
  et~al.}{2014}]{Arzoumanian2014}
{Arzoumanian} Z.,  et~al., 2014, \mn@doi [\apj] {10.1088/0004-637X/794/2/141},
  \href {https://ui.adsabs.harvard.edu/abs/2014ApJ...794..141A} {794, 141}

\bibitem[\protect\citeauthoryear{{Arzoumanian} et~al.,}{{Arzoumanian}
  et~al.}{2016}]{NG9}
{Arzoumanian} Z.,  et~al., 2016, \mn@doi [\apj] {10.3847/0004-637X/821/1/13},
  \href {https://ui.adsabs.harvard.edu/abs/2016ApJ...821...13A} {821, 13}

\bibitem[\protect\citeauthoryear{{Arzoumanian} et~al.,}{{Arzoumanian}
  et~al.}{2020a}]{Arzoumanian+2020}
{Arzoumanian} Z.,  et~al., 2020a, \mn@doi [\apj] {10.3847/1538-4357/ababa1},
  \href {https://ui.adsabs.harvard.edu/abs/2020ApJ...900..102A} {900, 102}

\bibitem[\protect\citeauthoryear{{Arzoumanian} et~al.,}{{Arzoumanian}
  et~al.}{2020b}]{Arzoumanian2020}
{Arzoumanian} Z.,  et~al., 2020b, \mn@doi [\apjl] {10.3847/2041-8213/abd401},
  \href {https://ui.adsabs.harvard.edu/abs/2020ApJ...905L..34A} {905, L34}

\bibitem[\protect\citeauthoryear{{Babak} et~al.,}{{Babak}
  et~al.}{2016}]{Babak+2016}
{Babak} S.,  et~al., 2016, \mn@doi [\mnras] {10.1093/mnras/stv2092}, \href
  {https://ui.adsabs.harvard.edu/abs/2016MNRAS.455.1665B} {455, 1665}

\bibitem[\protect\citeauthoryear{{Baier}, {Hazboun}  \& {Romano}}{{Baier}
  et~al.}{2025}]{Baier+2024}
{Baier} J.~G.,  {Hazboun} J.~S.,   {Romano} J.~D.,  2025, \mn@doi [Classical
  and Quantum Gravity] {10.1088/1361-6382/adbbab}, \href
  {https://ui.adsabs.harvard.edu/abs/2025CQGra..42g5008B} {42, 075008}

\bibitem[\protect\citeauthoryear{{Begelman}, {Blandford}  \& {Rees}}{{Begelman}
  et~al.}{1980}]{Begelman+1980}
{Begelman} M.~C.,  {Blandford} R.~D.,   {Rees} M.~J.,  1980, \mn@doi [\nat]
  {10.1038/287307a0}, \href
  {https://ui.adsabs.harvard.edu/abs/1980Natur.287..307B} {287, 307}

\bibitem[\protect\citeauthoryear{Bingham et~al.,}{Bingham
  et~al.}{2019}]{numpyro2}
Bingham E.,  et~al., 2019, J. Mach. Learn. Res., 20, 28:1

\bibitem[\protect\citeauthoryear{{CHIME/Pulsar Collaboration}
  et~al.,}{{CHIME/Pulsar Collaboration} et~al.}{2021}]{CHIME2021}
{CHIME/Pulsar Collaboration} et~al., 2021, \mn@doi [\apjs]
  {10.3847/1538-4365/abfdcb}, \href
  {https://ui.adsabs.harvard.edu/abs/2021ApJS..255....5C} {255, 5}

\bibitem[\protect\citeauthoryear{{Caballero} et~al.,}{{Caballero}
  et~al.}{2016}]{Caballero+2016}
{Caballero} R.~N.,  et~al., 2016, \mn@doi [\mnras] {10.1093/mnras/stw179},
  \href {https://ui.adsabs.harvard.edu/abs/2016MNRAS.457.4421C} {457, 4421}

\bibitem[\protect\citeauthoryear{{Caprini} \& {Figueroa}}{{Caprini} \&
  {Figueroa}}{2018}]{Caprini2018}
{Caprini} C.,  {Figueroa} D.~G.,  2018, \mn@doi [Classical and Quantum Gravity]
  {10.1088/1361-6382/aac608}, \href
  {https://ui.adsabs.harvard.edu/abs/2018CQGra..35p3001C} {35, 163001}

\bibitem[\protect\citeauthoryear{{Chalumeau} et~al.,}{{Chalumeau}
  et~al.}{2022}]{Chalumeau+2022}
{Chalumeau} A.,  et~al., 2022, \mn@doi [\mnras] {10.1093/mnras/stab3283}, \href
  {https://ui.adsabs.harvard.edu/abs/2022MNRAS.509.5538C} {509, 5538}

\bibitem[\protect\citeauthoryear{{Chamberlin}, {Creighton}, {Siemens},
  {Demorest}, {Ellis}, {Price}  \& {Romano}}{{Chamberlin}
  et~al.}{2015}]{Chamberlin+2015}
{Chamberlin} S.~J.,  {Creighton} J. D.~E.,  {Siemens} X.,  {Demorest} P.,
  {Ellis} J.,  {Price} L.~R.,   {Romano} J.~D.,  2015, \mn@doi [\prd]
  {10.1103/PhysRevD.91.044048}, \href
  {https://ui.adsabs.harvard.edu/abs/2015PhRvD..91d4048C} {91, 044048}

\bibitem[\protect\citeauthoryear{{Champion} et~al.,}{{Champion}
  et~al.}{2010}]{Champion+2010}
{Champion} D.~J.,  et~al., 2010, \mn@doi [\apjl]
  {10.1088/2041-8205/720/2/L201}, \href
  {https://ui.adsabs.harvard.edu/abs/2010ApJ...720L.201C} {720, L201}

\bibitem[\protect\citeauthoryear{{Chen} et~al.,}{{Chen}
  et~al.}{2021a}]{Chen+2021}
{Chen} S.,  et~al., 2021a, \mn@doi [\mnras] {10.1093/mnras/stab2833}, \href
  {https://ui.adsabs.harvard.edu/abs/2021MNRAS.508.4970C} {508, 4970}

\bibitem[\protect\citeauthoryear{{Chen} et~al.,}{{Chen}
  et~al.}{2021b}]{Chen2021}
{Chen} S.,  et~al., 2021b, \mn@doi [\mnras] {10.1093/mnras/stab2833}, \href
  {https://ui.adsabs.harvard.edu/abs/2021MNRAS.508.4970C} {508, 4970}

\bibitem[\protect\citeauthoryear{{Chen} et~al.,}{{Chen}
  et~al.}{2021c}]{EPTA_6psr}
{Chen} S.,  et~al., 2021c, \mn@doi [\mnras] {10.1093/mnras/stab2833}, \href
  {https://ui.adsabs.harvard.edu/abs/2021MNRAS.508.4970C} {508, 4970}

\bibitem[\protect\citeauthoryear{{Cordes} \& {Rickett}}{{Cordes} \&
  {Rickett}}{1998}]{CordesRickett1998}
{Cordes} J.~M.,  {Rickett} B.~J.,  1998, \mn@doi [\apj] {10.1086/306358}, \href
  {https://ui.adsabs.harvard.edu/abs/1998ApJ...507..846C} {507, 846}

\bibitem[\protect\citeauthoryear{{Desvignes} et~al.,}{{Desvignes}
  et~al.}{2016}]{EPTADR1}
{Desvignes} G.,  et~al., 2016, \mn@doi [\mnras] {10.1093/mnras/stw483}, \href
  {https://ui.adsabs.harvard.edu/abs/2016MNRAS.458.3341D} {458, 3341}

\bibitem[\protect\citeauthoryear{{Detweiler}}{{Detweiler}}{1979}]{Detweiler1979}
{Detweiler} S.,  1979, \mn@doi [\apj] {10.1086/157593}, \href
  {https://ui.adsabs.harvard.edu/abs/1979ApJ...234.1100D} {234, 1100}

\bibitem[\protect\citeauthoryear{Dickey}{Dickey}{1971}]{Dickey1971}
Dickey J.~M.,  1971, The Annals of Mathematical Statistics, pp 204--223

\bibitem[\protect\citeauthoryear{{EPTA Collaboration} et~al.,}{{EPTA
  Collaboration} et~al.}{2023a}]{EPTADR2}
{EPTA Collaboration} et~al., 2023a, \mn@doi [\aap]
  {10.1051/0004-6361/202346841}, \href
  {https://ui.adsabs.harvard.edu/abs/2023A&A...678A..48E} {678, A48}

\bibitem[\protect\citeauthoryear{{EPTA Collaboration} et~al.,}{{EPTA
  Collaboration} et~al.}{2023b}]{EPTAnoise2023}
{EPTA Collaboration} et~al., 2023b, \mn@doi [\aap]
  {10.1051/0004-6361/202346842}, \href
  {https://ui.adsabs.harvard.edu/abs/2023A&A...678A..49E} {678, A49}

\bibitem[\protect\citeauthoryear{{EPTA Collaboration} et~al.,}{{EPTA
  Collaboration} et~al.}{2023c}]{EPTADR2_gwb}
{EPTA Collaboration} et~al., 2023c, \mn@doi [\aap]
  {10.1051/0004-6361/202346844}, \href
  {https://ui.adsabs.harvard.edu/abs/2023A&A...678A..50E} {678, A50}

\bibitem[\protect\citeauthoryear{Ellis \& van Haasteren}{Ellis \& van
  Haasteren}{2017}]{ptmcmcsampler}
Ellis J.,  van Haasteren R.,  2017, jellis18/PTMCMCSampler: Official Release,
  \mn@doi{10.5281/zenodo.1037579}, \url
  {https://doi.org/10.5281/zenodo.1037579}

\bibitem[\protect\citeauthoryear{Ellis, Vallisneri, Taylor  \& Baker}{Ellis
  et~al.}{2020}]{enterprise}
Ellis J.~A.,  Vallisneri M.,  Taylor S.~R.,   Baker P.~T.,  2020, ENTERPRISE:
  Enhanced Numerical Toolbox Enabling a Robust PulsaR Inference SuitE, Zenodo,
  \mn@doi{10.5281/zenodo.4059815}, \url
  {https://doi.org/10.5281/zenodo.4059815}

\bibitem[\protect\citeauthoryear{{FERMI-LAT Collaboration} et~al.,}{{FERMI-LAT
  Collaboration} et~al.}{2022}]{FermiPTA_2021}
{FERMI-LAT Collaboration} et~al., 2022, \mn@doi [Science]
  {10.1126/science.abm3231}, \href
  {https://ui.adsabs.harvard.edu/abs/2022Sci...376..521F} {376, 521}

\bibitem[\protect\citeauthoryear{{Falxa} et~al.,}{{Falxa}
  et~al.}{2023}]{Falxa+2023}
{Falxa} M.,  et~al., 2023, \mn@doi [\mnras] {10.1093/mnras/stad812}, \href
  {https://ui.adsabs.harvard.edu/abs/2023MNRAS.521.5077F} {521, 5077}

\bibitem[\protect\citeauthoryear{{Ferranti} et~al.,}{{Ferranti}
  et~al.}{2025}]{Ferranti+2024}
{Ferranti} I.,  et~al., 2025, \mn@doi [\aap] {10.1051/0004-6361/202452805},
  \href {https://ui.adsabs.harvard.edu/abs/2025A&A...694A..38F} {694, A38}

\bibitem[\protect\citeauthoryear{{Gaia Collaboration} et~al.,}{{Gaia
  Collaboration} et~al.}{2021}]{Gaia2021}
{Gaia Collaboration} et~al., 2021, \mn@doi [\aap]
  {10.1051/0004-6361/202039657}, \href
  {https://ui.adsabs.harvard.edu/abs/2021A&A...649A...1G} {649, A1}

\bibitem[\protect\citeauthoryear{{Geiger} et~al.,}{{Geiger}
  et~al.}{2025}]{Geiger2024}
{Geiger} A.,  et~al., 2025, \mn@doi [\apj] {10.3847/1538-4357/add0b6}, \href
  {https://ui.adsabs.harvard.edu/abs/2025ApJ...986..191G} {986, 191}

\bibitem[\protect\citeauthoryear{{Gersbach}, {Taylor}, {Meyers}  \&
  {Romano}}{{Gersbach} et~al.}{2025}]{Gersbach+2024}
{Gersbach} K.~A.,  {Taylor} S.~R.,  {Meyers} P.~M.,   {Romano} J.~D.,  2025,
  \mn@doi [\prd] {10.1103/PhysRevD.111.023027}, \href
  {https://ui.adsabs.harvard.edu/abs/2025PhRvD.111b3027G} {111, 023027}

\bibitem[\protect\citeauthoryear{{Goncharov} \& {Sardana}}{{Goncharov} \&
  {Sardana}}{2025}]{GoncharovSardana2025}
{Goncharov} B.,  {Sardana} S.,  2025, \mn@doi [\mnras] {10.1093/mnras/staf190},
  \href {https://ui.adsabs.harvard.edu/abs/2025MNRAS.537.3470G} {537, 3470}

\bibitem[\protect\citeauthoryear{{Goncharov} et~al.,}{{Goncharov}
  et~al.}{2021a}]{Goncharov2021}
{Goncharov} B.,  et~al., 2021a, \mn@doi [\mnras] {10.1093/mnras/staa3411},
  \href {https://ui.adsabs.harvard.edu/abs/2021MNRAS.502..478G} {502, 478}

\bibitem[\protect\citeauthoryear{{Goncharov} et~al.,}{{Goncharov}
  et~al.}{2021b}]{PPTADR2}
{Goncharov} B.,  et~al., 2021b, \mn@doi [\apjl] {10.3847/2041-8213/ac17f4},
  \href {https://ui.adsabs.harvard.edu/abs/2021ApJ...917L..19G} {917, L19}

\bibitem[\protect\citeauthoryear{{Goncharov} et~al.,}{{Goncharov}
  et~al.}{2022}]{Goncharov+2022}
{Goncharov} B.,  et~al., 2022, \mn@doi [\apjl] {10.3847/2041-8213/ac76bb},
  \href {https://ui.adsabs.harvard.edu/abs/2022ApJ...932L..22G} {932, L22}

\bibitem[\protect\citeauthoryear{{Goncharov} et~al.,}{{Goncharov}
  et~al.}{2024}]{Goncharov+2024}
{Goncharov} B.,  et~al., 2024, \mn@doi [arXiv e-prints]
  {10.48550/arXiv.2409.03627}, \href
  {https://ui.adsabs.harvard.edu/abs/2024arXiv240903627G} {p. arXiv:2409.03627}

\bibitem[\protect\citeauthoryear{{Good} \& {International Pulsar Timing Array
  Team}}{{Good} \& {International Pulsar Timing Array Team}}{2023}]{Good+2023}
{Good} D.,  {International Pulsar Timing Array Team} 2023, in American
  Astronomical Society Meeting Abstracts. p. 438.02

\bibitem[\protect\citeauthoryear{{Hazboun}, {Romano}  \& {Smith}}{{Hazboun}
  et~al.}{2019}]{Hazboun+2019}
{Hazboun} J.~S.,  {Romano} J.~D.,   {Smith} T.~L.,  2019, \mn@doi [\prd]
  {10.1103/PhysRevD.100.104028}, \href
  {https://ui.adsabs.harvard.edu/abs/2019PhRvD.100j4028H} {100, 104028}

\bibitem[\protect\citeauthoryear{{Hazboun} et~al.,}{{Hazboun}
  et~al.}{2020}]{Hazboun+2020_slicing}
{Hazboun} J.~S.,  et~al., 2020, \mn@doi [\apj] {10.3847/1538-4357/ab68db},
  \href {https://ui.adsabs.harvard.edu/abs/2020ApJ...890..108H} {890, 108}

\bibitem[\protect\citeauthoryear{{Hazboun} et~al.,}{{Hazboun}
  et~al.}{2022}]{Hazboun+2022}
{Hazboun} J.~S.,  et~al., 2022, \mn@doi [\apj] {10.3847/1538-4357/ac5829},
  \href {https://ui.adsabs.harvard.edu/abs/2022ApJ...929...39H} {929, 39}

\bibitem[\protect\citeauthoryear{{Hazboun}, {Meyers}, {Romano}, {Siemens}  \&
  {Archibald}}{{Hazboun} et~al.}{2023}]{Hazboun2023}
{Hazboun} J.~S.,  {Meyers} P.~M.,  {Romano} J.~D.,  {Siemens} X.,   {Archibald}
  A.~M.,  2023, \mn@doi [\prd] {10.1103/PhysRevD.108.104050}, \href
  {https://ui.adsabs.harvard.edu/abs/2023PhRvD.108j4050H} {108, 104050}

\bibitem[\protect\citeauthoryear{{Hee}, {Handley}, {Hobson}  \&
  {Lasenby}}{{Hee} et~al.}{2016}]{Hee+2016}
{Hee} S.,  {Handley} W.~J.,  {Hobson} M.~P.,   {Lasenby} A.~N.,  2016, \mn@doi
  [\mnras] {10.1093/mnras/stv2217}, \href
  {https://ui.adsabs.harvard.edu/abs/2016MNRAS.455.2461H} {455, 2461}

\bibitem[\protect\citeauthoryear{{Hellings} \& {Downs}}{{Hellings} \&
  {Downs}}{1983}]{HellingsDowns1983}
{Hellings} R.~W.,  {Downs} G.~S.,  1983, \mn@doi [\apjl] {10.1086/183954},
  \href {https://ui.adsabs.harvard.edu/abs/1983ApJ...265L..39H} {265, L39}

\bibitem[\protect\citeauthoryear{{Hemberger} \& {Stinebring}}{{Hemberger} \&
  {Stinebring}}{2008}]{HembergerStinebring2008}
{Hemberger} D.~A.,  {Stinebring} D.~R.,  2008, \mn@doi [\apjl]
  {10.1086/528985}, \href
  {https://ui.adsabs.harvard.edu/abs/2008ApJ...674L..37H} {674, L37}

\bibitem[\protect\citeauthoryear{{Hobbs} et~al.,}{{Hobbs}
  et~al.}{2012}]{Hobbs2012}
{Hobbs} G.,  et~al., 2012, \mn@doi [\mnras] {10.1111/j.1365-2966.2012.21946.x},
  \href {https://ui.adsabs.harvard.edu/abs/2012MNRAS.427.2780H} {427, 2780}

\bibitem[\protect\citeauthoryear{{Hourihane}, {Meyers}, {Johnson},
  {Chatziioannou}  \& {Vallisneri}}{{Hourihane} et~al.}{2023}]{Hourihane2023}
{Hourihane} S.,  {Meyers} P.,  {Johnson} A.,  {Chatziioannou} K.,
  {Vallisneri} M.,  2023, \mn@doi [\prd] {10.1103/PhysRevD.107.084045}, \href
  {https://ui.adsabs.harvard.edu/abs/2023PhRvD.107h4045H} {107, 084045}

\bibitem[\protect\citeauthoryear{{Johnson}, {Vigeland}, {Siemens}  \&
  {Taylor}}{{Johnson} et~al.}{2022}]{Johnson+2022}
{Johnson} A.~D.,  {Vigeland} S.~J.,  {Siemens} X.,   {Taylor} S.~R.,  2022,
  \mn@doi [\apj] {10.3847/1538-4357/ac6f5e}, \href
  {https://ui.adsabs.harvard.edu/abs/2022ApJ...932..105J} {932, 105}

\bibitem[\protect\citeauthoryear{{Johnson} et~al.,}{{Johnson}
  et~al.}{2024}]{Johnson+2024}
{Johnson} A.~D.,  et~al., 2024, \mn@doi [\prd] {10.1103/PhysRevD.109.103012},
  \href {https://ui.adsabs.harvard.edu/abs/2024PhRvD.109j3012J} {109, 103012}

\bibitem[\protect\citeauthoryear{{Joshi} et~al.,}{{Joshi}
  et~al.}{2018}]{Joshi2018}
{Joshi} B.~C.,  et~al., 2018, \mn@doi [Journal of Astrophysics and Astronomy]
  {10.1007/s12036-018-9549-y}, \href
  {https://ui.adsabs.harvard.edu/abs/2018JApA...39...51J} {39, 51}

\bibitem[\protect\citeauthoryear{{Kaspi}, {Taylor}  \& {Ryba}}{{Kaspi}
  et~al.}{1994}]{Kaspi+1994}
{Kaspi} V.~M.,  {Taylor} J.~H.,   {Ryba} M.~F.,  1994, \mn@doi [\apj]
  {10.1086/174280}, \href
  {https://ui.adsabs.harvard.edu/abs/1994ApJ...428..713K} {428, 713}

\bibitem[\protect\citeauthoryear{{Keith} et~al.,}{{Keith}
  et~al.}{2013}]{Keith+2013}
{Keith} M.~J.,  et~al., 2013, \mn@doi [\mnras] {10.1093/mnras/sts486}, \href
  {https://ui.adsabs.harvard.edu/abs/2013MNRAS.429.2161K} {429, 2161}

\bibitem[\protect\citeauthoryear{{Laal}, {Taylor}, {van Haasteren}, {Lamb}  \&
  {Siemens}}{{Laal} et~al.}{2025}]{Laal+2024}
{Laal} N.,  {Taylor} S.~R.,  {van Haasteren} R.,  {Lamb} W.~G.,   {Siemens} X.,
   2025, \mn@doi [\prd] {10.1103/PhysRevD.111.063067}, \href
  {https://ui.adsabs.harvard.edu/abs/2025PhRvD.111f3067L} {111, 063067}

\bibitem[\protect\citeauthoryear{{Lam}, {McLaughlin}, {Cordes}, {Chatterjee}
  \& {Lazio}}{{Lam} et~al.}{2018a}]{Lam+2018_bandwidths}
{Lam} M.~T.,  {McLaughlin} M.~A.,  {Cordes} J.~M.,  {Chatterjee} S.,   {Lazio}
  T.~J.~W.,  2018a, \mn@doi [\apj] {10.3847/1538-4357/aac48d}, \href
  {https://ui.adsabs.harvard.edu/abs/2018ApJ...861...12L} {861, 12}

\bibitem[\protect\citeauthoryear{{Lam} et~al.,}{{Lam} et~al.}{2018b}]{Lam2018}
{Lam} M.~T.,  et~al., 2018b, \mn@doi [\apj] {10.3847/1538-4357/aac770}, \href
  {https://ui.adsabs.harvard.edu/abs/2018ApJ...861..132L} {861, 132}

\bibitem[\protect\citeauthoryear{{Lam} et~al.,}{{Lam} et~al.}{2019}]{Lam+2019}
{Lam} M.~T.,  et~al., 2019, \mn@doi [\apj] {10.3847/1538-4357/ab01cd}, \href
  {https://ui.adsabs.harvard.edu/abs/2019ApJ...872..193L} {872, 193}

\bibitem[\protect\citeauthoryear{{Lange}}{{Lange}}{2023}]{nautilus2023}
{Lange} J.~U.,  2023, \mn@doi [\mnras] {10.1093/mnras/stad2441}, \href
  {https://ui.adsabs.harvard.edu/abs/2023MNRAS.525.3181L} {525, 3181}

\bibitem[\protect\citeauthoryear{{Larsen} et~al.,}{{Larsen}
  et~al.}{2024}]{Larsen2024}
{Larsen} B.,  et~al., 2024, \mn@doi [\apj] {10.3847/1538-4357/ad5291}, \href
  {https://ui.adsabs.harvard.edu/abs/2024ApJ...972...49L} {972, 49}

\bibitem[\protect\citeauthoryear{{Lasky} et~al.,}{{Lasky}
  et~al.}{2016}]{Lasky+2016}
{Lasky} P.~D.,  et~al., 2016, \mn@doi [Physical Review X]
  {10.1103/PhysRevX.6.011035}, \href
  {https://ui.adsabs.harvard.edu/abs/2016PhRvX...6a1035L} {6, 011035}

\bibitem[\protect\citeauthoryear{{Lee}}{{Lee}}{2016}]{Lee2016}
{Lee} K.~J.,  2016, in {Qain} L.,  {Li} D.,  eds,  Astronomical Society of the
  Pacific Conference Series Vol. 502, Frontiers in Radio Astronomy and FAST
  Early Sciences Symposium 2015. p.~19

\bibitem[\protect\citeauthoryear{{Lentati}, {Alexander}, {Hobson}, {Taylor},
  {Gair}, {Balan}  \& {van Haasteren}}{{Lentati} et~al.}{2013}]{Lentati+2013}
{Lentati} L.,  {Alexander} P.,  {Hobson} M.~P.,  {Taylor} S.,  {Gair} J.,
  {Balan} S.~T.,   {van Haasteren} R.,  2013, \mn@doi [\prd]
  {10.1103/PhysRevD.87.104021}, \href
  {https://ui.adsabs.harvard.edu/abs/2013PhRvD..87j4021L} {87, 104021}

\bibitem[\protect\citeauthoryear{{Lentati} et~al.,}{{Lentati}
  et~al.}{2016}]{Lentati+2016}
{Lentati} L.,  et~al., 2016, \mn@doi [\mnras] {10.1093/mnras/stw395}, \href
  {https://ui.adsabs.harvard.edu/abs/2016MNRAS.458.2161L} {458, 2161}

\bibitem[\protect\citeauthoryear{{Loredo} \& {Hendry}}{{Loredo} \&
  {Hendry}}{2019}]{LoredoHendry2019}
{Loredo} T.~J.,  {Hendry} M.~A.,  2019, \mn@doi [arXiv e-prints]
  {10.48550/arXiv.1911.12337}, \href
  {https://ui.adsabs.harvard.edu/abs/2019arXiv191112337L} {p. arXiv:1911.12337}

\bibitem[\protect\citeauthoryear{{Madison}, {Cordes}  \&
  {Chatterjee}}{{Madison} et~al.}{2014}]{Madison2014}
{Madison} D.~R.,  {Cordes} J.~M.,   {Chatterjee} S.,  2014, \mn@doi [\apj]
  {10.1088/0004-637X/788/2/141}, \href
  {https://ui.adsabs.harvard.edu/abs/2014ApJ...788..141M} {788, 141}

\bibitem[\protect\citeauthoryear{{Manchester} et~al.,}{{Manchester}
  et~al.}{2013}]{PPTADR1}
{Manchester} R.~N.,  et~al., 2013, \mn@doi [\pasa] {10.1017/pasa.2012.017},
  \href {https://ui.adsabs.harvard.edu/abs/2013PASA...30...17M} {30, e017}

\bibitem[\protect\citeauthoryear{{Meyers}, {Chatziioannou}, {Vallisneri}  \&
  {Chua}}{{Meyers} et~al.}{2023}]{Meyers+2023}
{Meyers} P.~M.,  {Chatziioannou} K.,  {Vallisneri} M.,   {Chua} A. J.~K.,
  2023, \mn@doi [\prd] {10.1103/PhysRevD.108.123008}, \href
  {https://ui.adsabs.harvard.edu/abs/2023PhRvD.108l3008M} {108, 123008}

\bibitem[\protect\citeauthoryear{{Miles} et~al.,}{{Miles}
  et~al.}{2023}]{Miles2023}
{Miles} M.~T.,  et~al., 2023, \mn@doi [\mnras] {10.1093/mnras/stac3644}, \href
  {https://ui.adsabs.harvard.edu/abs/2023MNRAS.519.3976M} {519, 3976}

\bibitem[\protect\citeauthoryear{{Mingarelli} et~al.,}{{Mingarelli}
  et~al.}{2017}]{Mingarelli2017}
{Mingarelli} C. M.~F.,  et~al., 2017, \mn@doi [Nature Astronomy]
  {10.1038/s41550-017-0299-6}, \href
  {https://ui.adsabs.harvard.edu/abs/2017NatAs...1..886M} {1, 886}

\bibitem[\protect\citeauthoryear{{Ni{\c{t}}u} et~al.,}{{Ni{\c{t}}u}
  et~al.}{2024}]{Nitu2024}
{Ni{\c{t}}u} I.~C.,  et~al., 2024, \mn@doi [\mnras] {10.1093/mnras/stae220},
  \href {https://ui.adsabs.harvard.edu/abs/2024MNRAS.528.3304N} {528, 3304}

\bibitem[\protect\citeauthoryear{{Perera} et~al.,}{{Perera}
  et~al.}{2019}]{Perera+2019}
{Perera} B.~B.~P.,  et~al., 2019, \mn@doi [\mnras] {10.1093/mnras/stz2857},
  \href {https://ui.adsabs.harvard.edu/abs/2019MNRAS.490.4666P} {490, 4666}

\bibitem[\protect\citeauthoryear{Phan, Pradhan  \& Jankowiak}{Phan
  et~al.}{2019}]{numpyro1}
Phan D.,  Pradhan N.,   Jankowiak M.,  2019, arXiv preprint arXiv:1912.11554

\bibitem[\protect\citeauthoryear{{Rajagopal} \& {Romani}}{{Rajagopal} \&
  {Romani}}{1995}]{Rajagopal1995}
{Rajagopal} M.,  {Romani} R.~W.,  1995, \mn@doi [\apj] {10.1086/175813}, \href
  {https://ui.adsabs.harvard.edu/abs/1995ApJ...446..543R} {446, 543}

\bibitem[\protect\citeauthoryear{{Ransom} et~al.,}{{Ransom}
  et~al.}{2019}]{Ransom2019}
{Ransom} S.,  et~al., 2019, in Bulletin of the American Astronomical Society.
  p.~195 (\mn@eprint {arXiv} {1908.05356}), \mn@doi{10.48550/arXiv.1908.05356}

\bibitem[\protect\citeauthoryear{{Reardon} et~al.,}{{Reardon}
  et~al.}{2016}]{Reardon+2016}
{Reardon} D.~J.,  et~al., 2016, \mn@doi [\mnras] {10.1093/mnras/stv2395}, \href
  {https://ui.adsabs.harvard.edu/abs/2016MNRAS.455.1751R} {455, 1751}

\bibitem[\protect\citeauthoryear{{Reardon} et~al.,}{{Reardon}
  et~al.}{2023a}]{PPTADR3_gwb}
{Reardon} D.~J.,  et~al., 2023a, \mn@doi [\apjl] {10.3847/2041-8213/acdd02},
  \href {https://ui.adsabs.harvard.edu/abs/2023ApJ...951L...6R} {951, L6}

\bibitem[\protect\citeauthoryear{{Reardon} et~al.,}{{Reardon}
  et~al.}{2023b}]{Reardon2023}
{Reardon} D.~J.,  et~al., 2023b, \mn@doi [\apjl] {10.3847/2041-8213/acdd03},
  \href {https://ui.adsabs.harvard.edu/abs/2023ApJ...951L...7R} {951, L7}

\bibitem[\protect\citeauthoryear{{Reardon} et~al.,}{{Reardon}
  et~al.}{2024}]{Reardon+2024}
{Reardon} D.~J.,  et~al., 2024, \mn@doi [\apjl] {10.3847/2041-8213/ad614a},
  \href {https://ui.adsabs.harvard.edu/abs/2024ApJ...971L..18R} {971, L18}

\bibitem[\protect\citeauthoryear{{Sardesai}, {Vigeland}, {Gersbach}  \&
  {Taylor}}{{Sardesai} et~al.}{2023}]{Sardesai2023}
{Sardesai} S.~C.,  {Vigeland} S.~J.,  {Gersbach} K.~A.,   {Taylor} S.~R.,
  2023, \mn@doi [\prd] {10.1103/PhysRevD.108.124081}, \href
  {https://ui.adsabs.harvard.edu/abs/2023PhRvD.108l4081S} {108, 124081}

\bibitem[\protect\citeauthoryear{{Sazhin}}{{Sazhin}}{1978}]{Sazhin1978}
{Sazhin} M.~V.,  1978, \sovast, \href
  {https://ui.adsabs.harvard.edu/abs/1978SvA....22...36S} {22, 36}

\bibitem[\protect\citeauthoryear{{Shannon} \& {Cordes}}{{Shannon} \&
  {Cordes}}{2017}]{ShannonCordes2017}
{Shannon} R.~M.,  {Cordes} J.~M.,  2017, \mn@doi [\mnras]
  {10.1093/mnras/stw2449}, \href
  {https://ui.adsabs.harvard.edu/abs/2017MNRAS.464.2075S} {464, 2075}

\bibitem[\protect\citeauthoryear{{Shannon} et~al.,}{{Shannon}
  et~al.}{2015}]{Shannon2015}
{Shannon} R.~M.,  et~al., 2015, \mn@doi [Science] {10.1126/science.aab1910},
  \href {https://ui.adsabs.harvard.edu/abs/2015Sci...349.1522S} {349, 1522}

\bibitem[\protect\citeauthoryear{{Siemens}, {Ellis}, {Jenet}  \&
  {Romano}}{{Siemens} et~al.}{2013}]{siemens2013}
{Siemens} X.,  {Ellis} J.,  {Jenet} F.,   {Romano} J.~D.,  2013, \mn@doi
  [Classical and Quantum Gravity] {10.1088/0264-9381/30/22/224015}, \href
  {https://ui.adsabs.harvard.edu/abs/2013CQGra..30v4015S} {30, 224015}

\bibitem[\protect\citeauthoryear{{Sosa Fiscella} et~al.,}{{Sosa Fiscella}
  et~al.}{2024}]{Sosa+2024}
{Sosa Fiscella} S.~V.,  et~al., 2024, \mn@doi [\apj]
  {10.3847/1538-4357/ad2858}, \href
  {https://ui.adsabs.harvard.edu/abs/2024ApJ...966...95S} {966, 95}

\bibitem[\protect\citeauthoryear{{Speri}, {Porayko}, {Falxa}, {Chen}, {Gair},
  {Sesana}  \& {Taylor}}{{Speri} et~al.}{2023}]{Speri+2023}
{Speri} L.,  {Porayko} N.~K.,  {Falxa} M.,  {Chen} S.,  {Gair} J.~R.,  {Sesana}
  A.,   {Taylor} S.~R.,  2023, \mn@doi [\mnras] {10.1093/mnras/stac3237}, \href
  {https://ui.adsabs.harvard.edu/abs/2023MNRAS.518.1802S} {518, 1802}

\bibitem[\protect\citeauthoryear{{Srivastava} et~al.,}{{Srivastava}
  et~al.}{2023}]{Srivastava+2023}
{Srivastava} A.,  et~al., 2023, \mn@doi [\prd] {10.1103/PhysRevD.108.023008},
  \href {https://ui.adsabs.harvard.edu/abs/2023PhRvD.108b3008S} {108, 023008}

\bibitem[\protect\citeauthoryear{{Stappers} et~al.,}{{Stappers}
  et~al.}{2011}]{LOFAR2011}
{Stappers} B.~W.,  et~al., 2011, \mn@doi [\aap] {10.1051/0004-6361/201116681},
  \href {https://ui.adsabs.harvard.edu/abs/2011A&A...530A..80S} {530, A80}

\bibitem[\protect\citeauthoryear{{Susarla} et~al.,}{{Susarla}
  et~al.}{2024}]{Susarla+2024}
{Susarla} S.~C.,  et~al., 2024, \mn@doi [\aap] {10.1051/0004-6361/202450680},
  \href {https://ui.adsabs.harvard.edu/abs/2024A&A...692A..18S} {692, A18}

\bibitem[\protect\citeauthoryear{{Taylor}}{{Taylor}}{2021}]{Taylor2021}
{Taylor} S.~R.,  2021, \mn@doi [arXiv e-prints] {10.48550/arXiv.2105.13270},
  \href {https://ui.adsabs.harvard.edu/abs/2021arXiv210513270T} {p.
  arXiv:2105.13270}

\bibitem[\protect\citeauthoryear{Taylor, Baker, Hazboun, Simon  \&
  Vigeland}{Taylor et~al.}{2021}]{enterprise_extensions}
Taylor S.~R.,  Baker P.~T.,  Hazboun J.~S.,  Simon J.,   Vigeland S.~J.,  2021,
  enterprise\_extensions, \url
  {https://github.com/nanograv/enterprise\_extensions}

\bibitem[\protect\citeauthoryear{{Taylor}, {Simon}, {Schult}, {Pol}  \&
  {Lamb}}{{Taylor} et~al.}{2022}]{Taylor2022}
{Taylor} S.~R.,  {Simon} J.,  {Schult} L.,  {Pol} N.,   {Lamb} W.~G.,  2022,
  \mn@doi [\prd] {10.1103/PhysRevD.105.084049}, \href
  {https://ui.adsabs.harvard.edu/abs/2022PhRvD.105h4049T} {105, 084049}

\bibitem[\protect\citeauthoryear{{Thrane} \& {Talbot}}{{Thrane} \&
  {Talbot}}{2019}]{Thranetalbot2019}
{Thrane} E.,  {Talbot} C.,  2019, \mn@doi [\pasa] {10.1017/pasa.2019.2}, \href
  {https://ui.adsabs.harvard.edu/abs/2019PASA...36...10T} {36, e010}

\bibitem[\protect\citeauthoryear{{Turner} et~al.,}{{Turner}
  et~al.}{2021}]{Turner2021}
{Turner} J.~E.,  et~al., 2021, \mn@doi [\apj] {10.3847/1538-4357/abfafe}, \href
  {https://ui.adsabs.harvard.edu/abs/2021ApJ...917...10T} {917, 10}

\bibitem[\protect\citeauthoryear{{Vallisneri}, {Meyers}, {Chatziioannou}  \&
  {Chua}}{{Vallisneri} et~al.}{2023}]{Vallisneri+2023}
{Vallisneri} M.,  {Meyers} P.~M.,  {Chatziioannou} K.,   {Chua} A. J.~K.,
  2023, \mn@doi [\prd] {10.1103/PhysRevD.108.123007}, \href
  {https://ui.adsabs.harvard.edu/abs/2023PhRvD.108l3007V} {108, 123007}

\bibitem[\protect\citeauthoryear{{Valtolina} \& {van Haasteren}}{{Valtolina} \&
  {van Haasteren}}{2024}]{Valtolina2024}
{Valtolina} S.,  {van Haasteren} R.,  2024, \mn@doi [arXiv e-prints]
  {10.48550/arXiv.2412.11894}, \href
  {https://ui.adsabs.harvard.edu/abs/2024arXiv241211894V} {p. arXiv:2412.11894}

\bibitem[\protect\citeauthoryear{{Verbiest} et~al.,}{{Verbiest}
  et~al.}{2016}]{Verbiest+2016}
{Verbiest} J.~P.~W.,  et~al., 2016, \mn@doi [\mnras] {10.1093/mnras/stw347},
  \href {https://ui.adsabs.harvard.edu/abs/2016MNRAS.458.1267V} {458, 1267}

\bibitem[\protect\citeauthoryear{{Vigeland}, {Islo}, {Taylor}  \&
  {Ellis}}{{Vigeland} et~al.}{2018}]{Vigeland+2018}
{Vigeland} S.~J.,  {Islo} K.,  {Taylor} S.~R.,   {Ellis} J.~A.,  2018, \mn@doi
  [\prd] {10.1103/PhysRevD.98.044003}, \href
  {https://ui.adsabs.harvard.edu/abs/2018PhRvD..98d4003V} {98, 044003}

\bibitem[\protect\citeauthoryear{{Volonteri}, {Haardt}  \& {Madau}}{{Volonteri}
  et~al.}{2003}]{Volonteri+2003}
{Volonteri} M.,  {Haardt} F.,   {Madau} P.,  2003, \mn@doi [\apj]
  {10.1086/344675}, \href
  {https://ui.adsabs.harvard.edu/abs/2003ApJ...582..559V} {582, 559}

\bibitem[\protect\citeauthoryear{{You}, {Hobbs}, {Coles}, {Manchester}  \&
  {Han}}{{You} et~al.}{2007}]{You+2007}
{You} X.~P.,  {Hobbs} G.~B.,  {Coles} W.~A.,  {Manchester} R.~N.,   {Han}
  J.~L.,  2007, \mn@doi [\apj] {10.1086/522227}, \href
  {https://ui.adsabs.harvard.edu/abs/2007ApJ...671..907Y} {671, 907}

\bibitem[\protect\citeauthoryear{{Zarka}, {Girard}, {Tagger}  \&
  {Denis}}{{Zarka} et~al.}{2012}]{NenuFAR2012}
{Zarka} P.,  {Girard} J.~N.,  {Tagger} M.,   {Denis} L.,  2012, in {Boissier}
  S.,  {de Laverny} P.,  {Nardetto} N.,  {Samadi} R.,  {Valls-Gabaud} D.,
  {Wozniak} H.,  eds, SF2A-2012: Proceedings of the Annual meeting of the
  French Society of Astronomy and Astrophysics. pp 687--694

\bibitem[\protect\citeauthoryear{{Zhu} et~al.,}{{Zhu} et~al.}{2015}]{Zhu+2015}
{Zhu} W.~W.,  et~al., 2015, \mn@doi [\apj] {10.1088/0004-637X/809/1/41}, \href
  {https://ui.adsabs.harvard.edu/abs/2015ApJ...809...41Z} {809, 41}

\bibitem[\protect\citeauthoryear{{Zic} et~al.,}{{Zic} et~al.}{2022}]{Zic+2022}
{Zic} A.,  et~al., 2022, \mn@doi [\mnras] {10.1093/mnras/stac2100}, \href
  {https://ui.adsabs.harvard.edu/abs/2022MNRAS.516..410Z} {516, 410}

\bibitem[\protect\citeauthoryear{{Zic} et~al.,}{{Zic}
  et~al.}{2023a}]{PPTA2023_data}
{Zic} A.,  et~al., 2023a, \mn@doi [\pasa] {10.1017/pasa.2023.36}, \href
  {https://ui.adsabs.harvard.edu/abs/2023PASA...40...49Z} {40, e049}

\bibitem[\protect\citeauthoryear{{Zic} et~al.,}{{Zic} et~al.}{2023b}]{Zic+2023}
{Zic} A.,  et~al., 2023b, \mn@doi [\pasa] {10.1017/pasa.2023.36}, \href
  {https://ui.adsabs.harvard.edu/abs/2023PASA...40...49Z} {40, e049}

\bibitem[\protect\citeauthoryear{{van Haasteren}}{{van
  Haasteren}}{2024}]{vanHaasteren2024}
{van Haasteren} R.,  2024, \mn@doi [\apjs] {10.3847/1538-4365/ad530f}, \href
  {https://ui.adsabs.harvard.edu/abs/2024ApJS..273...23V} {273, 23}

\bibitem[\protect\citeauthoryear{{van Haasteren}}{{van
  Haasteren}}{2025}]{vanHaasteren2025}
{van Haasteren} R.,  2025, \mn@doi [\mnras] {10.1093/mnrasl/slae108}, \href
  {https://ui.adsabs.harvard.edu/abs/2025MNRAS.537L...1V} {537, L1}

\bibitem[\protect\citeauthoryear{{van Haasteren} \& {Levin}}{{van Haasteren} \&
  {Levin}}{2010}]{vanHaasterenLevin2010}
{van Haasteren} R.,  {Levin} Y.,  2010, \mn@doi [\mnras]
  {10.1111/j.1365-2966.2009.15885.x}, \href
  {https://ui.adsabs.harvard.edu/abs/2010MNRAS.401.2372V} {401, 2372}

\bibitem[\protect\citeauthoryear{{van Haasteren} \& {Levin}}{{van Haasteren} \&
  {Levin}}{2013}]{vanHaasterenLevin2013}
{van Haasteren} R.,  {Levin} Y.,  2013, \mn@doi [\mnras]
  {10.1093/mnras/sts097}, \href
  {https://ui.adsabs.harvard.edu/abs/2013MNRAS.428.1147V} {428, 1147}

\bibitem[\protect\citeauthoryear{{van Haasteren} \& {Vallisneri}}{{van
  Haasteren} \& {Vallisneri}}{2014}]{vanHaasterenVallisneri2014}
{van Haasteren} R.,  {Vallisneri} M.,  2014, \mn@doi [\prd]
  {10.1103/PhysRevD.90.104012}, \href
  {https://ui.adsabs.harvard.edu/abs/2014PhRvD..90j4012V} {90, 104012}

\makeatother
\end{thebibliography}


\appendix
\section{Figures of Merit for other Gravitational Wave searches}
\label{appendix:alt_FoM}

We highlight here some Figures of Merit that could be used if creating a Lite dataset optimized for other types of GW searches than a GWB. For example, a continuous GW from an isolated SMBHB in the white-noise dominated regime will have the FoM \citep{Arzoumanian2014, Mingarelli2017}
\begin{equation}
    \label{eq:FoM_CW}
    \mathrm{FoM}_\mathrm{CGW} = \left(\frac{T_{\mathrm{obs}}c}{\langle\sigma_{\mathrm{TOA}}\rangle^2}\right)^{1/2},
\end{equation}
where we've used $c = 1/\langle\Delta t\rangle$. Here Equation~\ref{eq:FoM_CW} ignores the contribution of the GW antenna beam pattern, as it may be not known a priori. However, for a targeted SMBHB search (e.g., \citealt{Arzoumanian+2020}), the contribution of the antenna beam may be included. For a GW burst with memory, we have \citep{Madison2014, vanHaasterenLevin2010}
\begin{equation}
    \label{eq:FoM_BwM}
    \mathrm{FoM}_\mathrm{BwM} = \left(\frac{T_{\mathrm{obs}}^3c}{\langle\sigma_{\mathrm{TOA}}\rangle^2}\right)^{1/2}.
\end{equation}
$\mathrm{FoM}_\mathrm{BwM}$ and $\mathrm{FoM}_\mathrm{CGW}$ depend more strongly than $\mathrm{FoM}_\mathrm{GWB}$ on $c/\langle\sigma_{\mathrm{TOA}}\rangle^2$, and $\mathrm{FoM}_\mathrm{BwM}$ depends more strongly than $\mathrm{FoM}_\mathrm{CGW}$ on $T_{\mathrm{obs}}$. This suggests different Lite datasets to be curated for different GW signals. Additional FoMs can be constructed for any type of signal affecting PTAs so long as a theoretical S/N can be defined.


\ \\
$^{1}$Department of Physics, Yale University, New Haven, CT 06520, USA \\
$^{2}$ Department of Physics and Astronomy, Widener University, One University Place, Chester, PA 19013, USA \\
$^{3}$ Department of Physics, Oregon State University, Corvallis, OR 97331, USA \\
$^{4}$ Shanghai Astronomical Observatory, Chinese Academy of Sciences, 80 Nandan Road, Shanghai 200030, China \\
$^{5}$ Key Laboratory of Radio Astronomy and Technology (Chinese Academy of Sciences), A20 Datun Road, Chaoyang District, Beijing, 100101, P. R. China \\
$^{6}$ Department of Physics and Astronomy, Vanderbilt University, 2301 Vanderbilt Place, Nashville, Tennessee 37235, USA \\
$^{7}$ Department of Astrophysical and Planetary Sciences, University of Colorado, Boulder, CO 80309, USA \\
$^{8}$ FORTH Institute of Astrophysics, N. Plastira 100, 70013, Heralkion, Greece, Max-Planck Institut für Radioastronomie \\
$^{9}$ Kavli Institute for Astronomy and Astrophysics, Peking University, Beijing 100871, China \\
$^{10}$ ASTRON, Netherlands Institute for Radio Astronomy, Oude Hoogeveensedijk 4, 7991 PD Dwingeloo, The Netherlands \\
$^{11}$ Department of Physics and Synergetic Innovation Center for Quantum Effects and Applications, Hunan Normal University, Changsha, Hunan 410081, China \\
$^{12}$ Institute of Interdisciplinary Studies, Hunan Normal University, Changsha, Hunan 410081, China \\
$^{13}$ LPC2E, OSUC, Univ. Orl\'eans, CNRS, CNES, Observatoire de Paris, Universit\'e PSL, F-45071 Orl\'eans, France \\
$^{14}$ Observatoire Radioastronomique de Nançay, Observatoire de Paris, Universit\'e PSL, Univ Orl\'eans, CNRS, 18330 Nançay, France \\
$^{15}$ The Institute of Mathematical Sciences, C.I.T. Campus, Taramani, Chennai 600113, India \\
$^{16}$ School of Physics and Astronomy, Monash University, Clayton VIC 3800, Australia \\
$^{17}$ Australian Research Council Centre of Excellence for Gravitational Wave Discovery (OzGrav), Clayton VIC 3800, Australia \\
$^{18}$ Australia Telescope National Facility, CSIRO, Space and Astronomy, PO Box 76, Epping, NSW 1710, Australia \\
$^{19}$ Department of Physics, Hillsdale College, 33 E. College Street, Hillsdale, MI 49242, USA \\
$^{20}$ Department of Physics and Astronomy, University of Nigeria, Nsukka 410105, Enugu, Nigeria \\
$^{21}$ Department of Astronomy, University of Maryland, College Park, MD, 20742, USA \\
$^{22}$ Center for Research and Exploration and Space Studies (CRESST), NASA/GSFC, Greenbelt, MD 20771, USA \\
$^{23}$ NASA Goddard Space Flight Center, Greenbelt, MD 20771, USA \\
$^{24}$ Department of Physics and Astronomy, University of Montana, 32 Campus Drive, Missoula, MT 59812 \\
$^{25}$ Max-Planck-Institut f\"ur Radioastronomie, Auf dem H\"ugel 69, 53121 Bonn, Germany \\
$^{26}$ Department of Physics and Astronomy, Macquarie University, Sydney, NSW 2109, Australia \\
$^{27}$ SETI Institute, 339 N Bernardo Ave Suite 200, Mountain View, CA 94043, USA \\
$^{28}$ School of Physics and Astronomy, Rochester Institute of Technology, Rochester, NY 14623, USA \\
$^{29}$ Laboratory for Multiwavelength Astrophysics, Rochester Institute of Technology, Rochester, NY 14623, USA \\
$^{30}$ Jet Propulsion Laboratory, California Institute of Technology, 4800 Oak Grove Drive, Pasadena, CA 91109, USA \\
$^{31}$ Department of Physics and Astronomy, West Virginia University, P.O. Box 6315, Morgantown, WV 26506, USA \\
$^{32}$ Department of Physics, Lafayette College, Easton, PA 18042, USA \\
$^{33}$ Florida Space Institute, University of Central Florida, 12354 Research Parkway, Orlando, FL 32826, USA \\
$^{34}$ IRFU, CEA, Universit\'e Paris-Saclay, F-91191 Gif-sur-Yvette, France \\
$^{35}$ NRAO, 520 Edgemont Rd., Charlottesville, VA 22903 \\
$^{36}$ Centre for Astrophysics and Supercomputing, Swinburne University of Technology, P.O. Box 218, Hawthorn, Victoria 3122, Australia \\
$^{37}$ CSIRO Scientific Computing, Australian Technology Park, Locked Bag 9013, Alexandria, NSW 1435, Australia \\
$^{38}$ Dipartimento di Fisica ``G. Occhialini'', Universit{\'a} degli Studi di Milano-Bicocca, Piazza della Scienza 3, I-20126 Milano, Italy \\
$^{39}$ INAF - Osservatorio Astronomico di Cagliari, via della Scienza 5, 09047 Selargius (CA), Italy \\
$^{40}$ INFN, Sezione di Milano-Bicocca, Piazza della Scienza 3, I-20126 Milano, Italy \\
$^{41}$ European Space Agency (ESA), European Space Research and Technology Centre (ESTEC), Keplerlaan 1, 2201 AZ Noordwijk, the Netherlands \\
$^{42}$ Department of Physics, GLA University, Mathura 281406, India \\
$^{43}$ Institute of Optoelectronic Technology, Lishui University, Lishui, Zhejiang 323000, People's Republic of China \\
$^{44}$ Ruhr University Bochum, Faculty of Physics and Astronomy, Astronomical Institute (AIRUB), 44780 Bochum, Germany \\
$^{45}$ National Astronomical Observatories, Chinese Academy of Sciences, A20 Datun Road, Chaoyang District, Beijing 100101, China

\bsp	
\label{lastpage}
\end{document}